ROYAUME DU MAROC
Université Mohammed V - RABAT - جامعة محمد الخامس - الرباط -
Faculté des sciences
كلية العلوم
CENTRE D'ÉTUDES DOCTORALES - SCIENCES ET TECHNOLOGIES

Order No.: **4103**

# THESIS

*submitted for the award of the degree of:* ***DOCTORATE***

**Research Structure :** High Energy Physics – Modeling and Simulation
**Discipline :** Physics
**Specialty :** Theoretical and Quantum Physics

***Presented and Defended on :*** *June 21, 2025*
***By :***

## *Jamal ELFAKIR*

# *Exploring the Geometric and Dynamical Properties of Spin Systems and Their Interplay with Quantum Entanglement*

***Before the Jury:***

| | | |
|---|---|---|
| Hamid EZ-ZAHRAOUY | PES, Faculty of Sciences, Mohammed V University in Rabat. | Chair / Rapporteur |
| Morad EL BAZ | PES, Faculty of Sciences, Mohammed V University in Rabat. | Rapporteur / Examiner |
| Mohammed EL FALAKI | PES, Faculty of Sciences, Chouaib Doukkali University in El Jadida. | Rapporteur / Examiner |
| Hanane EL HADFI | MCH, Faculty of Sciences, Mohammed V University in Rabat. | Rapporteur / Examiner |
| Abdallah SLAOUI | MC, Faculty of Sciences, Mohammed V University in Rabat. | Invited Member |
| Brahim AMGHAR | MC, Faculty of Sciences, Chouaib Doukkali University in El Jadida. | Co-Supervisor |
| Mohamed DAOUD | PES, Faculty of Sciences, Ibn Tofail University in Kénitra. | Co-Supervisor |
| Rachid AHL LAAMARA | PES, Faculty of Sciences, Mohammed V University in Rabat. | Thesis Supervisor |

**Academic Year : 2024 - 2025**

✉ Faculty of Sciences, Avenue Ibn Battouta, P.O. Box 1014 RP, Rabat – Morocco
☎ 00212 (05) 37 77 18 76 📠 00212(05) 37 77 42 61; http://www. fsr.um5.ac.ma

# Dedication

*To my beloved mother,* ***Habiba Elkajir****,*
*and to my father,* ***Said El Faker****, for their unconditional love, sacrifices, and guidance throughout my life.*

*To my dear sisters:* ***Rachida****,* ***Latifa****,* ***Malika****, and* ***Hanan****, whose constant encouragement has always been a source of strength.*

*To my nephews and nieces:* ***Hiba****,* ***Youness****,* ***Ayman****,* ***Imane****,* ***Adam****,* ***Rayan****,* ***Jannat****,* ***Yassmine****,* ***Aya****, and* ***Fatimazzahra****, who bring joy and hope to my life.*

*To my brothers-in-law:* ***Elhoussine Addaoui****,* ***Omar El Abbouz****, and* ***Ahmed Oukarroum****, for their support and kindness.*

*To my uncles and aunts:* ***Abdelkrim El Faker*** *(May ALLAH have mercy on him),* ***Hassan El Faker****,* ***Lahcen El Faker****,* ***Fatima El Faker****(May ALLAH have mercy on her),* ***Aziz El Faker****,* ***Said Elkajir****,* ***Mohamed Elkajir****,* ***Sadia Elkajir****,* ***Fadma Elkajir****,* ***Lahcen Elkajir****,* ***Ahmed Elkajir****,* ***Mostafa Elkajir****,* ***Brahim Elkajir****, and* ***Abderrahim Elkajir****.*

*Finally, to my childhood friends:* ***Mohammed Aajail****,* ***Elhabib Abahri****, and* ***Rachid Bouago****, whose friendship, loyalty, and support have accompanied me since the earliest days and continue to inspire me.*

*This work is dedicated with gratitude, respect, and love.*

# Acknowledgements

I would like to express my deep gratitude to the **High Energy Physics Laboratory - Modeling and Simulation** for having welcomed me during these years of research. This laboratory has provided a stimulating and enriching scientific environment that has greatly contributed to the quality of my work. I warmly thank all its members for their support, kindness, and the fruitful exchanges we have had. I extend my sincere thanks to Mr. **El Hassan SAIDI**, *PES at the Faculty of Sciences, Mohammed V University in Rabat*, and Mr. **Rachid AHL LAAMARA**, *PES at the Faculty of Sciences, Mohammed V University in Rabat*, for their leadership and for fostering such a dynamic and supportive research environment.

I express my most sincere gratitude to Mr. **Rachid AHL LAAMARA**, *PES at the Faculty of Sciences, Mohammed V University in Rabat*, my thesis supervisor, for his trust, scientific rigor, availability, and insightful advice throughout this journey. His guidance has been essential to the successful completion of this work.

I also thank Mr. **Brahim AMGHAR**, *MC at the Faculty of Sciences, Chouaib Doukkali University in El Jadida*, my co-supervisor, for his valuable insights, the quality of his teaching, and the relevance of his comments and support during this research. His contribution was instrumental in shaping the direction of this thesis.

I am equally grateful to Mr. **Mohamed DAOUD**, *PES at the Faculty of Sciences, Ibn Tofail University in Kenitra*, my co-supervisor, for his attentive supervision, his clear and constructive feedback, and the enriching discussions we had throughout this work.

I would like to respectfully thank Mr. **Hamid EZ-ZAHRAOUY**, *PES at the Faculty of Sciences, Mohammed V University in Rabat*, who honored me by serving as *president* and *reporter* of the thesis defense jury. I am particularly grateful for the interest he showed in my work and for his constructive remarks, which greatly enriched this thesis.

I would like to express my sincere gratitude to Mr. **Morad EL BAZ**, *PES at the Faculty of Sciences, Mohammed V University in Rabat*, for kindly accepting to be a reporter of my thesis. His meticulous review and valuable remarks helped highlight certain subtleties in the theoretical framework and inspired me to further refine key arguments.

I also warmly thank Mr. **Mohamed EL FALAKI**, *PES at the Faculty of Sciences, Chouaib Doukkali University in El Jadida*, for graciously accepting to serve as a reporter of my thesis.

His insightful observations and constructive comments greatly contributed to clarifying several technical aspects and enriching the depth of this manuscript.

My heartfelt thanks go to Ms. **Hanane EL HADFI**, *MCH at the Faculty of Sciences, Mohammed V University in Rabat*, for accepting to be a reporter of my thesis. Her thoughtful feedback and critical reading led to significant improvements in both the structure and clarity of the presentation.

I thank Mr. **Abdallah SLAOUI**, *MC at the Faculty of Sciences, Mohammed V University in Rabat*, for honoring my defense with his presence as a guest. His participation was a true pleasure for me and an additional source of motivation.

I would like to thank all the jury members who agreed to take part in the examination of this thesis, for their questions, critiques, and suggestions. Their expertise contributed to making the defense a moment rich in scientific exchange.

From the bottom of my heart, I wish to thank my family for their unwavering support, patience, and love throughout this journey. Without them, none of this would have been possible.

I will not forget my friends and colleagues, for their presence, their attentive listening, and their encouragement during moments of doubt as well as in times of joy. Thank you to all those who, directly or indirectly, contributed to this adventure.

# Résumé

La présente thèse intitulée "Exploration des propriétés géométriques et dynamiques des systèmes de spin et de leur interaction avec l'intrication quantique", vise à explorer l'intrication et l'évolution quantique à travers une double lecture : géométrique et dynamique. Le premier volet traite de l'espace des phases classique et de son rôle central en mécanique hamiltonienne, en mettant en valeur l'importance des structures symplectiques dans la description des états mécaniques. L'étude met en évidence l'analogie formelle entre l'espace des phases classiques et l'espace de Hilbert en mécanique quantique. Le second volet est consacré à la description géométrique des états quantiques à travers la structure projective de l'espace de Hilbert. L'accent est mis sur l'interprétation géométrique de l'évolution quantique, notamment via la métrique de Fubini-Study, les structures symplectiques associées, ainsi que la phase géométrique acquise lors des évolutions unitaires. Les deux derniers volets sont consacrés à l'étude des systèmes de spins (à deux et plusieurs particules), sous différents modèles d'interaction (Heisenberg anisotropique et Ising), où l'on analyse à la fois la dynamique (vitesse d'évolution, intrication, problème du brachistochrone quantique) et les structures géométriques et topologiques associées aux états du système.

# Abstract

This thesis, entitled Exploring the Geometric and Dynamical Properties of Spin Systems and Their Interplay with Quantum Entanglement, explores the quantum entanglement and evolution through both a geometric and dynamical perspective. The first part focuses on classical phase space and its central role in Hamiltonian mechanics, emphasizing the importance of symplectic structures in describing mechanical states. The study highlights the formal analogy between classical phase space and the Hilbert space used in quantum mechanics. The second part is devoted to the geometric description of quantum states through the projective structure of Hilbert space. Emphasis is placed on the geometric interpretation of quantum evolution, particularly via the Fubini-Study metric, associated symplectic structures, and the geometric phase acquired during unitary evolutions. The final two parts are dedicated to the study of spin systems (both two-body and many-body) under different interaction models (XXZ Heisenberg and all-range Ising). Both the dynamical aspects (evolution speed, entanglement, and the quantum brachistochrone problem) and the geometric and topological structures of the corresponding quantum states are analyzed.

# Résumé détaillé

Cette thèse s'inscrit dans le cadre général de la mécanique quantique et de ses interprétations géométriques et topologiques, en mettant particulièrement l'accent sur l'étude des systèmes de spins. L'objectif principal est de comprendre en profondeur l'évolution des systèmes quantiques à travers une double approche : d'une part, en comparant les espaces d'états classiques et quantiques afin de révéler leurs similitudes et leurs différences fondamentales ; et d'autre part, en développant et en appliquant des outils géométriques et topologiques pour caractériser la dynamique, la vitesse d'évolution et l'intrication quantique. Ce travail se situe ainsi à l'interface entre la mécanique classique, la physique quantique, et la géométrie différentielle moderne, en cherchant à établir un langage unifié pour décrire des phénomènes complexes.

L'étude débute par l'introduction de l'espace des phases classique, cadre naturel de la mécanique hamiltonienne. Dans cet espace, les coordonnées généralisées et leurs moments conjugués définissent une variété symplectique différentiable où chaque point représente un état possible du système mécanique. La structure symplectique, à travers la forme canonique et les crochets de Poisson, organise la dynamique et met en évidence la conservation de grandeurs physiques, tandis que les transformations canoniques révèlent les symétries du système. Ce formalisme géométrique permet de visualiser l'évolution des systèmes classiques sous forme de trajectoires dans l'espace des phases, tout en soulignant l'importance des invariants dynamiques. L'exemple d'oscillateurs couplés illustre la pertinence de ce cadre pour représenter les phénomènes collectifs et dégager des structures universelles.

Le passage à la mécanique quantique introduit un changement conceptuel majeur : l'espace de Hilbert devient l'arène mathématique où évoluent les états quantiques. Contrairement à l'espace de phase classique, l'espace de Hilbert est linéaire et infini-dimensionnel dans les cas généraux. Les vecteurs d'état, définis à un facteur de phase globale près, représentent les états physiques possibles. Cette redondance impose de considérer l'espace projectif de Hilbert, où chaque état est défini par une classe d'équivalence de vecteurs différant uniquement par une phase globale. Ce passage de $\mathcal{H}$ à $\mathcal{PH}$ révèle des structures géométriques riches : une structure complexe naturelle, une métrique riemannienne donnée par la distance de FubiniStudy, et une structure symplectique héritée du produit scalaire hermitien. Ces outils permettent de définir la notion de distance entre états, de vitesse d'évolution et de phases géométriques, en établissant un lien profond entre la géométrie de l'espace des états et la dynamique quantique.

Dans ce cadre, la phase géométrique apparaît comme une manifestation remarquable. Lorsqu'un système quantique effectue une évolution cyclique, l'état accumule non seulement une phase dynamique, dépendante de l'énergie et du temps d'évolution, mais également une phase purement géométrique, liée à la courbure de l'espace projectif parcouru. Ce phénomène, connu sous le nom de phase de Berry dans le cas adiabatique, a été généralisé aux régimes non adiabatiques et aux évolutions non cycliques. Il constitue un pont entre la géométrie différentielle et la physique quantique, tout en ayant des implications pratiques en métrologie quantique, en information quantique et dans l'étude des transitions de phase topologiques. L'examen des courbures et des invariants topologiques associés à l'espace des états met en lumière des propriétés universelles et invariantes, qui enrichissent notre compréhension des phénomènes quantiques collectifs.

Ces concepts sont ensuite appliqués de manière concrète à l'étude de systèmes de spins, qui représentent un terrain d'exploration privilégié. Dans le cas de deux spins couplés par une interaction de Heisenberg anisotrope, il est possible d'analyser explicitement l'évolution unitaire, les limites de vitesse quantique et la dynamique de l'intrication. L'équation de Schrödinger, combinée au formalisme géométrique, révèle que la vitesse d'évolution est directement liée aux incertitudes énergétiques du système, en accord avec les principes de la limite quantique de MandelstamTamm. La résolution du problème de la brachistochrone quantique, consistant à déterminer le temps minimal d'évolution entre deux états donnés, illustre la complémentarité entre analyse dynamique et interprétation géométrique. Ces résultats montrent que la manipulation de paramètres tels que l'anisotropie de l'interaction ou l'intensité du couplage peut réduire le temps requis pour réaliser une transformation quantique donnée.

L'analyse est ensuite généralisée à des systèmes à plusieurs corps. Le modèle d'Ising à interaction de portée infinie est étudié comme exemple de système collectif où chaque spin interagit avec tous les autres. Dans ce contexte, l'évolution temporelle, la dynamique de l'intrication et les vitesses quantiques sont explorées à travers une approche analytique et géométrique. L'étude est ensuite étendue aux systèmes de spin-$s$, mettant en évidence comment l'augmentation de la dimension locale influence la complexité des trajectoires dans l'espace projectif, ainsi que la structure des corrélations quantiques. Ces résultats confirment le rôle déterminant des symétries internes et de la nature des interactions dans la dynamique globale des systèmes de spins, tout en offrant des perspectives pour la simulation quantique et le contrôle de l'intrication.

Une analyse complémentaire s'attache à la dimension géométrique et topologique des états collectifs de spins. Les phases géométriques globales, la courbure de l'espace projectif et les invariants topologiques associés permettent d'interpréter l'intrication et la cohérence quantique dans un langage géométrique unifié. L'étude met en évidence que les corrélations quantiques peuvent être représentées par des structures géométriques bien définies, offrant ainsi une classification et une compréhension plus profondes des comportements collectifs. Ces résultats démontrent que les concepts de la géométrie différentielle et de la topologie ne sont pas de simples outils mathématiques, mais des éléments intrinsèques à la description des phénomènes quantiques.

L'ensemble de ce travail contribue à consolider l'idée que la mécanique quantique, loin de se réduire à une formulation abstraite, peut être comprise et enrichie par une approche géométrique et topologique. Les notions de métrique de Fubini-Study, de phase géométrique, de vitesse d'évolution et de courbure constituent des éléments essentiels pour décrire et analyser les systèmes de spins. Les résultats obtenus ouvrent de nouvelles perspectives pour le développement de l'information quantique, notamment en ce qui concerne la réalisation d'opérations optimales, la caractérisation géométrique de l'intrication et la compréhension des transitions de phase quantiques. Plus largement, ils témoignent du potentiel considérable qu'offre l'approche géométrique pour unifier des concepts classiques et quantiques, et pour éclairer les dynamiques complexes des systèmes quantiques à plusieurs corps.

# List of Publications

This thesis is based in part on the following published articles:

- Jamal Elfakir, Brahim Amghar, Abdallah Slaoui, Mohammed Daoud, *Complementarity between quantum entanglement and geometric and dynamical appearances in N spin-1/2 system under all-range Ising model*, The European Physical Journal Plus **139**(8), 1–17 (2024).
- Brahim Amghar, Abdallah Slaoui, Jamal Elfakir, Mohammed Daoud, *Geometrical, topological, and dynamical description of N interacting spin-s particles in a long-range Ising model and their interplay with quantum entanglement*, Physical Review A **107**(3), 032402 (2023).
- Jamal Elfakir, Brahim Amghar, Mohammed Daoud, *Geometrical and dynamical description of two interacting spins under the XXZ-type Heisenberg model*, International Journal of Geometric Methods in Modern Physics **20**(01), 2350006 (2023).

# List of Figures

# Contents

# General Introduction

Quantum mechanics emerged as a revolutionary shift in our understanding of the microscopic world, driven by the inability of classical physics to account for certain experimental observations. Pioneering work by W. Heisenberg, E. Schrödinger, and others in the early $20^{\text{th}}$ century laid the groundwork for this new theoretical framework [1]. Among the phenomena that catalyzed this transition were the discrete spectral lines observed as early as the late $18^{\text{th}}$ century [2], the photoelectric effect discovered by H. R. Hertz in 1886 [3], and the problem of blackbody radiation [4], which classical physics could not adequately explain. These anomalies spurred a reconsideration of fundamental assumptions about the nature of energy, matter, and light, culminating in a conceptual revolution that reshaped the foundations of physical theory.

At its core, quantum mechanics provides a consistent and comprehensive description of the behavior and evolution of the microscopic constituents of matter. Key features include quantum entanglement, the superposition principle, and interference phenomena effects with no counterpart in classical theory. These effects become especially pronounced at scales where the uncertainty in position $\Delta x$ and momentum $\Delta p$ cannot be simultaneously reduced beyond a certain limit. This is formalized by the Heisenberg uncertainty principle, expressed as $\Delta x \Delta p \geq \frac{\hbar}{2}$ [5], a relation that encapsulates the intrinsic limitations in measuring complementary observables simultaneously. This limitation is not due to experimental imperfections, but instead reflects a fundamental trait of quantum systems, marking a profound departure from the deterministic predictability of classical mechanics.

The novelty of quantum theory lies not only in its predictive power but also in its conceptual foundations. The superposition principle, quantum entanglement, and Bohr's notion of complementarity radically challenge classical intuitions, calling into question the very nature of physical reality. While classical mechanics describes the evolution of particles through definite trajectories in phase space [6, 7], quantum mechanics instead represents states as vectors in an abstract Hilbert space, where evolution proceeds through unitary operations [8]. This framework discards the notion of objective trajectories and replaces them with probabilistic amplitude distributions, encoded in wavefunctions or state vectors. As a consequence, physical observables correspond to Hermitian operators whose eigenvalues represent measurable outcomes, and measurements themselves influence the system's state in a fundamentally indeterministic fashion.

Geometry has long been instrumental in shaping physical theories. For instance, *Einstein's general relativity* interprets gravity through the curvature of spacetime [9], and modern gauge theories express fundamental interactions via geometric connections and fiber bundles [10]. Yet, the geometric content of quantum mechanics remained less apparent until the seminal work of M. V. Berry. His discovery of what is now called the *geometric phase*, commonly referred to as the Berry phase, demonstrated that, during a cyclic adiabatic evolution, quantum systems accumulate a phase factor that depends only on the shape of the evolution trajectory, and not on how quickly the path is followed [11,12]. This insight uncovered a rich interplay between geometry and quantum dynamics, emphasizing the significance of the structural properties of the quantum state space.

The geometric formulation of quantum mechanics provides an effective framework to examine both its conceptual foundations and practical applications. Central to this approach is the structure of Hamiltonian mechanics, which not only governs classical systems but also forms the mathematical basis for quantum dynamics. In classical mechanics, a system's state evolves within a symplectic manifold -phase space- where dynamics are governed by Hamilton's equations and symplectic geometry [13,14]. This elegant structure extends naturally into quantum theory, where the classical notion of phase space is substituted by a higher-dimensional Hilbert space, and observables are represented as operators on that space [15]. Even with this abstraction, symplectic geometry maintains a central role by shaping the framework of quantum transformations and providing a geometric understanding of processes such as time evolution, transitions between states, and measurement outcomes.

Symplectic geometry remains central in this quantum setting, underpinning the evolution of quantum states and enabling a geometric interpretation of measurement, entanglement, and coherence. In particular, when quantum states are viewed as rays in Hilbert space, they form a complex projective manifold known as the projective Hilbert space [16]. This space inherits both a symplectic structure and a Riemannian structure, encapsulated by the FubiniStudy metric and a compatible symplectic form [17,18]. These geometric tools provide a natural language to characterize the ability to distinguish between quantum states, to analyze the curvature associated with quantum evolution, and to interpret how geometric features influence physical quantities. For example, the Fubini-Study metric quantifies the distance between pure quantum states and thereby reflects their potential to interfere, which plays a key role in quantum information processing and coherent control.

One of the most significant implications of this geometric framework is the appearance of geometric phases, such as the Berry phase, which manifest when a quantum system follows a closed trajectory in its parameter space. These phases, determined exclusively by the shape and structure of the path taken in state space, produce measurable effects in interference setups and quantum control schemes. Unlike dynamical phases, which depend on the specific timing of the evolution, geometric phases retain information about the global evolution of the system and are therefore well-suited for applications demanding stability and coherence. The extension of Berry's result to include non-adiabatic, non-cyclic, and mixed-state evolutions has contributed to a deeper understanding of quantum holonomies and the geometric aspects that underlie contemporary quantum theory.

The implications of this geometric viewpoint extend far beyond conceptual elegance. In modern quantum technologies, geometric and topological features are harnessed for practical purposes. For example, geometric-phase-based gates have gained attention as a powerful tool in the realization of robust quantum logic operations [19]. Because these gates rely on global geometric properties, they exhibit a degree of immunity to specific classes of noise and imperfections, offering intrinsic fault-tolerance [20, 21]. Furthermore, topological methods play a pivotal role within the frameworks of quantum error correction and in shaping fault-tolerant quantum computing architectures [22–24]. Here, quantum information is encoded in global topological features of many-body states, which are inherently protected from local perturbations, enabling a new paradigm of quantum computation that is both theoretically profound and practically viable.

The study of geometric and topological aspects of quantum systems has thus become a major focus in current research, influencing a variety of domains including quantum computation, condensed matter physics, and quantum information science. Whether addressing the entanglement structure of multipartite systems, exploring quantum speed bounds, or formulating quantum control strategies, geometry offers a unifying framework for describing, analyzing, and steering quantum behavior. By situating the formal structure of quantum theory within a geometric setting, researchers are able to gain more profound insights into quantum phenomena and to develop methods that enrich both theoretical foundations and the advancement of quantum technologies.

This thesis consists of four core chapters, each contributing step by step to the construction of a coherent understanding of how geometry and quantum dynamics intertwine, particularly within spin systems.

**The first chapter** lays the foundational groundwork by introducing the structure of classical and quantum state spaces. It opens with a presentation of classical phase space, emphasizing its symplectic nature and central role in Hamiltonian mechanics, followed by a discussion on canonical transformations and symmetries. An illustrative example involving coupled pendulums underscores the geometric formulation of classical dynamics. The focus then transitions toward quantum mechanics, where the Hilbert space formalism is developed together with the concepts of unitary evolution and degrees of freedom in quantum systems. A spin system is employed as a concrete example to ground these abstract concepts. The chapter concludes with a detailed comparison between classical and quantum state spaces, emphasizing analogies such as observables correspondence, the role of Poisson brackets versus commutators, and their geometric formulations, particularly through symplectic geometry.

**Chapter two** delves into the geometric structures inherent to quantum mechanics. It begins with an analysis of projective Hilbert space, presenting its complex structure, Riemannian formu-lation (via the FubiniStudy metric), and symplectic properties. These geometric tools are shown to govern the kinematics and dynamics of quantum evolution. The notion of geometric phase is explored through both adiabatic and non-adiabatic cyclic evolutions, along with general cases, linking phase accumulation to the geometry of quantum state trajectories. The chapter further examines curvature and topological features of quantum geometries and their implications, before culminating in applications such as the time-optimal evolution problem and the geometric representation of quantum entanglement. These applications highlight how geometric intuition can reveal hidden structures within quantum dynamics.

***The third chapter*** transitions to a dynamical study of spin systems. Starting with a general overview of spin models, it focuses on the evolution of a system consisting of two coupled spins described by the anisotropic Heisenberg interaction. Key aspects such as unitary evolution, quantum speed limits, and the brachistochrone problem are investigated, alongside the time-dependent behavior of entanglement. The analysis then extends to many-body systemsfirst considering $N$ interacting spin-$\frac{1}{2}$ particles under an all-range Ising model, and then generalizing to spin-$s$ systems. For each case, the dynamical evolution, quantum speed limits, and entanglement dynamics are thoroughly studied. These investigations provide deep insights into how interactions and internal symmetries influence the evolution and correlation properties of quantum systems.

***In the final chapter***, the focus shifts to a geometrical and topological analysis of the same spin systems. Starting again with a pair of coupled spins evolving under the anisotropic Heisenberg interaction, the study explores the geometric structure of their joint quantum state, the phases arising from their evolution, and the geometric expression of quantum entanglement. This approach is then extended to $N$-particle systems with spin-$\frac{1}{2}$ under a uniform long-range Ising-type interaction, revealing how collective geometric phases and entanglement patterns emerge from global interactions. Finally, for general $N$-particle spin-$s$ systems, a thorough investigation is carried out into the underlying geometric and topological structures, phase evolution under cyclic and non-cyclic dynamics, and how entanglement is encoded geometrically. These analyses demonstrate how geometric and topological concepts offer a powerful language to understand and classify quantum correlations and dynamical behavior in many-body systems.

CHAPTER 1

# From Classical to Quantum States Space

## 1.1 Classical states space: *phase space*

### 1.1.1 Definition and properties

We will initially examine the classical side of mechanics before constructing its geometrical framework on the quantum side. Phase space is a fundamental tool in classical mechanics [25], allowing all possible states of a physical system to be represented by a single point in a multi-dimensional space. This geometric approach plays an essential role in understanding dynamical systems [13], particularly in visualizing particle trajectories and analyzing phenomena such as phase transitions, chaos [26], and energy conservation. In classical mechanics, phase space is a conceptual space where each point specifies the full state of the system, characterized by a set of generalized coordinates $\mathbf{x} \equiv (x_1, x_2, ..., x_N)$ and their conjugate moments $\mathbf{p} \equiv (p_1, p_2, ..., p_N)$. These pairs $(\mathbf{x}, \mathbf{p})$ define what is known as a phase point, and the collection of such points constitutes the phase space. The dimension of this space is $2N$, with $N$ denoting the number of degrees of freedom in the system.

Phase space provides an intuitive and powerful representation of how mechanical systems behave over time. The trajectories traced by points in phase space offer a clear depiction of how the states of a system evolve, starting from a specific initial condition. This framework proves useful in several respects. First, it enables a direct visualization of dynamical paths, emphasizing how systems change under mechanical influences. In addition, thanks to Liouville's theorem, it supports the study of conserved quantities, particularly the invariance of the volume occupied in phase space for isolated systems, which reflects the persistence of dynamic flows. Finally, when studying phase transitions or chaotic behavior, phase space exposes intricate trajectories that, although not always predictable, remain confined within defined boundariesthus deepening our insight into nonlinear dynamics. One essential aspect of this formulation is the introduction of generalized coordinates. How are these coordinates defined? For a system of $n$ particles, modeled as point masses, the description requires $3n$ coordinates. However, the existence of contact or binding forces between components induces relations among the coordinates due to constraint equations. These forces are often difficult to model, and even when a model is possible, their explicit values may remain undetermined. This complexity motivated the elimination of unknown forces in favor of a set of coordinates that fully capture the unconstrained aspects of motion. These are called generalized coordinates, which may take the form of lengths, angles, or other quantities depending on the system. When constraints are incorporated, they provide a tailored description of the mechanical configuration. To characterize the mechanical state of a system, the number of generalized coordinates needed equals the number of degrees of freedom. Each degree of freedom corresponds to an independent mode of motion after accounting for constraints. If the system is described by $n$ coordinates and subject to $k$ constraints, the number of degrees of freedom is given by $N = n - k$. Hence, the choice and count of generalized coordinates are determined by the specific setup and limitations inherent to the system.

### 1.1.2 Hamiltonian Mechanics

To better comprehend the subject matter of this thesis, we begin by exploring the Hamiltonian formalism of classical mechanics. For a solid foundational understanding of this framework, it is helpful to briefly revisit the Lagrangian formulation, which is based on a set of generalized coordinates and their time derivatives, represented by the pair $(x_k, \dot{x}_k)$, where $k = 1, \ldots, N$, corresponding to a system with $N$ degrees of freedom. Each mechanical system is described by a function that depends on the generalized coordinates, velocities, and time, known as the Lagrangian, defined as the difference between the system's kinetic and potential energies:

$$\mathcal{L}(x_k, \dot{x}_k, t) = \mathbf{T} - \mathbf{V}, \tag{1.1}$$

where $\mathbf{T}$ denotes the kinetic energy and $\mathbf{V}$ denotes the potential energy. As the system evolves between two time instants $t_1$ and $t_2$, the principle of least action asserts that the trajectory taken is the one that minimizes the action, expressed as

$$\mathcal{S} = \int_{t_1}^{t_2} \mathcal{L}(x_k, \dot{x}_k, t) dt, \tag{1.2}$$

is minimal. In accordance with the principle of least action, the variation $\delta\mathcal{S}$ of the action $\mathcal{S}$ during the system's evolution between instants $t_1$ and $t_2$, as it passes from coordinates $\mathbf{x}^A$ to $\mathbf{x}^B$ (see Figure 1.1), must remain zero for the entire duration of the evolution, i.e

$$\delta\mathcal{S} = 0 \tag{1.3}$$

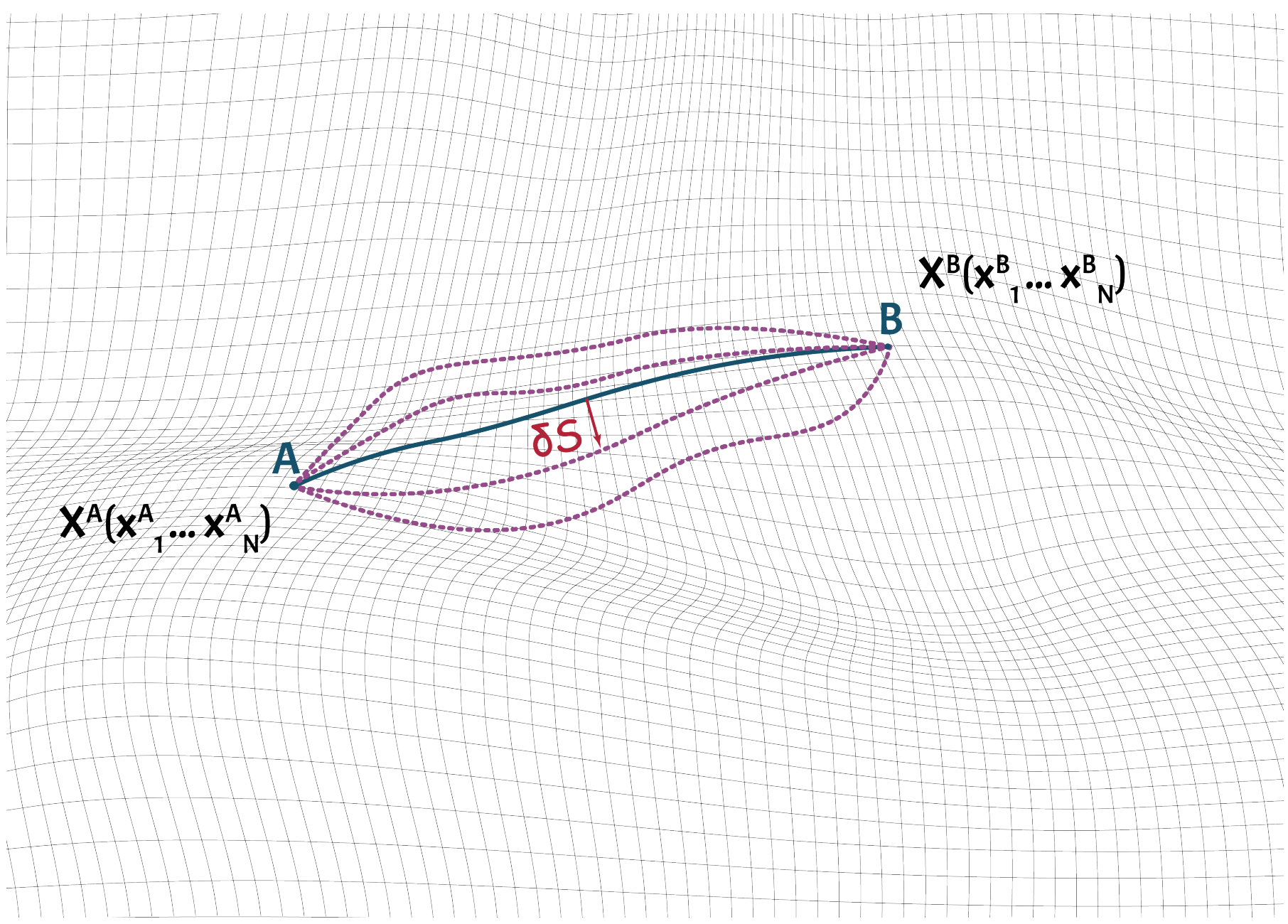

Figure 1.1: The system follows the minimum-action trajectory.

expanding the equation (1.2),

$$\begin{aligned}\delta\mathcal{S} &= \int_{t_1}^{t_2} \left(\mathcal{L}(x_k+\delta x_k, \dot{x}_k+\delta\dot{x}_k, t) - \mathcal{L}(x_k, \dot{x}_k, t)\right) dt \\ &= \sum_i \int_{t_1}^{t_2} \left(\frac{\partial\mathcal{L}}{\partial x_k}\delta x_k + \frac{\partial\mathcal{L}}{\partial\dot{x}_k}\frac{d}{dt}\delta x_k\right) dt\end{aligned}$$

By employing the term in conjunction with the factor $d/dt(\delta\dot{\mathbf{x}})$ and subsequently integrating it by parts, one arrives to the following equation

$$\delta\mathcal{S} = \sum_k \int_{t_1}^{t_2} \left(\frac{\partial\mathcal{L}}{\partial x_k} - \frac{d}{dt}\frac{\partial\mathcal{L}}{\partial\dot{x}_k}\right)\delta x_k\, dt + \left[\frac{\partial\mathcal{L}}{\partial\dot{x}_k}\delta x_k\right]_{t_1}^{t_2}. \tag{1.4}$$

Since all possible trajectories of the system are identical at the initial point $\mathbf{x}(1)$ and the final point $\mathbf{x}(1)$, we can conclude that the infinitesimal changes in the generalized coordinates, represented by $\delta\mathbf{x}(1)$ and $\delta\mathbf{x}(2)$, are null. Consequently, this leads us to

$$\left[\frac{\partial\mathcal{L}}{\partial\dot{x}_k}\delta x_k\right]_{t_1}^{t_2} = 0. \tag{1.5}$$

Given that the generalized coordinates are all independent, the remaining term is null only if the equations

$$\frac{\partial\mathcal{L}}{\partial x_k} - \frac{d}{dt}\frac{\partial\mathcal{L}}{\partial\dot{x}_k} = 0 \tag{1.6}$$

are satisfied. The resulting are $N$ independent equations, describing system motion or evolution, corresponding to the number of its degrees of freedom. In terms of the Lagrange function, generalized coordinates, and generalized velocities, their form is as follows

$$\frac{\partial \mathcal{L}}{\partial x_k} = \frac{d}{dt}\frac{\partial \mathcal{L}}{\partial \dot{x}_k} \tag{1.7}$$

where $(k = 1, ..., N)$, these are second-order differential equations whose solution determines the dynamics of the considered system.

Next, let's reformulate the equations of motion obtained from the Lagrange formalism, using a transformation known as Legendre transformation. Thanks to this transformation, the generalized velocities $\dot{x}_k$ are replaced by a new variables $p_k$, and the Lagrange function $\mathcal{L}(x_k, \dot{x}_k, t)$ by a new function $\mathbf{H}(x_k, p_k, t)$ as illustrated below

$$\begin{cases} \dot{x}_k \to p_k \\ \\ \mathcal{L}(x_k, \dot{x}_k, t) \to \mathbf{H}(x_k, p_k, t) = \sum_{k=1}^{N} \dot{x}_k p_k - \mathcal{L}(x_k, \dot{x}_k, t) \end{cases} \tag{1.8}$$

where $p_k$ denotes the conjugate moment of the generalized coordinate $x_k$, while $\mathbf{H}$ represents the system's Hamiltonian. This quantity encapsulates the total mechanical energy of the system, comprising the sum of kinetic and potential energy. Thus, we have moved from Lagrange's to Hamilton's formulation of mechanics. The latter is based on generalized coordinates and their conjugate moments instead of their first derivatives with respect to time, which will subsequently facilitate the transition to quantum mechanics. Let's obtain the equations of motion in terms of the new quantities that result from the aforementioned transformation, for this purpose, we consider the following expression for the total derivative of Hamilton's function with respect to time:

$$\frac{d\mathbf{H}}{dt} = \sum_k \left( \frac{\partial \mathbf{H}}{\partial x_k}\dot{x}_k + \frac{\partial \mathbf{H}}{\partial p_k}\dot{p}_k \right) + \frac{\partial \mathbf{H}}{\partial t}. \tag{1.9}$$

Otherwise, the total derivative of $\mathbf{H}$ with respect to time can be calculated from its expression obtained by the transformation (1.8), and provided as follows

$$\begin{aligned} \frac{d\mathbf{H}}{dt} &= \sum_k \left( p_k\dot{x}_k + p_k\ddot{x}_k - \frac{\partial \mathcal{L}_k}{\partial x_k}\dot{x}_k - \frac{\partial}{\partial x_k}\left(\mathcal{L}_k \dot{x}_k\right)\ddot{x}_k \right) - \frac{\partial \mathcal{L}}{\partial t} \\ &= \sum_k \left( -\frac{\partial \mathcal{L}_k}{\partial x_k}\dot{x}_k + \dot{p}_k\dot{x}_k \right) - \frac{\partial \mathcal{L}}{\partial t}. \end{aligned} \tag{1.10}$$

A comparison of the two equations term by term reveals the following relations

$$\dot{x}_k = \frac{\partial \mathbf{H}}{\partial p_k}, \quad \dot{p}_k = -\frac{\partial \mathbf{H}}{\partial q_k}. \tag{1.11}$$

This set of equations is known as Hamilton's canonical equations. Consequently, the dynamics of the system are now governed by $2N$ first-order differential equations, which represent a significant mathematical simplification in comparison to the $N$ second-order differential equations present in the Lagrange equations. In contrast to Lagrange's methodology, Hamilton's characterization of the system's evolution is founded upon first-order differential equations, which offers several benefits. The Hamiltonian approach allows for a more direct analysis of the system's dynamic properties and facilitates the conservation of physical quantities. Furthermore, it offers a more organic foundation for the transition to quantum mechanics, where dynamical observables are represented by operators operating within an extended phase space. This renders the Hamiltonian framework especially efficacious for the study of complex systems and the establishment of connections between classical and quantum mechanics.

### 1.1.3 Transformations, symmetries and symplectic geometry

The principle of least action, also known as the variational principle, serves as the foundation of Lagrange's formalism, which characterizes the dynamics of a system through differential equations. Consequently, a system with $n$ degrees of freedom is governed by $n$ second-order differential equations, the solution of which enables a complete determination of the system's dynamical behavior. Hamilton's framework, in contrast, reformulates the problem into $2n$ first-order differential equations and introduces the notion of phase space. The state of the system at a specific time $t$ is then represented as a point in this phase space. This formulation underlies the development of Hamilton-Jacobi theory. The idea is as follows: suppose two configurations of a system exist at times $t$ and $t'$, respectively. These configurations correspond to two points in phase space, denoted $(\mathbf{x}, \mathbf{p})$ and $(\mathbf{X}, \mathbf{P})$. The transition from one to the other can be analyzed either by solving differential equations or more directly by identifying a function that transforms the initial point $(\mathbf{x}, \mathbf{p})$ into the final point $(\mathbf{X}, \mathbf{P})$. In this section, we will explore this second strategy. As already indicated, our objective is to identify a transformation that maps one point in phase space to another. The most straightforward type of transformation is the so-called point transformation $(\mathbf{x} \to \mathbf{X}(x_k, t))$, such as switching from Cartesian to cylindrical or spherical coordinates. We now turn to more general transformations, known as contact transformations, in which $\mathbf{x}$ and $\mathbf{p}$ are treated symmetrically. These transformations are expressed as follows

$$\begin{cases} x_k \to X_k(x_i, p_i, t) \\ \\ p_k \to P_k(x_i, p_i, t) \end{cases} \tag{1.12}$$

So we obtain a new point with coordinates $(\mathbf{X}, \mathbf{P})$, and if the following equations

$$\dot{X}_k = \frac{\partial \mathbf{H}'}{\partial P_k}, \quad \dot{P}_k = -\frac{\partial \mathbf{H}'}{\partial X_k} \tag{1.13}$$

are satisfied,then the transformations (1.12) are called canonical [27], where $\mathbf{H}'$ represents the Hamiltonian function expressed in terms of the new variables and describes the same system under study in a consistent manner. Consequently, it must obey the principle of least action

$$\begin{aligned}\delta\mathcal{S} &= \delta\int_{t_1}^{t_2}\left(\sum_{k=1}^{N}p_k\dot{x}_k-\mathbf{H}\right)dt\\ &= \delta\int_{t_1}^{t_2}\left(\sum_{k=1}^{N}P_k\dot{X}_k-\mathbf{H}'\right)dt\end{aligned}. \tag{1.14}$$

It can be concluded that the two quantities which are the subject of integration in the initial and second lines differ only by a derivative with respect to time. Based on this, we can formulate a more precise definition than the previous one, as follows: A canonical transformation is any contact transformation that satisfies the following equation

$$\mathbf{H}'=\mathbf{H}+\sum_k\left(P_k\dot{X}_k-p_k\dot{x}_k\right)+\frac{d\mathbf{g}}{dt} \tag{1.15}$$

where $\mathbf{g}$ is called the generating function of the transformation. In the case of a system with one degree of freedom ($N=1$), four configurations of the generating function $\mathbf{g}$ are possible. The first of these is designated $\mathbf{g}_1(x,X,t)$ , and its associated equation is as follows

$$\left(\mathbf{H}'-\mathbf{H}-\frac{\partial\mathbf{g}_1}{\partial t}\right)dt=\left(\frac{\partial\mathbf{g}}{\partial x}-p\right)dx+\left(\frac{\partial\mathbf{g}}{\partial X}+P\right)dX, \tag{1.16}$$

which results in

$$\begin{cases} p=\dfrac{\partial\mathbf{g}_1}{\partial x}\\ \\ P=-\dfrac{\partial\mathbf{g}_1}{\partial X}\end{cases} \tag{1.17}$$

Therefore, as with the preceding approaches, $\mathbf{g}_1(x,X,t)$ and the last two equations fully determine the system dynamics with the new variables. Moreover, if $\mathbf{g}_1(x,X)$ is not explicitly time-dependent, then $\mathbf{H}'=\mathbf{H}$. The other three possible configurations for $\mathbf{g}$ are $\mathbf{g}_2(x,P,t)$ , $\mathbf{g}_3(p,X,t)$ and $\mathbf{g}_4(p,P,t)$ , and can be studied in a similar way to the previous case. However, in some configurations another variable may appear that depends on the two independent variables for the chosen function $\mathbf{g}$. In this case, a Legendre transformation is used to obtain an equation that contains only the two independent variables. This analysis can be generalized to systems with more degrees of freedom($N>1$). Having completed an investigation of phase space and developed an understanding of how to transition between its points, we now turn our attention to a deeper examination of its intrinsic characteristics. This will enable us to identify how the evolution of a system manifests within this conceptual framework. To initiate this analysis, we will define a quantity known as Poisson brackets, which is expressed between two functions $\mathbf{f}(x,p,t)$ and $\mathbf{h}(x,p,t)$ defined on phase space, by the following relation

$$\{\mathbf{f},\mathbf{h}\}=\sum_{k=1}^{N}\left(\frac{\partial\mathbf{f}}{\partial x_k}\frac{\partial\mathbf{h}}{\partial p_k}-\frac{\partial\mathbf{f}}{\partial p_i}\frac{\partial\mathbf{h}}{\partial x_k}\right). \tag{1.18}$$

It is a characteristic form of the phase space, with the following properties:

$$\begin{aligned}
&\{\mathbf{f},\mathbf{h}\} = -\{\mathbf{h},\mathbf{f}\} \\
&\{\mathbf{f},c\} = 0 \quad \text{(where } c \text{ is a constant)} \\
&\{\mathbf{f}_1+\mathbf{f}_2,\mathbf{h}\} = \{\mathbf{f}_1,\mathbf{h}\} + \{\mathbf{f}_2,\mathbf{h}\} \\
&\{\mathbf{f}_1\mathbf{f}_2,\mathbf{h}\} = \mathbf{f}_1\{\mathbf{f}_2,\mathbf{h}\} + \{\mathbf{f}_1,\mathbf{h}\}\mathbf{f}_2 \\
&\tfrac{\partial}{\partial t}\{\mathbf{f},\mathbf{h}\} = \left\{\tfrac{\partial \mathbf{f}}{\partial t},\mathbf{h}\right\} + \{\mathbf{f},\tfrac{\partial \mathbf{h}}{\partial t}\} \\
&\{\mathbf{f},x_i\} = -\tfrac{\partial \mathbf{f}}{\partial p_i} \\
&\{\mathbf{f},p_i\} = \tfrac{\partial \mathbf{f}}{\partial x_i},
\end{aligned} \tag{1.19}$$

As well as the following property

$$\{\mathbf{f},\{\mathbf{g},\mathbf{h}\}\} + \{\mathbf{h},\{\mathbf{f},\mathbf{g}\}\} + \{\mathbf{g},\{\mathbf{h},\mathbf{f}\}\} = 0 \tag{1.20}$$

known as *Jacobi identity*, thanks to which it is easy to demonstrate that the property of the partial derivative with respect to time remains valid for the total derivative. Also, Poisson brackets are not related to canonical variables, namely

$$\sum_{k=1}^{N}\left(\frac{\partial \mathbf{f}}{\partial x_k}\frac{\partial \mathbf{h}}{\partial p_k} - \frac{\partial \mathbf{f}}{\partial p_k}\frac{\partial \mathbf{h}}{\partial x_k}\right) = \sum_{k=1}^{N}\left(\frac{\partial \mathbf{f}}{\partial X_k}\frac{\partial \mathbf{h}}{\partial P_k} - \frac{\partial \mathbf{f}}{\partial P_k}\frac{\partial \mathbf{h}}{\partial X_k}\right), \tag{1.21}$$

this implies that if the transformation ($x_k \to X_k$, $p_k \to P_k$) is canonical, then the form of Poisson brackets is preserved, as well as all the properties previously mentioned. This makes Poisson brackets a robust mechanism for checking whether a transformation is canonical or not. We note that Poisson's brackets appear naturally when we are computing the total derivative of a function on phase space with respect to time, as shown below:

$$\frac{d\mathbf{f}}{dt} = \sum_{k=1}^{N}\left(\frac{\partial \mathbf{f}}{\partial x_k}\dot{x}_k + \frac{\partial \mathbf{f}}{\partial p_k}\dot{p}_k\right) + \frac{\partial \mathbf{f}}{\partial t}. \tag{1.22}$$

Using the equations of motion (1.11), we obtain

$$\frac{d\mathbf{f}}{dt} = \{\mathbf{f},\mathbf{H}\} + \frac{\partial \mathbf{f}}{\partial t}. \tag{1.23}$$

This equation describes the time evolution of the function $\mathbf{f}$ in phase space. If $\mathbf{f}$ is not explicitly time-dependent, then it can be a conserved quantity if $\{\mathbf{H}, \mathbf{f}\} = 0$. This provides us with a tool for determining whether a quantity is conserved or not. Let's return to the consideration of the generating function of the canonical transformation, since it is responsible for the transformation and its behavior, and determine its effect on a function $\mathbf{h}$ defined on the phase space. For this purpose we will use the identity generating function

$$\mathbf{g} = \sum_i x_i P_i. \tag{1.24}$$

Based on the fundamental equation of canonical transformations (1.15), we obtain the following relations

$$\begin{cases} p_i = \frac{\partial \mathbf{G}}{\partial x_i} = P_i \\ X_i = \frac{\partial \mathbf{G}}{\partial P_i} = x_i \end{cases} \tag{1.25}$$

So the function $\mathbf{g}$ is called an identity because it doesn't affect the canonical variable. We will now consider an infinitesimal transformation:

$$\begin{cases} x_i \to X_i = x_i + \delta x_i \\ p_i \to P_i = p_i + \delta p_i \end{cases} \tag{1.26}$$

results in an infinitely small increase of the generating function $\mathbf{g}$

$$\mathbf{g}(\mathbf{x}, \mathbf{P}) = \sum_i x_i P_i + \varepsilon \mathbf{f}, \tag{1.27}$$

Since $\varepsilon \to 0$, it is called a transformation parameter, and $\mathbf{f}$ is a function related to the type of transformation. It is clear that the effect of the generating function $\mathbf{g}$ is entirely reduced to the function $\mathbf{f}$. By using the equations (1.25) and (1.27), we can obtain the following results

$$\begin{cases} p_k = \frac{\partial \mathbf{g}}{\partial x_k} = P_k + \varepsilon \frac{\partial \mathbf{f}_k}{\partial x_k}, \\ \\ X_k = \frac{\partial \mathbf{g}}{\partial P_k} = x_i + \varepsilon \frac{\partial \mathbf{f}}{\partial P_k}, \\ \\ \mathbf{H}' = \mathbf{H}. \end{cases} \tag{1.28}$$

For the infinitesimal increase of the canonical variables $(\mathbf{x}, \mathbf{p})$ we obtain the following expressions

$$\begin{aligned} \delta p_k &= P_k - p_k = -\epsilon \frac{\partial \mathbf{f}}{\partial x_k} \\ \\ \delta x_k &= X_k - x_k = \epsilon \frac{\partial \mathbf{f}}{\partial P_k} \sim \epsilon \frac{\partial \mathbf{f}}{\partial p_k} \end{aligned} . \tag{1.29}$$

Thus, the effect of the generating function on the considered function $\mathbf{h}$ can be determined as follows

$$\begin{aligned}\delta\mathbf{h} &= \sum_k \left(\frac{\partial\mathbf{h}}{\partial x_k}\delta x_k + \frac{\partial\mathbf{h}}{\partial p_k}\delta p_k\right) \\ &= \varepsilon\sum_k \left(\frac{\partial\mathbf{h}}{\partial x_k}\frac{\partial\mathbf{f}}{\partial x_k} - \frac{\partial\mathbf{h}}{\partial p_k}\frac{\partial\mathbf{f}}{\partial x_k}\right) \\ &= \varepsilon\{\mathbf{h},\mathbf{f}\}. \end{aligned} \tag{1.30}$$

The significance of Poisson's brackets is once again evident in this context, as they facilitate the determination of an infinitesimal canonical transformation impact on a function defined in phase space. Our analysis of equation (1.23) revealed that if a quantity $\mathbf{K}$ is conserved, it satisfies $\{\mathbf{H},\mathbf{K}\} = 0$. This serves as an effective introduction to the discussion of symmetry in phase space. In this context, the first concept that is typically considered is that of Noether's theorem [28], which states:

**Noether's theorem**

If a mechanical system possesses a continuous symmetry, that is to say, if its Lagrangian remains invariant under a specific generalized coordinate transformation, then there exists a conserved quantity associated with this symmetry. In other words, for each transformation that does not affect the system's action, a physical quantity remains constant throughout the motion.

In the framework of canonical transformations, a significant generalization of Noether's theorem can be established through the use of Poisson brackets. In contrast with the classical version of the theorem, which relates continuous action symmetries to conserved quantities. The condition $\{\mathbf{H},\mathbf{g}\} = 0$, where $\mathbf{H}$ is the Hamiltonian of the system and $\mathbf{g}$ is a generating function of a canonical transformation, must be satisfied. This condition ensures that the canonical transformation leaves the Hamiltonian unaltered ($\delta\mathbf{H} = 0$), which implies that $\mathbf{g}$ is a prime integral of the system. In other words, the value of $\mathbf{g}$ remains conserved over time. Therefore, in this framework, any canonical transformation that commutes with the Hamiltonian leads directly to a conserved quantity. This observation establishes a direct link between the generators of symmetry transformations and prime integrals, thus unifying the notions of symmetry and conservation in phase space. Furthermore, it demonstrates the significance of Poisson brackets, which enable the concise expression of the effect of an infinitesimal transformation on a function defined in phase space. Consequently, this generalization extends beyond the domain of continuous action symmetries to identify conserved quantities, thereby enhancing our comprehension of the interrelationship between symmetry and conservation in Hamiltonian mechanics. In the same context, it can be observed that the system's Hamiltonian remains unchanged over time, unless it is explicitly time-dependent. To illustrate, consider an infinitesimal canonical transformation generated by $\mathbf{H}$. This would consequently result in the following increase in canonical variables

$$\begin{aligned} \delta p_k &= -\epsilon \frac{\partial \mathbf{H}}{\partial x_k} = \epsilon \dot{p}_k \\ \delta x_k &= \epsilon \frac{\partial \mathbf{H}}{\partial p_k} = \epsilon \dot{x}_k \end{aligned} . \tag{1.31}$$

This means that $\varepsilon$ in this case corresponds to $dt$, and that the Hamiltonian generates an infinitesimal canonical transformation, representing a translation in time between the instants $t$ and $t + dt$.

In conclusion, Hamiltonian mechanics provides a robust and elegant methodology for elucidating how dynamical systems evolve via canonical transformations. The phase space, described by the variables $x$ and $p$, forms the basis on which the system's dynamic variables develop under the influence of the Hamiltonian, which acts as the generator of infinitesimal transformations over time. The use of Poisson brackets elucidates the fundamental structure of this formalization, enabling the definition of relationships between physical quantities and the identification of conserved quantities through a generalized form of Noether's theorem. This framework shows that the motion of a mechanical system can be seen as a continuous sequence of canonical transformations that preserve the essential structure of the phase space. From a geometric perspective, phase space is viewed as a symplectic manifold, endowed with a profound mathematical structure captured by the symplectic 2-form ($\omega = dx \wedge dp$). This structure is tightly linked to Poisson brackets, which characterize how functions on the space relate to each other. Symplecticity guarantees the conservation of phase space volume and underpins the Hamiltonian approach, tying together the principles of dynamical evolution and symmetry in a rigorous geometric setting.

### 1.1.4 Application example: *Coupled pendulum*

Consider a system consisting of two pendulums connected by a torsion wire [29]. The angular positions of the pendulums are described by $\theta_1$ and $\theta_2$, which represent their deviations from the vertical. The pendulums have masses $m_1$ and $m_2$ and lengths $l_1$ and $l_2$ respectively. The coupling between the pendulums is introduced by a torsion wire with a stiffness constant $k$, which exerts a restoring torque proportional to the relative angular displacement ($\theta_1 - \theta_2$). This description captures the set of physical quantities that define the motion and mutual interaction of the coupled pendulums.

Torsional interaction introduces a link between the dynamics of both pendulums. This interaction can be modeled by an additional term in the total energy representing the torsional energy stored

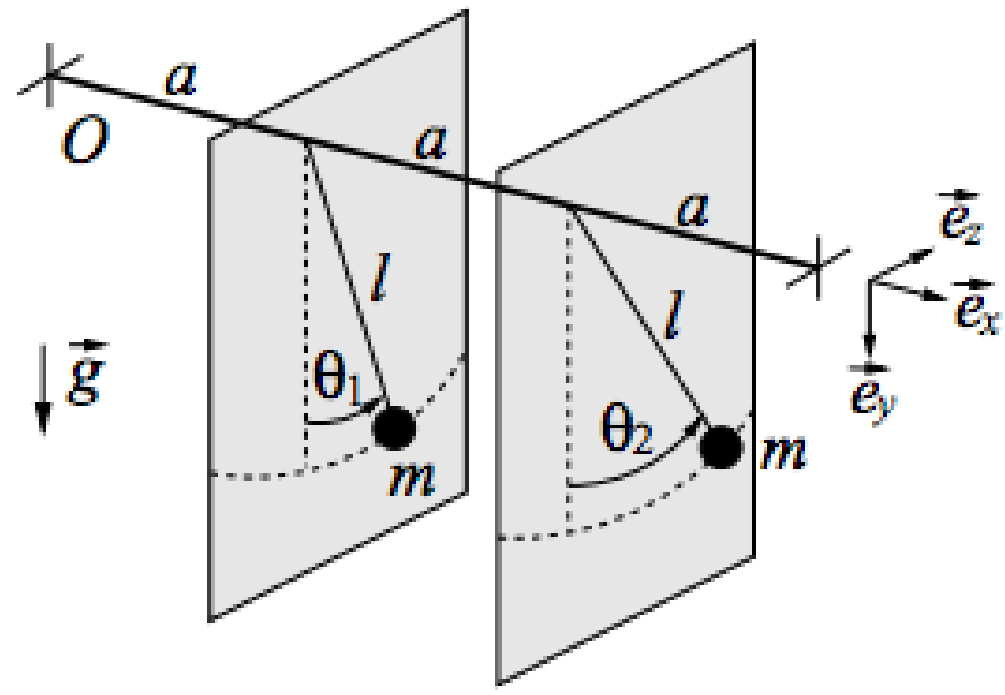

Figure 1.2: 2 pendulums coupled by a torsion wire

in the wire, as illustrated in the following energy components:

$$\begin{cases} \mathbf{T}_1 = \frac{1}{2}m_1 l_1^2 \dot{\theta}_1^2 \text{ (kinetic energy of pendulum 1)} \\ \mathbf{T}_2 = \frac{1}{2}m_2 l_2^2 \dot{\theta}_2^2 \text{ (kinetic energy of pendulum 2)} \\ \mathbf{V}_1 = m_1 g l_1 (1 - \cos(\theta_1)) \text{ (potential energy of pendulum 1)} \\ \mathbf{V}_2 = m_2 g l_2 (1 - \cos(\theta_2)) \text{ (potential energy of pendulum 2)} \\ \mathbf{E}_{\text{coupling}} = \frac{k}{2}(\theta_1 - \theta_2)^2 \text{ (torsional coupling energy)} \end{cases} \tag{1.32}$$

where $g$ is the gravitational acceleration. The modeled system possesses two degrees of freedom since its state is entirely defined by the independent angular coordinates $\theta_1$ and $\theta_2$. These angular variables determine the position of each pendulum relative to the vertical. Although the coupling induces an interaction between them, it does not reduce the number of independent coordinates. Therefore, both $\theta_1$ and $\theta_2$ are essential to fully characterize the system's state. The two degrees of freedom result in rich dynamical behavior, including synchronized motion and energy exchange between the pendulums.

The behavior of the coupled pendulum setup can be effectively modeled through the Lagrangian approach. The Lagrangian $\mathcal{L}$, defined in (1.1), is constructed as the difference between the system's total kinetic energy ($\mathbf{T}$) and its total potential energy ($\mathbf{V}$). For the system under consideration, the kinetic energy includes both pendulums:

$$\mathbf{T} = \frac{1}{2}m_1 l_1^2 \dot{\theta}_1^2 + \frac{1}{2}m_2 l_2^2 \dot{\theta}_2^2, \tag{1.33}$$

while the potential energy contains both the gravitational contributions and the interaction due to the torsion wire:

$$\mathbf{V} = m_1 g l_1 (1 - \cos\theta_1) + m_2 g l_2 (1 - \cos\theta_2) + \frac{k}{2}(\theta_1 - \theta_2)^2, \tag{1.34}$$

yielding the Lagrangian expression:

$$\mathcal{L} = \frac{1}{2}m_1 l_1^2 \dot{\theta}_1^2 + \frac{1}{2}m_2 l_2^2 \dot{\theta}_2^2 - \left[ m_1 g l_1 (1-\cos\theta_1) + m_2 g l_2 (1-\cos\theta_2) + \frac{k}{2}(\theta_1-\theta_2)^2 \right]. \tag{1.35}$$

This Lagrangian provides a full account of the system's dynamics, combining the individual contributions of kinetic energy, gravitational potential, and the torsional interaction. To switch to Hamiltonian mechanics, we perform a Legendre transformation:

$$\mathbf{H} = \sum_{i=1}^{2} p_i \dot{\theta}_i - \mathcal{L}, \tag{1.36}$$

with the canonical momenta $p_i$ corresponding to the generalized coordinates $\theta_i$, defined as:

$$p_i = \frac{\partial \mathcal{L}}{\partial \dot{\theta}_i} = m_i l_i^2 \dot{\theta}_i, \quad i = 1, 2. \tag{1.37}$$

Substituting into the Hamiltonian yields:

$$\mathbf{H} = \frac{p_1^2}{2m_1 l_1^2} + \frac{p_2^2}{2m_2 l_2^2} + m_1 g l_1 (1-\cos\theta_1) + m_2 g l_2 (1-\cos\theta_2) + \frac{k}{2}(\theta_1-\theta_2)^2, \tag{1.38}$$

The Hamiltonian reflects the total energy of the system: the sum of kinetic energy (now written in terms of momenta) and all potential energy contributions (gravitational and torsional). From a geometric point of view, the system's evolution unfolds in a four-dimensional phase space with coordinates $(\theta_1, p_1, \theta_2, p_2)$. This space carries a natural symplectic structure characterized by the fundamental 2-form.

$$\omega = \mathrm{d}\theta_1 \wedge \mathrm{d}p_1 + \mathrm{d}\theta_2 \wedge \mathrm{d}p_2, \tag{1.39}$$

which guarantees the conservation of fundamental Poisson relations:

$$\{\theta_1, p_1\} = 1, \quad \{\theta_2, p_2\} = 1, \quad \{\theta_1, \theta_2\} = 0, \quad \{p_1, p_2\} = 0.$$

The time evolution of the system is governed by the Hamiltonian, and controlled by the following equations

$$\dot{\theta}_1 = \frac{\partial H}{\partial p_1}, \quad \dot{p}_1 = -\frac{\partial H}{\partial \theta_1}, \quad \dot{\theta}_2 = \frac{\partial H}{\partial p_2}, \quad \dot{p}_2 = -\frac{\partial H}{\partial \theta_2}. \tag{1.40}$$

The coupling term $\frac{k}{2}(\theta_1-\theta_2)^2$ introduces a non-trivial interaction, linking the trajectories of the two pendulums in phase space. This leads to coupled equations of motion, implying complex dynamics characterized by coupled oscillations. To simplify these dynamics, we can introduce normal mode coordinates $(Q, P, q, p)$ reached by the following canonical transformations:

$$\begin{cases} \theta_1 \to Q = \frac{\theta_1+\theta_2}{2} \\ \\ p_1 \to P = \frac{\theta_1+p_2}{2} \\ \\ \theta_2 \to q = \frac{\theta_1-\theta_2}{2} \\ \\ p_2 \to p = \frac{\theta_1-p_2}{2} \end{cases} \tag{1.41}$$

In these new variables, $Q$ and $P$ describe symmetrical oscillations (center of mass), while $q$ and $p$ describe antisymmetrical oscillations (relative mode). This canonical transformation preserves the symplectic structure, enabling a clearer analysis of the dynamics. The conservation of energy encoded in the Hamiltonian implies that the system's trajectories are contained in constant-level surfaces of $\mathbf{H}$ in phase space. Moreover, the symplectic structure ensures the preservation of volumes in phase space, a direct consequence of Liouville's theorem. In summary, geometric evolution in phase space is characterized by complex trajectories due to torsional coupling, but can be simplified by the use of suitable canonical transformations, such as those for identifying normal modes.

## 1.2 Quantum states space

### 1.2.1 Introduction to Quantum States and Hilbert Space

As previously mentioned, classical Hamiltonian mechanics is predicated on a description of the physical system's state using canonical variables, namely generalized coordinates and their corresponding conjugate momenta. However, this approach faces intrinsic limitations when applied within the framework of quantum mechanics, as emphasized by Heisenberg's uncertainty principle [30]. This principle states that one cannot simultaneously determine both the position and momentum of a particle with unlimited accuracy. Consequently, representing a quantum state through both conjugate coordinates and momenta in a precise and simultaneous fashion, as done in classical mechanics, becomes unfeasible. In the quantum framework, the system's state is described by a wave function [31], which is a mathematical object encapsulating the entire information about the system in configuration space. The wave function is interpreted through probabilities, in the sense that the likelihood of locating a particle in a specific region or state corresponds to the square modulus of the wave function. Unlike classical mechanics where variables evolve deterministically, quantum states are probabilistic by nature and are described in a complex-valued formalism. Instead of relying on the classical phase space, which combines coordinates and momenta, quantum theory operates in Hilbert space [32], a complex vector space that may be infinite-dimensional [33], where states are represented by vectors and observables by linear operators. The formulation in Hilbert space provides the mathematical structure needed to express quantum features like superposition and entanglement. One of the most defining aspects of quantum theory, setting it apart from its classical counterpart, is the phenomenon of quantum entanglement, which allows for the description of multipartite systems in unified quantum states where the system's global behavior is not reducible to that of its individual components. This interconnection between parts of the system is not susceptible to independent classical representation, and it is one of the pillars of quantum formalism. This development paves the way to phenomena such as non-local correlations between distant particles. Quantum mechanics, with its utilization of Hilbert space and the wave function, provides a theoretical and mathematical framework that surpasses the limitations of classical description. This framework is indispensable for comprehending intrinsically quantum phenomena, such as entanglement, which find no equivalent in classical physics.

### Hilbert vector space

Quantum mechanics formulates the description of physical states through vectors defined in a complex Hilbert space [34]. Following Dirac's notation, these vectors are represented as kets and denoted as $|\psi\rangle$. The ket $|\psi\rangle$ carries the complete information about the system's state, allowing the derivation of any measurable quantity.

A key characteristic of such vectors is their ability to be added together to form other valid states. If $|\psi\rangle$ and $|\phi\rangle$ are two state vectors, their sum defines a new state:

$$|\psi\rangle + |\phi\rangle = |\chi\rangle. \tag{1.42}$$

This is referred to as the principle of vector superposition. Moreover, multiplying a ket by a complex scalar results in another legitimate ket:

$$c|\psi\rangle = |\psi\rangle c. \tag{1.43}$$

This illustrates that scalars commute with kets, reinforcing the vector space structure.

### Dual space and inner product

For every ket $|\psi\rangle$, there is an associated bra $\langle\psi|$ belonging to the dual space. This dual space is composed of linear functionals acting on the elements of the original space, with the basis $\{\langle\phi_i|\}$ being in correspondence with the basis $\{|\phi_i\rangle\}$ of the Hilbert space. This defines a one-to-one mapping:

$$|\psi\rangle \;\leftrightarrow\; \langle\psi|, \quad |\phi_i\rangle, |\phi_j\rangle, \ldots \;\leftrightarrow\; \langle\phi_i|, \langle\phi_j|, \ldots \tag{1.44}$$

This pairing respects the structure of linearity, so that linear combinations of kets correspond to analogous combinations of bras:

$$|\psi\rangle + |\chi\rangle \;\leftrightarrow\; \langle\psi| + \langle\chi|, \quad c|\psi\rangle \;\leftrightarrow\; c^*\langle\psi|. \tag{1.45}$$

Here, $c^*$ denotes the complex conjugate of the scalar $c$, which ensures compatibility with the properties of the inner product.

The operation of taking the inner product between a bra $\langle\phi|$ and a ket $|\psi\rangle$ is a central concept in quantum mechanics:

$$\langle\phi|.|\psi\rangle \equiv \langle\phi|\psi\rangle. \tag{1.46}$$

This operation returns a complex number and adheres to the following essential properties:

- **Complex conjugation**: The inner product satisfies

$$\langle\phi|\psi\rangle = \langle\psi|\phi\rangle^*, \tag{1.47}$$

which guarantees that $\langle\psi|\psi\rangle$ is always a real quantity.

- **Positive definiteness**: The scalar product of a ket with itself fulfills the condition

$$\langle\psi|\psi\rangle \geq 0, \tag{1.48}$$

with equality if and only if $|\psi\rangle$ is the zero vector. This property is essential for assigning probabilistic meaning to quantum states.

### Operators

Operators in a Hilbert space play a central role in quantum mechanics, as they are used to represent physical observables and govern the evolution of quantum states. An operator $\mathbf{O}$ is a mathematical object that acts on kets to produce new kets:

$$\mathbf{O}|\psi\rangle = |\phi\rangle. \tag{1.49}$$

A fundamental aspect of quantum theory is that certain operators are associated with measurable quantities. For example, the Hamiltonian operator $\mathbf{H}$ is used to represent the total energy of a system, while the momentum operator $\mathbf{p}$ corresponds to the observable of momentum.

A particularly significant category of vectors in Hilbert space are eigenvectors of an operator $\mathbf{A}$, which satisfy the relation:

$$\mathbf{A}|a\rangle = a|a\rangle. \tag{1.50}$$

Here, $a$ is called the eigenvalue, and the collection of all such eigenvalues constitutes the spectrum of $\mathbf{A}$. The presence of eigenvalues is essential because physical measurements in quantum theory always correspond to the eigenvalues of Hermitian operators.

Operators obey a variety of algebraic rules:

- **Addition**: It is possible to add operators, and operator addition is both commutative and associative:

$$\mathbf{O}+\mathbf{P} = \mathbf{P}+\mathbf{O}, \quad \mathbf{O}+(\mathbf{P}+\mathbf{Q}) = (\mathbf{O}+\mathbf{P})+\mathbf{Q}. \tag{1.51}$$

- **Linear transformation**: An operator $\mathbf{O}$ is linear if it satisfies the following condition:

$$\mathbf{O}(\lambda_1|\psi_1\rangle + \lambda_2|\psi_2\rangle) = \lambda_1\mathbf{O}|\psi_1\rangle + \lambda_2\mathbf{O}|\psi_2\rangle. \tag{1.52}$$

- **Non-commutativity**: In general, the product of two operators is not commutative:

$$\mathbf{OP} \neq \mathbf{PO}. \tag{1.53}$$

However, multiplication is associative:

$$\mathbf{O}(\mathbf{PQ}) = (\mathbf{OP})\mathbf{Q} = \mathbf{OPQ}. \tag{1.54}$$

One of the core ideas in quantum mechanics is the Hermitian adjoint of an operator $\mathbf{O}$, denoted $\mathbf{O}^\dagger$. It is defined by the following inner product relation:

$$\langle\psi|\mathbf{O}|\phi\rangle = \langle\phi|\mathbf{O}^\dagger|\psi\rangle^*. \tag{1.55}$$

An operator is Hermitian if it satisfies $\mathbf{O} = \mathbf{O}^\dagger$, in which case its eigenvalues are real. This makes Hermitian operators ideal for representing physical quantities. Another essential operation is the outer product, written as $|\phi\rangle\langle\psi|$. Unlike the inner product, which yields a scalar, the outer product gives rise to an operator:

$$(|\phi\rangle\langle\psi|)|\chi\rangle = \langle\psi|\chi\rangle|\phi\rangle. \tag{1.56}$$

This construct is frequently used in quantum theory, especially in forming projection operators. It is important to note that some combinations, such as $|\psi\rangle\mathbf{O}$ or $\mathbf{O}^\dagger\langle\psi|$, are not well-defined in the formalism. These do not correspond to any physically meaningful mathematical entities.

### 1.2.2 Unitary Evolution of the Quantum State

In quantum mechanics, the time evolution of a quantum state is determined by the Schrödinger equation. This evolution is described by a unitary evolution operator $\mathbf{U}_{t,t_0}$, which maps the initial state $|\psi(0)\rangle$ to the state at a later time $|\psi(t)\rangle$ as follows:

$$|\psi(t)\rangle = \mathbf{U}_{t,t_0}|\psi(0)\rangle. \tag{1.57}$$

To ensure that probability is preserved over time, the evolution operator must be unitary:

$$\mathbf{U}^\dagger_{t,t_0}\mathbf{U}_{t,t_0} = 1. \tag{1.58}$$

Additionally, the operator satisfies a composition rule:

$$\mathbf{U}(t_2, t_0) = \mathbf{U}(t_2, t_1)\mathbf{U}(t_1, t_0), \tag{1.59}$$

which expresses that the evolution between two times can be factored through an intermediate time.

The differential equation governing the evolution operator is given by the Schrödinger equation:

$$i\hbar\frac{\partial}{\partial t}\mathbf{U}_{t,t_0} = \mathbf{H}\mathbf{U}_{t,t_0}, \tag{1.60}$$

where $\mathbf{U}_{t,t_0}$ denotes the evolution operator and $\mathbf{H}$ refers to the Hamiltonian of the system. The Schrödinger equation for a quantum state follows by applying the previous equation to $|\psi(t)\rangle$:

$$i\hbar\frac{\partial}{\partial t}|\psi(t)\rangle = H|\psi(t)\rangle. \tag{1.61}$$

For systems with a time-independent Hamiltonian, the evolution operator takes the form:

$$\mathbf{U}_{t,t_0} = e^{-\frac{i}{\hbar}\mathbf{H}(t-t_0)}. \tag{1.62}$$

The time evolution of an initial state $|\psi_0\rangle$ is then written as:

$$|\psi_0\rangle = \sum_a c_a^0|a\rangle \exp\left(-\frac{i}{\hbar}E_a t\right), \tag{1.63}$$

where $|a\rangle$ are eigenvectors of $\mathbf{H}$ with corresponding eigenvalues $E_a$. The coefficients $c_a^0$ describe the decomposition of the initial state in the eigenbasis of $\mathbf{H}$.

Based on equation M, the evolution operator between two infinitesimally close times, $t$ and $t+dt$, can be expressed via a Taylor expansion:

$$\mathbf{U}_{t_0+\mathrm{d}t,t_0} = \mathbf{U}_{t_0,t_0} - \frac{i}{\hbar}H\mathbf{U}_{t_0,t_0}\mathrm{d}t + \mathcal{O}(\mathrm{d}t^2). \tag{1.64}$$

This leads to the following differential relation:

$$d\ln(\mathbf{U}) \approx -\frac{i}{\hbar}\mathbf{H}dt. \tag{1.65}$$

Clearly, the Hamiltonian plays a central role in determining how the quantum system evolves over time. Similar to the classical formulation discussed in (1.1.3), the quantum system's evolution is described by a sequence of unitary transformations driven by the Hamiltonian.

To compute the evolution operator when the Hamiltonian varies with time, one must integrate the governing differential equation. This integration yields the following expression:

$$\mathbf{U}_{t,t_0} = \exp\left(-\frac{i}{\hbar}\int_{t_0}^{t}\mathbf{H}(t')dt'\right). \tag{1.66}$$

This result depends on how the Hamiltonian behaves over time and enables a complete description of the system's evolution.

In the quantum framework, the evolution of an observable operator $\mathbf{A}$ is governed by the Heisenberg equation of motion [5], which takes the form of a differential equation involving the commutator with the Hamiltonian:

$$i\hbar\frac{\mathrm{d}\mathbf{A}}{\mathrm{d}t} = [\mathbf{A},\mathbf{H}] + i\hbar\frac{\partial\mathbf{A}}{\partial t}, \tag{1.67}$$

Here, the first term on the right-hand side describes the internal dynamics induced by the Hamiltonian, while the second term accounts for the explicit time-dependence of the observable. This equation highlights how the time evolution of a measurable quantity is tightly connected to its commutation relation with the Hamiltonian.

### 1.2.3 Degrees of freedom in quantum dynamics

In general, physical theories are based on a well-defined mathematical structure. In quantum mechanics, each system is associated with a Lie group $\mathcal{G}$, referred to as the dynamical group, which is connected to a Lie algebra $\mathfrak{g}$ [21]. This association stems from the mathematical formulation of quantum systems, which are described by an algebra of operators that act on a Hilbert space $\mathcal{H}$. The group $\mathcal{G}$ is defined through operators $T_k$ satisfying specific algebraic relations known as commutation relations:

$$[T_k, T_l] = C_{kl}^m T_m, \tag{1.68}$$

where $C_{kl}^m$ denotes the structure constants of the Lie algebra $\mathfrak{g}$.

In this setting, both the Hamiltonian $\mathbf{H}$ and the transition operators $\mathbf{A}$ of a system can be formulated in terms of the generators $T_k$ of $\mathfrak{g}$:

$$\mathbf{H} = \mathbf{H}(T_k), \quad \mathbf{A} = \mathbf{A}(T_k). \tag{1.69}$$

The Hilbert space $\mathcal{H}$ may be decomposed as a direct sum of invariant subspaces corresponding to the irreducible representations of the group $\mathcal{G}$:

$$\mathcal{H} = \bigoplus_{\lambda} Y_\lambda \mathcal{H}_\lambda, \tag{1.70}$$

where $\lambda$ indexes the irreducible representations and $Y_\lambda$ denotes their multiplicity in $\mathcal{H}$. If the subspaces $\mathcal{H}_\lambda$ are orthogonal and non-overlapping, then the system's dynamical behavior can be studied independently within each $\mathcal{H}_\lambda$.

Consider now a group $\mathcal{G}$ of dimension $n$ and rank $r$. Suppose this group possesses $\mu$ distinct subgroup chains $\{\mathcal{G}^{(1)}, \mathcal{G}^{(2)}, \ldots, \mathcal{G}^{(\mu)}\}$, where each chain is a nested sequence:

$$\mathcal{G} \supset \mathcal{G}_S^{(k)} \supset \mathcal{G}_{S-1}^{(k)} \supset \cdots \supset \mathcal{G}_1^{(k)}, \quad \forall k = 1, \ldots, \mu. \tag{1.71}$$

Each subgroup chain $\mathcal{G} \supset \mathcal{G}^{(k)}$ defines a complete set of commuting operators $\mathcal{C}^{(k)}$:

$$\mathcal{C}^{(k)} = \{\mathbf{Q}_i^{(k)} \in \mathcal{G}^{(k)} \mid [\mathbf{Q}_i^{(k)}, \mathbf{Q}_j^{(k)}] = 0, \ \forall i, j\}. \tag{1.72}$$

These operators are used to construct a basis $\{|\gamma^{(k)}\rangle\}$ for $\mathcal{H}_\lambda$, where $\gamma^{(k)} = (\gamma_1^{(k)}, \gamma_2^{(k)}, \ldots, \gamma_d^{(k)})$. In this basis, each operator in $\mathcal{C}^{(k)}$ acts as

$$\mathbf{Q}_i^{(k)} |\gamma^{(k)}\rangle = \gamma_i^{(k)} |\gamma^{(k)}\rangle, \quad i = 1, \ldots, d^{(k)}. \tag{1.73}$$

The dimension $d^{(k)}$ of this set is

$$d^{(k)} = r + \frac{n-r}{2}. \tag{1.74}$$

Importantly, the value of $d^{(k)}$ is independent of the subgroup chain selected, hence $d^{(k)} = d$ for all $k = 1, \dots, \mu$. The set $\mathcal{C}^{(k)}$ can be partitioned into two subsets: the totally degenerate operators $\mathcal{C}_T^{(k)}$, defined as:

$$\{\mathbf{Q}_i^{(k)} \in \mathcal{C}^{(k)} \mid \mathbf{Q}_i^{(k)}|\gamma^{(k)}\rangle = c|\gamma^{(k)}\rangle,\ \forall|\gamma^{(k)}\rangle \in \mathcal{H}_\lambda\}, \tag{1.75}$$

where $c$ is a constant not depending on the index $i$, and the non-totally degenerate operators $\mathcal{C}_N^{(k)}$, which do not satisfy this condition. Thus, one writes:

$$\mathcal{C}^{(k)} = \mathcal{C}_T^{(k)} + \mathcal{C}_N^{(k)}. \tag{1.76}$$

If $\mathcal{H}_\lambda$ corresponds to a non-degenerate irreducible representation of $\mathcal{G}$, then $\mathcal{C}_T^{(k)}$ contains $r$ Casimir operators of $\mathcal{G}$. The dimension of $\mathcal{C}_N^{(k)}$, denoted $M_\lambda$, is given by:

$$M_\lambda = d - r = \frac{n-r}{2}. \tag{1.77}$$

This quantity $M_\Lambda$ is independent of the chosen subgroup chain and uniquely identifies the Hilbert space $\mathcal{H}_\Lambda$. For degenerate irreducible representations, $M_\Lambda$ is strictly less than $\frac{n-r}{2}$ but remains constant regardless of the subgroup chain employed. Another approach to determining $M_\Lambda$ involves studying the complete set of commuting observables of the quantum system. Let us define $S = \{\mathbf{Q}_j \mid [\mathbf{Q}_j, \mathbf{Q}_{j'}] = 0,\ j, j' = 1, \dots, N\}$ as the set of all mutually commuting observables. In a Hilbert space basis, each vector can be uniquely labeled by quantum numbers $(\alpha_i,\ i = 1, \dots, M)$, corresponding to eigenvalues of those operators in $S$ that are not totally degenerate. Accordingly, $M$, or $M_\Lambda$, is interpreted as the number of degrees of freedom for quantum dynamics (QDDF). This analysis confirms that QDDF is a fundamental and unique characteristic of any quantum system.

### 1.2.4 Example: *Spin system*

Every irreducible (irrep) representation of $su(2)$, denoted $\mathcal{V}^{(2j+1)}$, is characterized by its corresponding complete set of commuting operators (CSCO) that follows the chain of subalgebras

$$su(2) \supset U(1), \tag{1.78}$$

where the total spin operator $\mathbf{J}^2$ along with its projection $\mathbf{J}_z$ constitute the CSCO:

$$\{\mathbf{J}^2, \mathbf{J}_z\}. \tag{1.79}$$

In this context, the total spin operator takes the form $\mathbf{J}^2 = \mathbf{J}_z + i(\mathbf{J}_+\mathbf{J}_- + \mathbf{J}_-\mathbf{J}_+)$, where $\mathbf{J}_+$ and $\mathbf{J}_-$ represent the raising and lowering operators, respectively. Since $su(2)$ is a Lie algebra of rank one, there exists only one family of irreducible representations, each indexed by the total spin $j$.

The orthonormal basis of $\mathcal{V}^{(2j+1)}$ is formed by the eigenvectors $\{|j,m\rangle\}$, where $m = -j, -j+1, \ldots, j-1, j$. These basis elements satisfy the spectral equations:

$$\begin{cases} \mathbf{J}^2|j,m\rangle = j(j+1)|j,m\rangle, \\ \mathbf{J}_z|j,m\rangle = m|j,m\rangle \end{cases} \tag{1.80}$$

As a result, the only non-degenerate operator in the CSCO is $\mathbf{J}_z$, and hence the quantum dynamical degrees of freedom (QDDF) for the considered spin system is **1**. Given that the universal covering group $SU(2)$ is compact, it is natural to select the lowest-weight state of $\mathcal{V}^{(2j+1)}$ as the reference vector:

$$|j,-j\rangle.$$

In this setting, the fundamental excitation operator for the spin system is $\mathbf{J}_+$. Through successive applications of $\mathbf{J}_+$, one can systematically construct the full set of basis vectors $\{|j,m\rangle\}$. For instance,

$$\mathbf{J}_+|j,-j\rangle \propto |j,-j+1\rangle,$$

and more generally, any arbitrary state $|\psi\rangle$ in $\mathcal{V}^{(2j+1)}$ may be expanded as

$$|\psi\rangle = \sum_{m=-j}^{j} c_m|j,m\rangle,$$

where the coefficients $c_m$ are determined via polynomial actions of $\mathbf{J}_+$ on $|j,-j\rangle$.

This approach reveals how spin excitations are governed by the repeated action of the raising operator $\mathbf{J}_+$, and confirms that the system's dynamics can be completely described by the operators $\mathbf{J}^2$ and $\mathbf{J}_z$.

## 1.3 Analogies between classical and quantum states space

Investigating the parallels between classical and quantum mechanics has long served as a productive avenue of study, offering valuable insights into the foundational principles of both frameworks. In this section, we will examine the structural correspondences between classical and quantum state spaces, focusing on how observables relate, the connection between Poisson brackets and quantum commutators, and the analogy between trajectories in classical phase space and paths in quantum theory. These similarities emphasize the profound links between the two theories and offer a conceptual perspective on the transition from classical to quantum frameworks. Exploring these links goes beyond simple comparison; it also enhances our understanding of how quantum theory reinterprets classical ideas through a fundamentally probabilistic and dynamical formalism.

### 1.3.1 Correspondence between Classical and Quantum Observables

In classical mechanics, observables such as position, momentum, and energy are typically expressed as smooth functions defined over the phase space. These functions follow the structure of Poisson algebra, which encodes the system's dynamics and provides a framework to describe the time evolution of observables. In this context, the Liouville equation (1.23) is of central importance, as it governs the temporal evolution of a function $\mathbf{f}(\mathbf{x}, \mathbf{p})$ on the phase space through the Poisson bracket.

In quantum mechanics, observables are instead represented by Hermitian operators acting within a Hilbert space. Their evolution is described by the Heisenberg equation (1.67), which serves as a quantum counterpart to the classical Liouville equation. This formulation replaces the classical Poisson bracket with the quantum commutator $\frac{1}{i\hbar}[\mathbf{A}, \mathbf{H}]$, emphasizing the structural continuity between classical and quantum formulations. According to the correspondence principle formulated by Niels Bohr, quantum behavior approaches classical predictions as quantum numbers become large, thereby establishing a bridge between both descriptions.

To make this link concrete, consider the operators associated with position and momentum in quantum theory. These play the role of their classical counterparts, and the connection is rigorously formalized through procedures such as Weyl quantization. Additionally, canonical transformations in classical mechanics correspond to unitary transformations in quantum theory, reflecting a shared framework rooted in conservation laws and symmetry principles that govern dynamical systems.

### 1.3.2 Role of Poisson Brackets and Commutators

The poisson brackets in classical mechanics and the commutators in quantum mechanics play similar, central roles in the description of system dynamics. In classical mechanics, the temporal evolution of an observable is determined by its Poisson bracket with the Hamiltonian. In addition, Poisson bracket reflects uncertainty relations between observables, such as the canonical relation between position and momentum. In quantum mechanics, on the other hand, the temporal evolution of an observable is defined by its commutator with the Hamiltonian. This commutator reflects the non-commutative nature of quantum observables, leading to uncertainty relations analogous to those in classical mechanics.

The classical Poisson bracket can be seen as the classical analogue of the quantum commutator, with Planck's constant serving as the transition scale between the two frameworks. This correspondence is precisely described in the framework of deformation quantization, where the Poisson bracket is used to deform the classical algebra of observables into a quantum algebra of operators. For example, the standard Poisson bracket between position and momentum, $\{x, p\} = 1$, corresponds to the standard commutator relation in quantum mechanics, $[\mathbf{X}, \mathbf{P}] = i\hbar$.

### 1.3.3 Analogy in Phase Space Trajectories and Quantum Paths

In classical mechanics, the trajectory $\mathcal{C}$ of a system is represented in phase space, a mathematical construct composed of generalized coordinates $x$ and their conjugate momenta $p$. These trajectories obey Hamilton's equations (1.11).

In quantum mechanics, system behavior differs fundamentally. Instead of following a single deterministic path, a quantum system evolves as a coherent superposition of multiple possible trajectories connecting an initial state $|\psi_i\rangle$ to a final state $|\psi_f\rangle$. This framework is formalized through the path integral formulation [35, 36], where the quantum transition amplitude is expressed as

$$\langle\psi_f|\psi_i\rangle = \int \mathcal{D}[\psi(t)]\, e^{\frac{i}{\hbar}S[\psi(t)]}. \tag{1.81}$$

Here, $\mathcal{D}[\psi(t)]$ denotes the functional measure over all possible trajectories $\psi(t)$, and each path contributes with a complex phase $e^{\frac{i}{\hbar}S[\psi(t)]}$, where $S[\psi(t)]$ is the action corresponding to that trajectory.

A connection between classical and quantum descriptions becomes evident in the semi-classical regime, where the Planck constant becomes negligible. In this approximation, quantum trajectories interfere constructively near the classical path, since the latter renders the action stationary. This behavior is captured by applying the saddle-point approximation to the path integral [37]

$$S[\psi(t)] \approx S[\mathcal{C}(t)] + \frac{1}{2}\delta^2 S[\mathcal{C}(t)], \tag{1.82}$$

with $\mathcal{C}(t)$ denoting the classical trajectory and $\delta^2 S[\mathcal{C}(t)]$ representing the variation around it. In this context, the classical path dominates the contribution, while the other non-classical paths cancel due to destructive interference.

A relevant illustration is the case of a particle subject to a potential $V(x)$. In classical theory, its motion is described by a unique trajectory governed by Hamilton's equations and the potential function. In contrast, quantum mechanics accounts for transitions between spatial states via a summation over all trajectories in configuration space, where each path contributes a phase that depends on its action. When the Planck constant approaches zero, the classical trajectory prevails in this summation, underscoring the emergence of classical behavior from quantum dynamics.

### 1.3.4 Analogy Based on the Symplectic Geometry of the State Space

Symplectic geometry offers a powerful framework for understanding the structure of state spaces in both classical and quantum mechanics. In classical mechanics, as discussed in subsection (1.1.3), the state space, or phase space, is equipped with a symplectic structure characterized by a closed, non-degenerate 2-form $\omega$. This structure enables a geometric formulation of motion and symmetries via transformations that preserve $\omega$. In quantum mechanics, although the mathematical formalism is different and relies on Hilbert spaces, meaningful analogies can be drawn with the symplectic formulation of classical theory. Such parallels arise, for instance, in the Wigner function formalism and the semi-classical description of quantum systems, where classical-like phase-space structures emerge effectively.

One fundamental connection lies in the resemblance between classical symplectic mappings and their quantum counterparts, typically represented by unitary operators in Hilbert space. These unitary maps maintain essential features of quantum states, particularly their norm, much like symplectic transformations preserve the classical symplectic form. Furthermore, the algebra of observables exhibits structural similarity: quantum observables, given by Hermitian operators, parallel classical observables through an analogous role played by commutators in place of Poisson brackets.

It is worth emphasizing that the study of symplectic geometry within the quantum framework transcends simple classical analogies. The passage from classical to quantum descriptions, often facilitated by approaches such as geometric quantization and Kähler geometry, reveals uniquely quantum geometric traits. Among these, Kähler manifolds provide a natural setting for incorporating both Riemannian and symplectic aspects. These features, along with differential geometric tools like curvature and connections, will be examined in detail in the next chapter. This upcoming analysis will show how such structures underpin the geometric organization and dynamical behavior of quantum states, reinforcing the deep interplay between classical symplectic concepts and their quantum extensions.

## 1.4 Summary

This chapter provides a comparative study of classical phase space and quantum state space, emphasizing their fundamental structures and analogies. In classical mechanics, the concept of phase space serves as a mathematical arena in which the full specification of a system's state is described through canonical coordinates $(\mathbf{x}, \mathbf{p})$. It offers a geometric viewpoint of system evolution, where motion unfolds along trajectories governed by Hamilton's equations.

This formulation underscores the relevance of symplectic structure, responsible for conserving the volume in phase space and encoding fundamental invariants of motion. The geometrical view is essential in exploring integrable systems, chaotic behavior, and thermodynamic equilibrium. In contrast to classical systems where states are precisely localized in phase space, quantum mechanics represents physical states as vectors in a complex Hilbert space $\mathcal{H}$, whose evolution is dictated by the Schrödinger equationmirroring the role of Hamiltonian dynamics.

However, in quantum theory, global phases have no physical significance, implying that the Hilbert space description is redundant unless considered up to a complex scalar factor. This distinction gives rise to the need for projective space when interpreting quantum states. Despite this difference, the chapter draws a parallel between classical and quantum formulations: classical observables as functions on phase space correspond to operators in the quantum setting, and Poisson brackets mirror quantum commutators. The Hamiltonian formalism bridges both paradigms, providing a common language for discussing dynamics. In doing so, this chapter sets the stage for understanding how structural parallels and conceptual gaps characterize the shift from classical mechanics to quantum theory.

CHAPTER 2

# Geometric structures in quantum mechanics

Quantum mechanics, the theoretical framework underlying the description of microscopic physical phenomena, is predicated on a precise mathematical representation of quantum states. These states, characterized by their inherent properties, are mathematically represented as vectors in a Hilbert space, with physical observables corresponding to operators acting on this space. However, an essential feature of quantum states is their intrinsically projective nature, which manifests as follows: state vectors associated with the same wave function, differing only by a global phase factor, in fact describe the same physical state. This redundancy in state description suggests that, while Hilbert space is indispensable, it is not entirely suitable for capturing the true dynamics of quantum systems.

To overcome this limitation, we introduce projective Hilbert space, a geometric model that identifies physically equivalent state vectors. Projective Hilbert space has become a central construct in the study of the geometric and topological aspects of quantum mechanics. It provides a rich set of tools for understanding phenomena such as geometric phases, the structure of entangled states, and the interplay between geometry and quantum dynamics. In this chapter, we undertake a comprehensive examination of these geometric structures and their role in quantum mechanics, underscoring their significance in both theoretical and applied contexts.

## 2.1 Geometry of the Projective Hilbert Space

### 2.1.1 Definition and properties

Although all the possible states of a quantum system can be gathered within a Hilbert space, this representation is not fully precise, since states differing only by a global phase appear as distinct elements, even though they describe the same physical reality. Therefore, it is necessary to adopt a refined framework, where elements are groups of vectors equivalent under multiplication by a phase factor. The most suitable space for this purpose is the projective Hilbert space, formed by classes of vectors in the Hilbert space that differ by a non-zero complex scalar.

We begin by introducing the complex projective space. By definition, $\mathbb{C}\mathcal{P}^n$ represents the set of lines through the origin in $\mathbb{C}^{n+1}$ [38], or, alternatively, the set of equivalence classes of $(n+1)$-tuples of complex numbers, not all zero, modulo scalar multiplication:

$$(\mathcal{Z}^0, \mathcal{Z}^1, ..., \mathcal{Z}^n) \equiv \lambda(\mathcal{Z}^0, \mathcal{Z}^1, ..., \mathcal{Z}^n) \qquad ; \qquad \lambda \in \mathbb{C}^*, \tag{2.1}$$

where the $\mathcal{Z}^i$ are known as homogeneous coordinates. This space can also be described using affine coordinate charts, such as:

$$\mathcal{Z}^\mu \equiv (1, z^1, ..., z^n) \equiv (z^{n'}, 1, z^{1'}, ..., z^{(n-1)'}), \tag{2.2}$$

This leads to two important observations. First, a given coordinate chart fails to cover the subset of points where the first homogeneous coordinate vanishes:

$$\mathcal{Z}^\mu \equiv (0, \mathcal{Z}^1, ..., \mathcal{Z}^n) \tag{2.3}$$

and such a subset is itself homeomorphic to a lower-dimensional projective space $\mathbb{C}\mathcal{P}^{n-1}$. From a topological viewpoint, we can view $\mathbb{C}\mathcal{P}^n$ as a union of $\mathbb{C}^n$ and $\mathbb{C}\mathcal{P}^{n-1}$ at infinity. Repeating this reasoning gives rise to the following decomposition:

$$\mathbb{C}\mathcal{P}^n = \mathbb{C}^n \cup \mathbb{C}^{n-1} \cup ... \cup \mathbb{C}^0. \tag{2.4}$$

It is clear that $\mathbb{C}\mathcal{P}^0$ reduces to a single point, and $\mathbb{C}\mathcal{P}^1$ corresponds to a two-dimensional sphere. For $n > 1$, the structure becomes more involved and less intuitive.

Second, in regions where coordinate patches intersect, the coordinates are related by:

$$z^{a'} = \frac{\mathcal{Z}^{a+1}}{\mathcal{Z}^1} = \frac{z^{a+1}}{z^1}, \tag{2.5}$$

which are clearly analytic functions on the overlap. This confirms that $\mathbb{C}\mathcal{P}^n$ is a complex manifold. The linear subspaces of $\mathbb{C}\mathcal{P}^n$ play an essential role. These are defined as the projective images of linear subspaces of $\mathbb{C}^{n+1}$, or equivalently as the solutions of homogeneous linear equations. In particular, a hyperplane is an $(n-1)$-dimensional submanifold of $\mathbb{C}\mathcal{P}^n$ defined by:

$$\mathcal{P}_\mu \mathcal{Z}^\mu = 0 \tag{2.6}$$

for a fixed set of $n+1$ complex numbers $\mathcal{P}_\mu$. This definition remains unchanged under rescaling of either the homogeneous coordinates or the coefficients $\mathcal{P}_\mu$. Therefore, the $\mathcal{P}_\mu$ themselves serve as homogeneous coordinates for another projective space, indicating that the space of hyperplanes is itself dual to $\mathbb{C}\mathcal{P}^n$.

If we impose $n-k$ independent linear conditions on the coordinates $\mathcal{Z}_\mu$, the resulting solution defines a $k$-dimensional complex linear subspace, also denoted as a $k$-plane, which corresponds to a $\mathbb{C}\mathcal{P}^k$. Geometrically, such spaces can be interpreted as intersections of multiple hyperplanes. The set of all such $k$-planes is called a Grassmannian. However, only in the cases $k=0$ and $k=n-1$ (points and hyperplanes) is the Grassmannian itself a projective space. One-dimensional subspaces correspond to complex projective lines, two-dimensional ones to complex projective planes, and so forth.

For instance, $\mathbb{C}\mathcal{P}^1$ is both a sphere topologically and behaves like a line in the sense that any pair of distinct points define a unique line, analogous to the two-dimensional subspace defined by two vectors in $\mathbb{C}^{n+1}$. Furthermore, any two projective lines intersect at a single point if they are not disjoint. The intersection of two linear subspaces $A$ and $B$ is called their meet $A \cap B$, while their span defines their join $A \cup B$ by extending the corresponding subspaces in $\mathbb{C}^{n+1}$ and projecting back.

Combining these two operationsintersection and spancreates a partially ordered set (a mesh) where any two elements have both a greatest lower bound and a least upper bound. This structure underpins the mathematical formulation of quantum logic. A group-theoretic interpretation of these transformations was introduced by Felix Klein [39]. Any statement about linear subspaces remains invariant under general linear transformations of $\mathbb{C}^{n+1}$. Hence, they form the group $GL(n+1,\mathbb{C})$, but the appropriate symmetry group for $\mathbb{C}\mathcal{P}^n$ is the projective linear group, namely $SL(n+1,\mathbb{C})/\mathbb{Z}_{n+1}$, where the effect of global complex scalars is modded out.

### 2.1.2 Riemannian Structure: Fubini-Study Metric

Now we will introduce the concept of distance, and more specifically a Riemannian metric, on the projective space $\mathbb{C}\mathcal{P}^n$. By selecting an arbitrary pair of points, the distance that separates them is given by the length of the geodesic curve that connects them. In the setting of complex projective geometry, the Fubini-Study metric provides a natural framework for defining a distance function on $\mathbb{C}\mathcal{P}^n$. It is a Riemannian metric that incorporates the intrinsic curvature of complex projective space. The Fubini-Study distance between two elements in $\mathbb{C}\mathcal{P}^n$ effectively measures the geodesic separation along the shortest curved path between them.

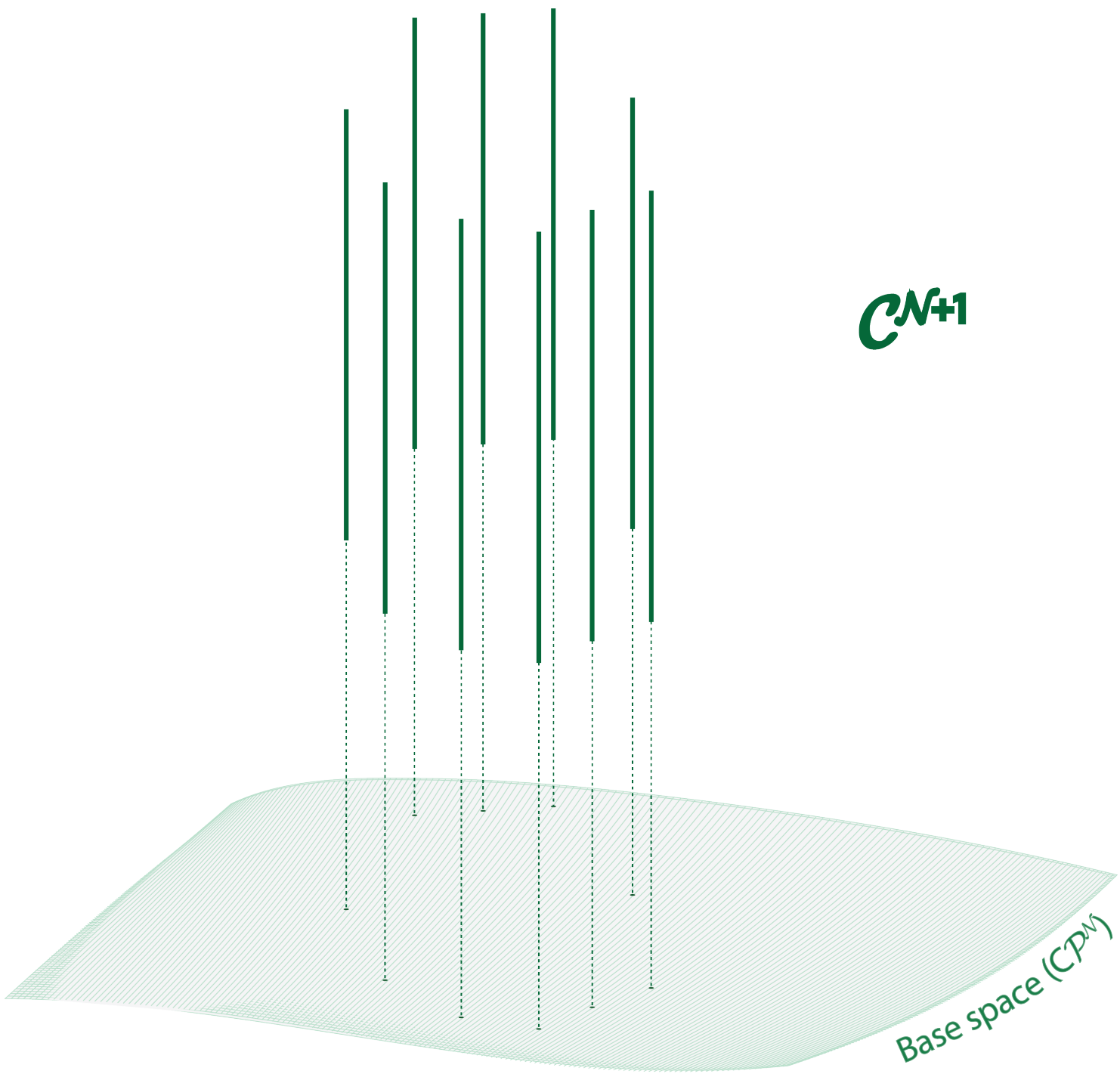

Figure 2.1: The Projective Space $\mathbb{C}\mathcal{P}^n$ as a Base of a Fiber Bundle over $\mathbb{C}^{n+1}$

Let us consider two points described in homogeneous coordinates, $\mathcal{Z}^\mu$ and $\mathcal{Y}^\mu$. The Fubini-Study distance [40] between these two points is given by

$$\mathbf{s}^2_{\text{FS}}(\mathcal{Z}^\mu, \mathcal{Y}^\mu) = \frac{|\mathcal{Z}^\mu \mathcal{Y}^{*\mu} - \mathcal{Y}^\mu \mathcal{Z}^{*\mu}|^2}{(1 + \mathcal{Z}^\mu \mathcal{Z}^{*\mu})(1 + \mathcal{Y}^\mu \mathcal{Y}^{*\mu})} \tag{2.7}$$

where the star (*) denotes complex conjugation, and Einstein's summation over repeated indices is assumed for all homogeneous coordinates.

To compute the metric tensor explicitly, we examine the infinitesimal distance between a point $\mathcal{Z}$ and a neighboring point $\mathcal{Z} + d\mathcal{Z}$. This means that the infinitesimal version of the Fubini-Study distance is governed by the Fubini-Study metric. The squared infinitesimal distance takes the form

$$d\mathbf{s}^2_{\text{FS}} = g_{\mu\nu} d\mathcal{Z}^\mu d\mathcal{Z}^{*\nu}, \tag{2.8}$$

where $g_{\mu\nu}$ denotes the components of the Fubini-Study metric on $\mathbb{C}\mathcal{P}^n$, given by

$$g_{\mu\nu} = \frac{\partial^2 \ln \mathcal{K}}{\partial \mathcal{Z}^\mu \partial \mathcal{Z}^{*\nu}} \tag{2.9}$$

where $\mathcal{K}$ is the so-called Bergman kernel, which is related to the geometry of the space and expressed in homogeneous coordinates as

$$\mathcal{K} = 1 + \mathcal{Z}^0 \mathcal{Z}^{*0} + \ldots + \mathcal{Z}^n \mathcal{Z}^{*n}, \tag{2.10}$$

This allows us to express the components of the Fubini-Study metric in terms of homogeneous coordinates as

$$g_{\mu\nu} = \frac{\delta_{\mu\nu}}{1 + \mathcal{Z}^{\rho}\mathcal{Z}^{*\rho}} - \frac{\mathcal{Z}^{\mu}\mathcal{Z}^{*\nu}}{(1 + \mathcal{Z}^{\rho}\mathcal{Z}^{*\rho})^2} \tag{2.11}$$

The metric defined above, from which the Fubini-Study distance is derived, originates from the Kähler potential corresponding to the space $\mathbb{C}\mathcal{P}^n$ [41]. Its components capture the intrinsic curvature of the complex projective space and are fundamental in analyzing the Riemannian structure of $\mathbb{C}\mathcal{P}^n$.

### 2.1.3 Riemannian Structure in context of quantum states

The evolution of a quantum system is governed by a Hamiltonian $\mathrm{H}(\boldsymbol{\lambda})$, which depends continuously on a set of parameters $\boldsymbol{\lambda} = (\lambda_1, ..., \lambda_{n'}) \in \mathcal{M}$, where $\mathcal{M}$ is the $n'$-dimensional parameter space. Let $\|\|$ and $(,)$ denote the norm and inner product defined on the corresponding parametric Hilbert space $\mathcal{H}(\lambda)$. The infinitesimal distance between two neighboring quantum states $\psi(\boldsymbol{\lambda})$ and $\psi(\boldsymbol{\lambda} + d\boldsymbol{\lambda})$ is given by [17, 42, 43]

$$d\mathbf{s}^2 = \|\psi(\boldsymbol{\lambda} + d\boldsymbol{\lambda}) - \psi(\boldsymbol{\lambda})\|^2 = \langle\delta\psi|\delta\psi\rangle = \langle\partial_\mu\psi|\partial_\nu\psi\rangle d\lambda^\mu d\lambda^\nu. \tag{2.12}$$

By decomposing the Hermitian inner product into its real and imaginary components, $\langle\partial_\mu\psi|\partial_\nu\psi\rangle = (\gamma_{\mu\nu} + i\omega_{\mu\nu})$, Eq. (2.12) transforms into

$$d\mathbf{s}^2 = (\gamma_{\mu\nu} + i\omega_{\mu\nu}) d\lambda^\mu d\lambda^\nu. \tag{2.13}$$

The Hermitian symmetry of the Hilbert space ensures that the tensors $\gamma_{\mu\nu}$ and $\omega_{\mu\nu}$ are symmetric and antisymmetric, respectively. Consequently, the geometric distance simplifies to

$$d\mathbf{s}^2 = \gamma_{\mu\nu} d\lambda^\mu d\lambda^\nu. \tag{2.14}$$

However, this metric tensor is not the optimal choice for the projective Hilbert space $\mathcal{PH}$, as it lacks gauge invariance under the $U(1)$ transformation:

$$|\psi(\boldsymbol{\lambda})\rangle \to |\psi(\boldsymbol{\lambda})\rangle' = e^{i\alpha}\,|\psi(\boldsymbol{\lambda})\rangle\,, \qquad \boldsymbol{\alpha} \in [0, 2\pi]. \tag{2.15}$$

Under this gauge transformation, the metric tensor $\gamma_{\mu\nu}$ transforms as

$$\gamma'_{\ \mu\nu} = \gamma_{\mu\nu} - \tilde{\beta}_\mu \partial_\nu \boldsymbol{\alpha} - \tilde{\beta}_\nu \partial_\mu \boldsymbol{\alpha} + \partial_\mu \boldsymbol{\alpha} \partial_\nu \boldsymbol{\alpha} \neq \gamma_{\mu\nu}, \tag{2.16}$$

where $\tilde{\beta}_\mu(\boldsymbol{\lambda}) = i\,\langle\psi(\boldsymbol{\lambda})\,|\partial_\mu\psi(\boldsymbol{\lambda})\rangle$ is the Berry connection on $\mathcal{PH}$ [44]. To preserve the $U(1)$ gauge invariance of the metric Eq. (2.14), the Berry connection must transform as

$$\tilde{\beta}_\mu \to \tilde{\beta}'_\mu = \tilde{\beta}_\mu - \partial_\mu \boldsymbol{\alpha}. \tag{2.17}$$

This leads to the Fubini-Study metric

$$d\mathbf{s}^2 = g_{\mu\nu} d\lambda^\mu d\lambda^\nu = \left(\gamma_{\mu\nu}(\boldsymbol{\lambda}) - \tilde{\beta}_\mu(\boldsymbol{\lambda})\tilde{\beta}_\nu(\boldsymbol{\lambda})\right) d\lambda^\mu d\lambda^\nu. \tag{2.18}$$

As a result, the Fubini-Study metric serves as the geometric tensor on the projective Hilbert space $\mathcal{PH}$, defining the geodesic distance between quantum states, i.e., complex lines. This metric is unitary, symmetric, and positive-definite due to the Cauchy-Schwarz inequality [45], and remains invariant under $U(1)$ gauge transformations. An alternative approach to deriving the Fubini-Study metric on the manifold of quantum states $\mathcal{PH}$ involves evaluating the transition probability between neighboring quantum states within the framework of perturbation theory [109]. This method interprets the Fubini-Study metric as a measure of quantum fluctuations in the vicinity of each point on $\mathcal{PH}$. The inner product between two neighboring states $|\psi(\boldsymbol{\lambda})\rangle$ and $|\psi(\boldsymbol{\lambda} + d\boldsymbol{\lambda})\rangle$ to second order in $\boldsymbol{\lambda}$ is given by [17]

$$\langle\psi(\boldsymbol{\lambda})|\psi(\boldsymbol{\lambda} + d\boldsymbol{\lambda})\rangle \simeq 1 + i\beta_\mu(\boldsymbol{\lambda})d\lambda^\mu + \frac{1}{2}\langle\psi(\boldsymbol{\lambda})|\partial_\mu\partial_\nu\psi(\boldsymbol{\lambda})\rangle d\lambda^\mu d\lambda^\nu. \tag{2.19}$$

Since $\langle\psi|\partial_\mu\psi\rangle$ and $\langle\partial_\mu\psi|\partial_\nu\psi\rangle + \langle\psi|\partial_\mu\partial_\nu\psi\rangle$ are purely imaginary, it follows that

$$\mathrm{Re}\langle\psi|\partial_\mu\partial_\nu\psi\rangle = -\mathrm{Re}\langle\partial_\mu\psi|\partial_\nu\psi\rangle = -\gamma_{\mu\nu}. \tag{2.20}$$

Substituting Eq. (2.19) and Eq. (2.20) yields the explicit form of the Fubini-Study metric:

$$d\mathbf{s}^2 = g_{\mu\nu}(\lambda) d\lambda^\mu d\lambda^\nu = 2\left(1 - |\langle\psi(\boldsymbol{\lambda})|\psi(\boldsymbol{\lambda} + d\boldsymbol{\lambda})\rangle|^2\right). \tag{2.21}$$

This formulation highlights how the Fubini-Study metric quantifies quantum fluctuations around each point in $\mathcal{PH}$. If the system depends solely on time, $\mathbf{H}(\boldsymbol{\lambda} = t)$, its evolution is governed by the Schrödinger equation (1.61). Substituting Eq. (1.61) into Eq. (2.21), the Fubini-Study metric takes the form

$$d\mathbf{s}^2 = 2\left(1 - |\langle\psi(t)|\psi(t + dt)\rangle|^2\right) = \frac{(\Delta E(t))^2}{\hbar^2} dt^2, \tag{2.22}$$

where $\Delta E(t)^2 = \langle\psi|\mathbf{H}(t)^2|\psi\rangle - \langle\psi|\mathbf{H}(t)|\psi\rangle^2$ is the energy uncertainty of the Hamiltonian $\mathbf{H}(t)$. This allows us to define the quantum speed of evolution as

$$\mathbf{s} = \frac{ds}{dt} = 2\Delta E(t), \tag{2.23}$$

with $\hbar = 1$. This result directly relates the quantum evolution speed to the system's energy uncertainty: greater energy fluctuations correspond to faster evolution, and vice versa. This is in line with the Aharonov-Anandan theorem [16, 25], which establishes that energy fluctuations drive quantum evolution. Moreover, the speed of evolution, being dependent on $\mathbf{H}(t)$, is experimentally measurable, which is crucial for quantum information theory applications [46–51].

### 2.1.4 Complex structure

Hilbert projective space $\mathbb{C}\mathcal{P}^n$ not only has a rich geometric and topological structure, but also naturally features a ***complex structure*** [52], which is essential to its description in mathematics and quantum physics. The complex structure allows the introduction of holomorphic and anti-holomorphic coordinates, providing a framework for the study of complex manifolds and facilitating calculations in quantum mechanics and field theory. Let's delve into the definition and implications of the complex structure on $\mathbb{C}\mathcal{P}^n$. The homogeneous coordinates $(\mathcal{Z}^0, \mathcal{Z}^1, \ldots, \mathcal{Z}^n)$ are generally used to describe points, but these coordinates are not unique due to their equivalence when scaled by a complex number. To define a complex structure, it's convenient to switch to *local affine coordinates*, such as

$$z^i = \frac{\mathcal{Z}^i}{\mathcal{Z}^0}, \quad i = 1, 2, \ldots, n, \tag{2.24}$$

on the map where $\mathcal{Z}^0 \neq 0$. These local coordinates $z^i$ are holomorphic, and the transition functions between overlapping coordinate maps are also holomorphic. This guarantees that $\mathbb{C}\mathcal{P}^n$ is a ***complex manifold***.

The complex structure on $\mathbb{C}\mathcal{P}^n$ is encapsulated by a tensor $\mathcal{J}^\mu_\nu$, which satisfies the following key properties:

- *Linearity :* $\mathcal{J}^\mu_\nu$ acts linearly on tangent vectors.
- *Complex evolution :* $(\mathcal{J}^\mu_\nu)^2 = -\delta^\mu_\nu$, guaranteeing that the manifold behaves like a complex space.
- *Metric compatibility :* The Fubini-Study metric $g_{\mu\nu}$ is Hermitian with respect to the complex structure, i.e.,

$$g_{\mu\nu} = \mathcal{J}^\rho_\mu g_{\rho\sigma} \mathcal{J}^\sigma_\nu. \tag{2.25}$$

Tangent and cotangent spaces at any point can be decomposed into holomorphic and anti-holomorphic parts. For a local coordinate system $(z^i, z^{*i})$, the basis vectors $\partial/\partial z^i$ and $\partial/\partial z^{*i}$ cover the holomorphic and anti-holomorphic tangent spaces respectively. This decomposition aligns with the complex structure tensor, which maps

$$\mathcal{J}\left(\frac{\partial}{\partial z^i}\right) = i\frac{\partial}{\partial z^i}, \quad \mathcal{J}\left(\frac{\partial}{\partial z^{*i}}\right) = -i\frac{\partial}{\partial z^{*i}}. \tag{2.26}$$

The complex structure of $\mathbb{C}\mathcal{P}^n$ has far-reaching consequences and applications in many fields. In quantum mechanics, the space of pure states is naturally represented by $\mathbb{C}\mathcal{P}^n$, and its complex structure is fundamental in the definitions of coherent states, geometric phases and quantization processes. In projective geometry and field theory, this structure makes it possible to define holomorphic sections of line bundles on $\mathbb{C}\mathcal{P}^n$, which is important in analyzing properties of the complex manifold. Furthermore, in representation theory, the symmetries of $\mathbb{C}\mathcal{P}^n$ under the action of $SL(n+1,\mathbb{C})/\mathbb{Z}_{n+1}$ are closely related to the complex structure and form the basis of many mathematical and physical representations. Thus, the complex structure of $\mathbb{C}\mathcal{P}^n$ is fundamental as it restricts the geometry, topology, quantum physics and differential geometry of the object. Such structures provide an integrated framework for analyzing the complex interrelationships between metric, symplectic and holomorphic properties within the framework of projective Hilbert space.

### 2.1.5 Symplectic Structure

In addition to the previous structures, the projecive Hilbert space $\mathbb{C}\mathcal{P}^n$ is also endowed with a Symplectic structure analogously to the classical phase space, arises naturally from its complex structure, and together with the Riemannian structure, it forms a Kähler manifold. The symplectic form $\omega$ for $\mathbb{C}\mathcal{P}^n$ is derived from the Kähler potential $\mathcal{K}$, just as the Fubini-Study metric is. In terms of homogeneous coordinates $\mathcal{Z}^\mu$, the symplectic form is expressed as

$$\omega = \frac{i}{2}\partial\bar{\partial}\ln\mathcal{K}, \tag{2.27}$$

where $\partial$ and $\bar{\partial}$ are the Dolbeault operators acting on the complex coordinates. Expanding the expression, the symplectic form in terms of coordinates becomes:

$$\omega = \frac{i}{2}\left(\frac{\delta_{\mu\nu}}{1+\mathcal{Z}^\rho\mathcal{Z}^{*\rho}} - \frac{\mathcal{Z}^\mu\mathcal{Z}^{*\nu}}{(1+\mathcal{Z}^\rho\mathcal{Z}^{*\rho})^2}\right)d\mathcal{Z}^\mu\wedge d\mathcal{Z}^{*\nu}. \tag{2.28}$$

It is a closed differential form since its external derivative is null

$$d\omega = 0, \tag{2.29}$$

as a fundamental propriety of symplectic forms. This closedness guarantees the local existence of a potential function for the symplectic structure, the Kähler potential $\mathcal{K}$. Moreover, this form is non-degenerate, meaning that the matrix formed by $\omega_{\mu\nu}$ has full rank everywhere on $\mathbb{C}\mathcal{P}^n$. This ensures that the symplectic structure can be used to define dynamics and geometric properties. Naturally, $\omega$ is compatible with the complex structure of $\mathbb{C}\mathcal{P}^n$, forming part of the Kähler structure. This compatibility ensures that $\mathbb{C}\mathcal{P}^n$ is not just a symplectic manifold but also a Kähler manifold.

From quantum mechanical point of view, the symplectic structure of the projective Hilbert space underpins the phase space description of quantum systems. The symplectic form plays a role analogous to the Poisson brackets in classical mechanics, facilitating the definition of commutation relations and the study of quantum evolution. Specifically, the symplectic form provides a geometric framework for the quantum state space dynamics:

- ***Hamiltonian Flow***: The symplectic form allows the definition of Hamiltonian vector fields, which describe the evolution of quantum states under a given Hamiltonian.

- ***Geodesics and Unitarity***: The symplectic structure ensures that the geodesic flow in $\mathbb{C}\mathcal{P}^n$ corresponds to unitary transformations in the underlying Hilbert space.

## 2.2 Geometrical Phases in Quantum Mechanics

This section explores the concept of geometric phases, initially conceived in the context of isolated systems described by pure states. Historically, these phases for which Michael Berry's landmark discovery in 1984 [11] laid the theoretical foundations, emerged from the analysis of adiabatic quantum systems undergoing cyclic evolution. Berry's seminal work, which highlighted the acquisition of an additional phase of purely geometric origin known as ***the Berry phase***, established a profound link between the topology of parameter space and the quantum properties of the system. This advance served as an interpretative foundation for the notion, and many of its features - such as independence from specific temporal dynamics - remain indispensable pillars for any subsequent generalization. The study of this paradigmatic case, although formally restricted to isolated systems, provides a didactic introduction to assimilate key concepts (holonomy, geometric connection) and cultivate a robust physical intuition. The structure of the section follows a chronological progression reflecting the evolution of formalisms, from Berry's early work to non-adiabatic extensions. It examines the various mathematical expressions describing the geometric phase accumulated by the state of a system, including formulations in terms of parallel transport over states manifold, their distinctive properties (gauge invariance, universality) and their operational interpretations, while making explicit the restrictive assumptions associated with each formalism (adiabaticity, cyclicity, etc.).

### 2.2.1 Cyclic Evolution: Adiabatic Case

The Berry phase, is a profound manifestation of the adiabatic theorem in quantum mechanics. In his pioneering work of 1984 [11], Michael Berry demonstrated that an eigenstate $|\psi_n(R(t))\rangle$ of a Hamiltonian $\mathbf{H}(R(t))$ slowly varying according to a multidimensional parameter $R(t)$ describing a closed cycle $\mathcal{C}$ in parameter space [53], acquires an additional geometric phase, independent of specific time dynamics. This phase, intrinsically linked to the curvature of the parameter space, emerges as a topological signature of the path traversed by $R(t)$. The Hamiltonian $\mathbf{H}(R(t))$ depends explicitly on time via the multidimensional parameter $R = (R_1, R_2, \dots)$ and satisfy

$$\mathbf{H}(R(t))|\psi_n(R(t))\rangle = E_n(R(t))|\psi_n(R(t))\rangle, \tag{2.30}$$

where the eigenvalues $E_n$ are assumed to be ordered $E_1 < E_2 < \dots$ and noun degenerate. Furthermore, the time evolution of the system state $|\psi(t)\rangle$ is governed by the Schrödinger equation:

$$i\hbar\frac{d}{dt}|\psi(t)\rangle = \mathbf{H}(R(t))|\psi(t)\rangle. \tag{2.31}$$

Expanding $|\psi(t)\rangle$ on the basis of instantaneous eigenstates:

$$|\psi(t)\rangle = \sum_n c_n(t)|\psi_n(R(t))\rangle, \tag{2.32}$$

the Schrödinger equation decomposes into a set of differential equations for complex coefficients $c_n(t)$,

$$i\hbar\dot{c}_n(t) = (E_n - i\hbar\langle\psi_n|\dot{\psi}_n\rangle)c_n(t) - i\hbar\sum_{m\neq n}\langle\psi_n|\dot{\psi}_m\rangle c_m(t). \tag{2.33}$$

By deriving the equation for $|\psi_n\rangle$, the terms $\langle\psi_n|\dot{\psi}_m\rangle$ can be expressed in terms of the matrix elements $\langle\psi_n|\dot{\mathbf{H}}|\psi_m\rangle$ of $\dot{\mathbf{H}}$. Substituting this relationship into the differential equations of $c_n(t)$, we obtain

$$\dot{c}_n(t) = -i(E_n - i\hbar\langle\psi_n|p\dot{s}i_n\rangle)c_n(t) - \sum_{m\neq n}\frac{\langle\psi_n|\dot{\mathbf{H}}|\psi_m\rangle}{E_m - E_n}c_m(t). \tag{2.34}$$

If the summation term is zero, the solutions $c_n(t)$ are simply the initial values $c_n(0)$ multiplied by a phase factor, which implies $|c_n(t)| = |c_n(0)|$. In this case, a system initially prepared in an eigenstate $|\psi_n(R(0))\rangle$ remains in the corresponding eigenstate $|\psi_n(R(t))\rangle$ at all subsequent times (to within one phase). Transitions between energy levels therefore originate from the summation term, but can be neglected in the adiabatic regime, where the variation of $\mathbf{H}(R(t)))$ is slow compared with the system characteristic scales $T \sim \hbar/(E_m - E_n)$. In this adiabatic regime, the coefficients $c_n(t)$ approximately verify

$$\dot{c}_n(t) = -i(E_n - \langle\psi_n|\dot{\psi}_n\rangle)c_n(t), \tag{2.35}$$

and the evolution of a Hamiltonian eigenstate is written

$$|\psi(t)\rangle = e^{-\frac{i}{\hbar}\int_0^t E_n(R(t'))dt'}e^{i\phi_n^a(t)}|\psi_n(R(t))\rangle. \tag{2.36}$$

The first exponential factor represents the accumulation of dynamic phase $(1/\hbar)\int_0^t E_n(R(t'))dt'$, while the second factor, studied by Berry, represents the geometric phase.

When the parameter $R(t)$ describes a closed path $C$ in parameter space, i.e. $R(0) = R(T)$, all functional quantities of $R$ inherit this periodicity. Thus, $\mathbf{H}(R(0)) = \mathbf{H}(R(T))$, as do the eigenstates and energies, so the dynamic phase is periodic. However, Berry's geometric phase $\phi_n^a$ is not necessarily so $\phi_n^a(0) \neq \phi_n^a(T)$. It obeys

$$\dot{\phi}_n^a(t) = i\langle\psi_n(R(t))|\nabla_R|\psi_n(R(t))\rangle \cdot \dot{R}(t). \tag{2.37}$$

The Berry phase accumulated on the closed circuit $C$ is expressed as a curvilinear integral:

$$\phi_n^a(C) = i\oint_C\langle\psi_n(R)|\nabla_R|\psi_n(R)\rangle \cdot dR, \tag{2.38}$$

where integration is performed on the path $C$ in parameter space. This expression is independent of the speed with which the circuit is traversed, as long as the assumptions of the adiabatic regime are satisfied. However, if the parameter $R$ does not vary slowly enough to neglect non-adiabatic transitions between energy levels, the evolution of the system no longer follows the previous equation. In this case, Aharonov and Anandan have shown that it is possible to define a geometric phase independent of the system's rate of evolution.

### 2.2.2 Cyclic Evolution: Non-Adiabatic Case

In the Hilbert space $\mathcal{H}$ associated with a quantum system, we are interested in a subspace $\mathcal{H}_N$ of unit-norm vectors. It's worth recalling here that in $\mathcal{H}_N$, two vectors $|\psi\rangle$ and $e^{i\alpha}|\psi\rangle$ (where $\alpha \in [0, 2\pi)$ are physically equivalent, since a global phase factor affects neither the probabilities nor the mean values of observables. Thus, a physical state is described by an equivalence class of vectors of the associated Hilbert vector space, where all vectors of the same class differ only by a global phase factor defined as

$$\tilde{\psi} = \{e^{i\alpha}|\psi\rangle \mid 0 \leq \alpha < 2\pi\}. \tag{2.39}$$

In other words, rays group together equivalent vectors in the physical sense. These rays are the essential structure for describing physical states within the framework of the projective Hilbert space $\mathcal{PH}$. Aharonov and Anandan, using a different notion of cyclic evolution from that introduced by Berry, defined an identical geometric phase for all trajectories $C \in \mathcal{H}$ in Hilbert space that project onto the same curve $\tilde{C} \in \mathcal{PH}$ in projective Hilbert space. Their proposal is based on the following elements: the Schrödinger equation defines a curve

$$C : t \in [0, T] \to \mathcal{H} \tag{2.40}$$

in Hilbert space. Thanks to an application $\Pi$

$$\mathcal{H}_N \to \mathcal{P}(\mathcal{H}), \tag{2.41}$$

this trajectory can be projected into a curve $\tilde{C} = \Pi(C)$ in ray space. From this observation, it is possible to define a cyclic evolution without reference to the system's Hamiltonian, which could even be independent of time. An evolution is said to be cyclic if the curve $\tilde{C}$ in ray space is closed, i.e. if

$$\tilde{\psi}(0) = \tilde{\psi}(T). \tag{2.42}$$

In this case, the Hilbert space vectors representing the initial and final states, $|\psi(0)\rangle$ and $|\psi(T)\rangle$, differ only by a total phase factor, which is written

$$|\psi(T)\rangle = e^{i\phi}|\psi(0)\rangle. \tag{2.43}$$

In addition, each vector belonging to the same equivalence class $\tilde{\psi}$ is related to the class representative by an appropriate phase factor [54]

$$|\psi(t)\rangle = e^{if(t)}|\tilde{\psi}(t)\rangle. \tag{2.44}$$

Substituting this relation into the Schrödinger equation, we obtain a differential equation for the phase factor $f(t)$

$$\hbar\dot{f}(t) = -\langle\tilde{\psi}(t)|\mathbf{H}|\tilde{\psi}(t)\rangle + i\hbar\langle\tilde{\psi}(t)|\dot{\tilde{\psi}}(t)\rangle, \tag{2.45}$$

which must satisfy the condition $f(T) - f(0) = \phi$. Integrating this equation gives the total phase

$$\phi = -\frac{1}{\hbar}\int_0^T \langle\tilde{\psi}(t)|\mathbf{H}|\tilde{\psi}(t)\rangle dt + \int_0^T i\langle\tilde{\psi}(t)|\dot{\tilde{\psi}}(t)\rangle dt. \tag{2.46}$$

The first term of this expression,

$$-\frac{1}{\hbar}\int_0^T \langle\tilde{\psi}(t)|\mathbf{H}|\tilde{\psi}(t)\rangle dt = -\frac{1}{\hbar}\int_0^T \langle\psi(t)|\mathbf{H}|\psi(t)\rangle dt, \tag{2.47}$$

corresponds to the dynamic phase, which depends explicitly on the system Hamiltonian. Subtracting this term from the total phase gives the geometric phase proposed by Aharonov and Anandan

$$\phi_{AA} = \phi + \frac{1}{\hbar}\int_0^T \langle\psi(t)|\mathbf{H}|\psi(t)\rangle dt = \int_0^T i\langle\tilde{\psi}(t)|\dot{\tilde{\psi}}(t)\rangle dt. \tag{2.48}$$

This geometric phase depends solely on the curve $\tilde{C}$ in the Hilbert projective space, without any reference to the Hamiltonian. Considering the special case of a slowly varying Hamiltonian, and expressing the state $|\psi(t)\rangle$ in the eigenstates basis of the instantaneous Hamiltonian, we recover Berry's result.

### 2.2.3 General Evolution: Phase-Distance Relation

The basic cases we have dealt with, such as Berry and Aharonov-Anandan phases, require explicit knowledge of the Hamiltonian of the quantum system. However, it is possible to generalize the concept of geometric phase so that it is independent of the dynamical details of the system. This generalization relies on the geometric properties of the projection $\tilde{\Gamma}$ of a curve $\Gamma$ into the projective space $\mathcal{PH}$, as we shall see in this section. In order to establish an expression for the geometric phase in the non-cyclic case that does not depend on the dynamic quantities of the system, a key notion known as ***reference section*** was introduced.

Consider a principal fibered space $\Pi : \mathcal{H}_N \to \mathcal{PH}$ with a group structure $U(1)$. Let $\Gamma$ be a curve in $\mathcal{H}_N$, defined as :

$$\Gamma : [0,t] \subset \mathbb{R} \to \mathcal{L}, \quad t \mapsto |\psi(t)\rangle. \tag{2.49}$$

The tangent vector to $\Gamma$ at the point $|\psi\rangle$ can be uniquely decomposed into two components. A vertical component $V_{|\psi\rangle}\mathcal{H}_N$, defined by

$$V_{|\psi\rangle}\mathcal{H}_N := \left\{\frac{D}{dt}|\psi\rangle \in T_{|\psi\rangle}\mathcal{H}_N \,|\, \Pi_*\frac{D}{dt}|\psi\rangle = 0\right\}. \tag{2.50}$$

And a horizontal component $H_{|\psi\rangle}\mathcal{H}_N$, defined as

$$H_{|\psi\rangle}\mathcal{H}_N := \left\{\frac{\delta}{dt}|\psi\rangle \in T_{|\psi\rangle}\mathcal{H}_N \,|\, \langle\psi|\frac{\delta}{dt}|\psi\rangle = 0\right\}. \tag{2.51}$$

A curve is said to be ***vertical*** if the projection of the point $|\psi(t)\rangle$ into $\mathcal{PH}$ remains constant. A vector tangent to such a curve is called a ***vertical vector***. A vector is ***horizontal*** if it is perpendicular to the fiber at a given point. The set of horizontal vectors at a point $|\psi\rangle$ constitutes the horizontal space $H_{|\psi\rangle}\mathcal{H}_N$.

To define parallel transport in a principal fiber space, a connection is introduced. It smoothly associates with each point a horizontal subspace $H_{|\psi\rangle}\mathcal{H}_N \subset T_{|\psi\rangle}\mathcal{H}_N$, satisfying the following conditions

- $T_{|\psi\rangle}\mathcal{H}_N = H_{|\psi\rangle}\mathcal{H}_N \oplus V_{|\psi\rangle}\mathcal{H}_N$, $\forall|\psi\rangle \in \mathcal{H}_N$.
- The connection is invariant under the action of $U(1)$.

Now consider a curve $\hat{\Gamma}$ in $\mathcal{PH}$ and its horizontal bearing $\bar{\Gamma}$ in $\mathcal{H}_N$, defined as $\bar{\Gamma} : t \mapsto |\bar{\psi}(t)\rangle$. Since $\bar{\psi}(t)\rangle$ is horizontal, it satisfies

$$\langle\bar{\psi}(t)|\frac{d}{dt}\bar{\psi}(t)\rangle = 0. \tag{2.52}$$

In the case of an evolution governed by a Hamiltonian $H(t)$, the horizontal state $|bar\psi(t)\rangle$ is linked to $|\psi(t)\rangle$ by a dynamic phase:

$$|\bar{\psi}(t)\rangle = e^{\frac{i}{\hbar}\int_0^t \langle\psi(t')|H(t')|\psi(t')\rangle dt'}|\psi(t)\rangle.$$

The geometric phase is then defined by :

$$e^{i[\Phi_g]_0^t} = \frac{\langle\bar{\psi}(0)|\bar{\psi}(t)\rangle}{|\langle\bar{\psi}(0)|\bar{\psi}(t)\rangle|}.$$

In the case of an evolution governed by a Hamiltonian $\mathbf{H}(t)$, the horizontal state $|\bar{\psi}(t)\rangle$ is linked to $|\psi(t)\rangle$ by a dynamic phase:

$$|\bar{\psi}(t)\rangle = e^{\frac{i}{\hbar}\int_0^t \langle\psi(t')|H(t')|\psi(t')\rangle dt'}|\psi(t)\rangle. \tag{2.53}$$

The geometric phase is then defined by

$$e^{i[\Phi_g]_0^t} = \frac{\langle\bar{\psi}(0)|\bar{\psi}(t)\rangle}{|\langle\bar{\psi}(0)|\bar{\psi}(t)\rangle|}. \tag{2.54}$$

In the cyclic case, this phase becomes $\Phi_g = \beta(C)$, corresponding to the holonomy associated with parallel transport along a closed curve in $\mathcal{PH}$.

On an open curve, the geometric phase can be expressed in terms of a so-called ***reference section***. Consider a section defined by [55, 56]

$$|\chi_0(t)\rangle = \eta(t,0)|\psi(t)\rangle, \quad \text{with } \eta(t,0) = \frac{\langle\psi(t)|\psi(0)\rangle}{|\langle\psi(t)|\psi(0)\rangle|}. \tag{2.55}$$

Differentiating this section, we find

$$i\langle\chi_0(t)|\dot{\chi}_0(t)\rangle = i\eta^*(t,0)\frac{d\eta}{dt}(t,0) + \frac{1}{\hbar}\langle\psi(t)|\mathbf{H}(t)|psi(t)\rangle. \tag{2.56}$$

After integration, the geometric phase becomes

$$[\Phi_g]_0^t = i \int_{\hat{\Gamma}} \langle \chi_0(t) | \dot{\chi}_0(t) \rangle dt, \tag{2.57}$$

where integration is performed on the curve $\hat{\Gamma}$ in $\mathcal{PH}$. This expression shows that the geometric phase depends only on the curve and not on the Hamiltonian used to generate the evolution. It is therefore a purely geometric property. Taking a connection defined by a 1-form $\mathcal{A}_\mu = \langle \chi_0(\lambda) | \partial_\mu \chi_0(\lambda) \rangle$, the geometric phase can be written as

$$[\Phi_g]_0^t = \int_{\hat{\Gamma}} \mathcal{A}_\mu \, d\lambda^\mu = \int_{\hat{\Gamma}} \mathcal{A}(\lambda). \tag{2.58}$$

Note The linear integral defining the geometric phase cannot be transformed directly into a surface integral according to Stokes' theorem, unless the open curve $\hat{\Gamma}$ is closed by a geodesic. In this case, the geometric phase can be seen as the holonomy of a connection on the principal fiber bundle.

It appears from the title of this subsection that it is concerned with the link between phase, in its two components, and the distance between states in projective Hilbert space. We begin by redefining the distance induced by the Fubini-Study metric, introduced mathematically at the beginning of this chapter (2.8) in the context of quantum states. The Hilbert projective space $\mathcal{PH}$ is defined from the Hilbert space $\mathcal{H}^*$, obtained by excluding the null vector, i.e. $\mathcal{H}^* = \mathcal{H} \setminus \{0\}$. Consequently, $\mathcal{H}^*$ constitutes a principal fiber bundle manifold over $\mathcal{PH}$, with a structural group given by $\mathbb{C}^* = \mathbb{C} \setminus \{0\}$. Consider two vectors $|\psi_1\rangle$ and $|\psi_2\rangle$ belonging to $H^*$. The distance $D(\psi_1, \psi_2)$ in $\mathcal{PH}$ can be defined from their scalar product. However, since the vectors $|\psi_1\rangle$ and $e^{i\phi_1}|\psi_1\rangle$ (as well as $|\psi_2\rangle$ and $e^{i\phi_2}|\psi_2\rangle$) represent the same quantum state, the distance must be invariant under these phase transformations. We introduce the following distance function

$$D(\psi_1, \psi_2) = \left\| e^{i\phi_1} \frac{|\psi_1\rangle}{|\psi_1\|} - e^{i\phi_2} \frac{|\psi_2\rangle}{\|\psi_2\|} \right\| \tag{2.59}$$

where the phases $\phi_1$ and $\phi_2$ are defined by :

$$e^{i\phi_1} = \frac{\langle \psi_0 | \psi_1 \rangle}{|\langle \psi_0 | \psi_1 \rangle|}, \quad e^{i\phi_2} = \frac{\langle \psi_0 | \psi_2 \rangle}{|\langle \psi_0 | \psi_2 \rangle|} \tag{2.60}$$

with $|\psi_0\rangle$ a reference state. These choices guarantee that $D(\psi_1, \psi_2)$ is well defined on the projective space $\mathcal{PH}$. To analyze the infinitesimal evolution of distance, let's consider a state $|\psi(t)\rangle$ evolving in time. At a given instant, the state becomes $|\psi(t + dt)\rangle$, and the associated phase evolves into $\phi(t + dt)$. Expanding to first order

$$dD^2 = \left\| |\hat{\psi}(t)\rangle - e^{id\phi(t)} |\hat{\psi}(t + dt)\rangle \right\|^2 \tag{2.61}$$

where $|\hat{\psi}(t)\rangle$ is the normalized state vector. We then obtain

$$dD^2 = \langle d\psi | d\psi \rangle + \dot{\phi}^2 dt^2 - 2i\dot{\phi} \langle \psi | d\psi \rangle dt^2. \tag{2.62}$$

Invariance under a local gauge transformation dictates that the time derivative of the phase $\dot{\phi}$ transforms as a gauge potential:

$$\dot{\phi} \to \dot{\phi} - \dot{\alpha}(t), \tag{2.63}$$

when the state undergoes the local transformation

$$|i\psi(t)\rangle \to e^{i\alpha(t)}|i\psi(t)\rangle. \tag{2.64}$$

The appropriate choice of phase

$$\phi(t) = \frac{1}{2}\frac{\langle\psi(0)|\psi(t)\rangle}{\langle\psi(t)|\psi(0)\rangle} \tag{2.65}$$

ensures that $\dot{\phi}$ behaves like a gauge potential, guaranteeing the invariance of $dD^2$ under these transformations. The total distance covered by the state $|\psi(t)\rangle$ during its evolution is given by

$$D = \int u(t)dt, \tag{2.66}$$

where the evolution density $u(t)$ is defined by

$$u(t) = \left[\langle d\psi|d\psi\rangle + \dot{\phi}^2 - 2i\dot{\phi}\langle\psi|d\psi\rangle\right]^{1/2}. \tag{2.67}$$

The minimum distance between two quantum states is defined as

$$S_{min} = \min_{\alpha\in\mathbb{R}} \left\| |\psi_1\rangle - e^{i\alpha}|\psi_2\rangle \right\|, \tag{2.68}$$

which corresponds to the distance induced by the Fubini-Study metric:

$$S_{min} = S_{FS} = \left(2 - 2|\langle\psi_1|\psi_2\rangle|\right)^{1/2}. \tag{2.69}$$

The distance $S_{FS}$ satisfies several fundamental properties:

- *Positive* :

$$S(\psi_1, \psi_2) \geq 0, \quad \text{with } S = 0 \text{ if and only if } |\psi_1\rangle \text{ and } |\psi_2\rangle \text{ represent the same state.}$$

- *Symmetry* :

$$S(\psi_1, \psi_2) = S(\psi_2, \psi_1).$$

- *Triangular inequality* :

$$S(\psi_1, \psi_2) + S(\psi_2, \psi_3) \geq S(\psi_1, \psi_3), \quad \text{for all distinct triplets } \psi_1, \psi_2, \psi_3.$$

- *Invariance under generalized conformal transformation* : The distance $S_{min}$ remains unchanged under the transformations :

$$|\psi(t)\rangle \to \eta(t)|\psi(t)\rangle = |\eta(t)|e^{i\alpha(t)}|\psi(t)\rangle,$$

  where $\eta(t)$ is a non-zero complex function.

Thus, the distance defined in the projective Hilbert space $\mathcal{PH}$ can be used to measure the separation between two quantum states, while respecting the fundamental symmetries of quantum theory.

Suppose a quantum system depends continuously on a single parameter $\lambda$. Its state, denoted $|\psi(\lambda)\rangle$, can be obtained from an initial state $|\psi(0)\rangle$ by a unitary transformation given by :

$$|\psi(\lambda)\rangle = \exp\left(i\int_0^\lambda A(\lambda')d\lambda'\right)|\psi(0)\rangle \tag{2.70}$$

where $A(\lambda)$ is a Hermitian operator acting as a generator associated with the parameter $\lambda$. From this equation, we can extract a parametric unitary evolution equation, given by

$$i\frac{d}{d\lambda}|\psi(\lambda)\rangle = A(\lambda)|\psi(\lambda)\rangle. \tag{2.71}$$

The uncertainty of operator $A$, denoted $\Delta A$, is defined by

$$(\Delta A)^2 = \langle\psi|A^2|\psi\rangle - \langle\psi|A|\psi\rangle^2. \tag{2.72}$$

Now consider two sufficiently close states $|\psi(\lambda)\rangle$ and $|\psi(\lambda + d\lambda)\rangle$, such that the latter can be expressed as a Taylor expansion of order two

$$|\psi(\lambda + d\lambda)\rangle = |\psi(\lambda)\rangle + \frac{d}{d\lambda}|\psi(\lambda)\rangle d\lambda + \frac{1}{2}\frac{d^2}{d\lambda^2}|\psi(\lambda)\rangle d\lambda^2 + o(d\lambda^3). \tag{2.73}$$

The previously defined distance can thus be written in the following form:

$$dS^2 = 2 - 2|\langle\psi(\lambda)|\psi(\lambda + d\lambda)\rangle| = (\Delta A)^2 d\lambda^2. \tag{2.74}$$

Thus, the distance on quantum state manifold is expressed by

$$S = \int \Delta A(\lambda)d\lambda. \tag{2.75}$$

Next, consider a quantum evolution governed by the Schrödinger equation, where the Hamiltonian $\mathbf{H}(t)$ is arbitrary and time-dependent, and $|\psi(t)\rangle$ is a vector in Hilbert space representing the state of the system at time $t$. After an infinitesimal time interval $dt$, the system state becomes $|\psi(t+dt)\rangle$. The corresponding infinitesimal distance is given by

$$dS^2 = 2 - 2|\langle\psi(t)|\psi(t + dt)\rangle|. \tag{2.76}$$

Identifying $\lambda = t$ and $A(\lambda) = H(t)/\hbar$, we obtain the distance expression in the projective Hilbert space $\mathcal{PH}$ :

$$dS = \frac{\Delta E(t)}{\hbar}dt, \quad S = \int \frac{\Delta E(t)}{\hbar}dt, \tag{2.77}$$

where $\Delta E(t)$ is the energy uncertainty defined by as

$$\Delta E(t) = \sqrt{\langle\psi|H^2|\psi\rangle - \langle\psi|H|\psi\rangle^2}. \tag{2.78}$$

The quantity $dS$ measures the distance between points $\Pi(|\psi(t)\rangle)$ and $\Pi(|\psi(t + dt)\rangle)$ on the curve $\hat{C}$ of $\mathcal{PH}$ manifold, measured previously by the Fubini-Study metric (2.8).

### Relation between dynamic phase and distance

The results of minimizing the distance function are not only limited to extracting the distance between two quantum states, but there is also a result that is considered just as important concerning the phase. Indeed, minimization by a phase parameter leads to a minimal phase $\alpha_{\min}$ given by :

$$e^{i\alpha_{\min}} = \frac{\langle\psi_1|\psi_2\rangle}{|\langle\psi_1|\psi_2\rangle|}. \tag{2.79}$$

In the case where $\psi_1 = \psi(t)$ and $\psi_2 = \psi(t+dt)$, this relation becomes

$$e^{id\alpha_{\min}} = \frac{\langle\psi(t)|\psi(t+dt)\rangle}{|\langle\psi(t)|\psi(t+dt)\rangle|}. \tag{2.80}$$

By applying a Taylor expansion and using Schrödinger equation, we obtain

$$\alpha_{\min} = -\frac{1}{\hbar}\langle\psi(t)|H(t)|\psi(t)\rangle dt. \tag{2.81}$$

Integrating between $t = 0$ and $t = T$, gives the dynamic phase

$$\alpha_{\min}(T) = \varphi_{\text{dyn}} = -\frac{1}{\hbar}\int_0^T \langle\psi(t)|H(t)|\psi(t)\rangle dt. \tag{2.82}$$

In this way, minimization of the distance function is perfectly in line with the principle of least action, highlighting a natural geometric interpretation of the evolution of the quantum state. This approach not only allows us to characterize the optimal trajectory followed by the system, but also to better understand the fundamental role played by phase in quantum dynamics.

### Remarks

+ ***Geometrical interpretation of the time-energy uncertainty relation:*** To separate two states on the evolution trajectory of a quantum system, these states must be orthogonal, as they correspond to states of the same observable with different eigenvalues. The minimum distance between two orthogonal states is the geodesic distance equal to $\pi$, which leads to

$$S \geq \pi. \tag{2.83}$$

  This allows us to write the uncertainty relation as

$$\hbar(\Delta E)(\Delta t) \geq \frac{\hbar}{4}. \tag{2.84}$$

  where $\Delta E$ is the time-averaged energy uncertainty over the interval $\Delta t$.

+ ***System evolution velocity:*** The distance $S$ provides an additional geometric interpretation of the energy uncertainty $\Delta E$ of a quantum system. Indeed, the quantity

$$\frac{d\mathbf{s}}{dt} = \frac{\Delta E(t)}{\hbar} \tag{2.85}$$

  expresses the system velocity on projective Hilbert space.

### Relation between geometrical phase and distance

Having defined the distance function $D$ and its minimum value within a quantum evolution, and having established its link with the dynamical phase, a fundamental question arises: is there a connection between this distance and the geometric phase? To answer this question, we will rely on the Aharonov - Anandan phase, introduced in Sec. (2.2.2), and define a positive quadratic function $\Omega^2(t)$ of the following form [55]

$$\Omega^2(t) = \langle \dot{\psi}(t)|\dot{\psi}(t)\rangle - \beta(t)^2, \tag{2.86}$$

where $\beta(t)$ is an arbitrary real function of time. As with the minimization of distance function $D$, we seek to minimize $\Omega(t)$ with respect to $\beta(t)$. The value that minimizes this function is given by

$$\beta_{\min}(t) = i\langle \psi(t)|\dot{\psi}(t)\rangle. \tag{2.87}$$

This quantity corresponds to Berry connection, and by time integration, we find

$$\beta = i\int_0^T \langle \psi(t)|\dot{\psi}(t)\rangle dt, \tag{2.88}$$

which is none other than Aharonov and Anandan **geometric phase**. Let us now introduce a new geometric quantity: the length of a curve $|\psi(t)\rangle$, which is a differentiable trajectory in the Hilbert space $\mathcal{H}$. Let $|\tilde{\psi}(t)\rangle$ be a differentiable section of this curve. The length travelled by the system between the initial state $|\tilde{\psi}(0)\rangle$ and the final state $|\tilde{\psi}(T)\rangle$ is defined by

$$L = \int_0^T \sqrt{\langle \dot{\tilde{\psi}}(t)|\dot{\tilde{\psi}}(t)\rangle} dt. \tag{2.89}$$

The velocity vector $|\dot{\tilde{\psi}}(t)\rangle$ then represents the tangent vector to the trajectory in Hilbert projective space. This length $L$ is invariant under a temporal reparametrization $t \to \tau(t)$. The infinitesimal length element is then given as

$$dL = \sqrt{\langle \dot{\tilde{\psi}}(t)|\dot{\tilde{\psi}}(t)\rangle} dt. \tag{2.90}$$

The $|\tilde{\psi}(t)\rangle$ state is linked to the $|\psi(t)\rangle$ state by a phase transformation

$$|\tilde{\psi}(t)\rangle = e^{-if(t)}|\psi(t)\rangle, \tag{2.91}$$

and its differentiation gives

$$dL^2 = \left(\dot{f}dt\right)^2 + \frac{1}{\hbar^2}\left(\langle H^2\rangle - \langle H\rangle^2\right)dt^2 + 2\dot{f}\frac{\langle H\rangle}{\hbar}dt^2. \tag{2.92}$$

By choosing

$$\dot{f} = -\frac{1}{\hbar}\langle\psi|H|\psi\rangle + i\langle\tilde{\psi}|\dot{\tilde{\psi}}\rangle, \tag{2.93}$$

we get

$$dL^2 = \frac{1}{\hbar^2}\left(\langle H^2\rangle - \langle H\rangle^2\right)dt^2 + (i\langle\tilde{\psi}|\dot{\tilde{\psi}}\rangle)^2 dt^2. \tag{2.94}$$

Comparing with the infinitesimal distance $dS$, defined by

$$dS^2 = \frac{1}{\hbar^2}\left(\langle H^2\rangle - \langle H\rangle^2\right)dt^2, \tag{2.95}$$

we find

$$dL^2 - dS^2 = (i\langle\tilde{\psi}|\dot{\tilde{\psi}}\rangle dt)^2. \tag{2.96}$$

Given that $\langle\tilde{\psi}|\dot{\tilde{\psi}}\rangle$ is a pure imaginary number, this implies that $dL^2 - dS^2$ is always positive. Thus, the geometric phase can be expressed in integral form as following

$$\beta = i\int_0^T \langle\tilde{\psi}|\dot{\tilde{\psi}}\rangle dt = \int_0^T \sqrt{dL^2 - dS^2}. \tag{2.97}$$

This allows us to interpret the geometric phase as the integral of a **contracted length**, denoted $dL_c$, defined by

$$dL_c = \sqrt{1 - \frac{v_H^2}{u_H^2}}dL. \tag{2.98}$$

where we introduce :

- $v_H = dS/dt$, which represents the transport velocity of the system in Hilbert projective space,
- $u_H = dL/dt$, which is the rate of change of length $L$.

Thus, in a non-adiabatic cyclic evolution, the geometric phase is interpreted as the integral of **contracted length** of the curve representing the quantum evolution on the quantum state manifold.

## 2.3 Curvature and Topology

### 2.3.1 Curvature in Quantum Geometries

Curvature is a fundamental concept in differential geometry, describing how a geometric space deviates from flatness. In Riemannian geometry, curvature is studied through various tensors, including the Riemann curvature tensor, Ricci curvature and scalar curvature. Of these, *Gaussian* *curvature* is of particular importance, as it is an intrinsic property of surfaces, independent of the embedding space. This section presents an overview of Gaussian curvature, its mathematical definition, its expressions in different formulations, and its geometric meaning. Given a Riemannian manifold $M$ with a metric $g$, the Riemann curvature tensor $R$ is defined as :

$$R(X,Y)Z = \nabla_X \nabla_Y Z - \nabla_Y \nabla_X Z - \nabla_{[X,Y]} Z, \tag{2.99}$$

where $\nabla$ is the Levi-Civita connection. The sectional curvature of a plane generated by two tangent vectors $X$ and $Y$ at a point $p \in M$ is given by :

$$K(X,Y) = \frac{\langle R(X,Y)Y, X\rangle}{\|X\|^2\|Y\|^2 - \langle X,Y\rangle^2}. \tag{2.100}$$

In the special case of a 2-dimensional Riemannian manifold with a metric $g$, the *Gaussian curvature* $K$ is an intrinsic measure of curvature, defined directly from the metric tensor [57, 58]. Consider a local parametrization with coordinates $(u, v)$, where the metric is written :

$$ds^2 = E\,du^2 + 2F\,du\,dv + G\,dv^2, \tag{2.101}$$

where the coefficients $E$, $F$, $G$ are smooth functions of the coordinates. The Gaussian curvature $K$ can be expressed entirely in terms of these coefficients and their partial derivatives:

$$K = -\frac{1}{2\sqrt{EG - F^2}} \left[ \frac{\partial}{\partial u} \left( \frac{G_u - F_v}{\sqrt{EG - F^2}} \right) + \frac{\partial}{\partial v} \left( \frac{F_u - E_v}{\sqrt{EG - F^2}} \right) \right], \tag{2.102}$$

where $E_u = \partial E/\partial u$, $G_v = \partial G/\partial v$, and so on. This so-called Liouville formula shows that $K$ is an intrinsic property that depends solely on the metric $g$. Gaussian curvature can also be written as follows

$$K = \frac{1}{\sqrt{det|g|}} \left[ \frac{\partial}{\partial v} \left( \frac{\sqrt{det|g|}}{E} \Gamma^2_{11} \right) - \frac{\partial}{\partial u} \left( \frac{\sqrt{det|g|}}{E} \Gamma^2_{12} \right) \right], \tag{2.103}$$

where $\Gamma^k_{ij}$ are Christoffel's symbols and $det|g| = EG - F^2$. Although Gaussian curvature can be interpreted extrinsically as the product of the principal curvatures $\kappa_1\kappa_2$ (Gauss's *Theorema Egregium* theorem) [59], it remains an intrinsic quantity calculable from the metric alone. As such, it can be used to determine the geometric behavior of the surface under study at any given point as follows

- If $K > 0$, the point is **elliptic**.
- If $K < 0$, the point is **hyperbolic**.
- If $K = 0$, the point is **parabolic** or **planar** (according the shape operator value).

To conclude, Gaussian curvature provides in-depth information on the geometry of surfaces, influencing many areas of mathematics and physics. It remains invariant under isometries, making it a crucial intrinsic property of surfaces in differential geometry.

### 2.3.2 Topological Aspects and Their Implications

Topology, the branch of mathematics that studies the properties of spaces preserved under continuous deformations, plays a crucial role in quantum mechanics and quantum computing. In the context of quantum evolution and entanglement, topological aspects offer a robust framework for understanding the resilience of quantum states to external perturbations. This subsection explores the role of topology in quantum evolution, with particular emphasis on its implications for entanglement and fault-tolerant quantum computing.

Quantum systems can exhibit exotic phases of matter that are better characterized by topological invariants than by traditional local order parameters. These phases, called **topological phases**, possess universal properties that remain unchanged under local perturbations. The evolution of quantum states in these topological systems is governed by geometric and adiabatic phases, with direct implications for quantum computing and quantum information processing. A fundamental example is the **geometric Berry phase**, acquired by a quantum system undergoing cyclic evolution. In topological quantum systems, this geometric phase generalizes in terms of **Chern invariants** in two-dimensional quantum Hall systems and **topological invariants** in more complex contexts, such as topological insulators and superconductors. These invariants play a key role in the stability of quantum states and the protection of quantum information against decoherence.

The deep connection between geometry and topology in quantum systems is elegantly illustrated by the **GaussBonnet theorem** [60, 61], which allows one to compute the integer Euler characteristic $\chi(M)$ of a compact, oriented two-dimensional Riemannian manifold $M$, representing the quantum state space.

**GaussBonnet theorem**

For a compact, oriented two-dimensional Riemannian manifold $M$, the Euler characteristic $\chi(M)$ is given by the sum of the integral of the Gaussian curvature $K$ over $M$ and the integral of the geodesic curvature $k_g$ along its boundary $\partial M$, normalized by $2\pi$:

$$\frac{1}{2\pi}\left[\int_M K\, dS + \oint_{\partial M} k_g\, dl\right] = \chi(M), \tag{2.104}$$

where $dS$, $dl$, and $k_g$ denote the surface element, line element, and geodesic curvature, respectively.

In other words, this equality identifies a profound link between local geometric quantities (curvatures) and a global topological invariant (Euler characteristic), encoding the topology of the quantum state space through geometric and boundary contributions. The two terms on the left-hand side represent the bulk and boundary contributions to the Euler characteristic, identifying the topological nature of the quantum state space. Since the Gauss curvature $K$ has been previously defined, this formulation offers a concrete link between local geometric quantities and global topological invariants. In quantum mechanics, particularly when studying projective Hilbert spaces or Bloch-like manifolds, this theorem helps classify the topological structure of the state space and understand how geometry influences entanglement and quantum phase transitions.

An interesting topological feature arises in quantum systems through the emergence of topological phases [62, 63], non-local properties of the state space that remain invariant under smooth deformations of the system's parameters. Unlike dynamical phases, which depend on the details of time evolution, topological phases are geometric in nature and reflect the underlying topology of the manifold that encodes the quantum states. These phases are associated with closed loops in the system's parameter space, where a quantum system undergoing cyclic evolution acquires a phase dependent solely on the topology and geometry of the path, independent of how fast the path is traversed. Such phases are characterized by topological invariantsglobal quantities like the Euler characteristic or Chern numberswhich classify the topology of the quantum state space and exhibit robustness against local perturbations. In the context of quantum computing, this robustness provides a natural foundation for fault-tolerant operations: by engineering quantum evolution paths that exploit topological protection, quantum information can be stored and manipulated in ways that are inherently resistant to decoherence and noise. This deep interplay between topology and quantum evolution highlights the potential of geometric and topological methods to design resilient quantum protocols, where the behavior of the system is governed more by global structure than by local interactions.

A particularly remarkable example of the influence of topology in quantum systems is the emergence of [64, 65]. These quasiparticles appear at the ends of certain superconducting nanowires and obey non-Abelian statistics [66, 67], meaning that the exchange of two such particles leads to a transformation of the quantum state dependent on the order of the exchanges. This property enables them to serve as topologically protected qubits for quantum computing. The non-local nature of Majorana modes has profound implications for quantum entanglement. Unlike traditional quantum qubits, where the entanglement is localized within a given subsystem, Majorana-based qubits exhibit topological entanglement [68], meaning that quantum information is delocalized over spatially distinct regions. This non-locality enhances their robustness against local noise and decoherence, making topological quantum computing a promising approach for fault-tolerant quantum computing [23].

In classical quantum computing architectures, qubits are extremely sensitive to environmental noise, which requires intensive error correction protocols. However, topological quantum computing exploits the intrinsic robustness of topological states to mitigate these problems. The energy gap associated with topological states protects against minor perturbations [69], thus dramatically reducing error rates. The use of anyonic statistics in topological quantum computing opens up a novel way of implementing quantum gates [70, 71]. Rather than relying on fragile unitary operations, topological quantum gates are realized by non-abelian anyon braids, which are intrinsically fault-tolerant [72]. This approach fits into the broader framework of geometric quantum evolution, where computational operations are encoded in the overall topology of the system rather than in local microscopic details.

The integration of topological concepts in quantum mechanics provides a powerful tool to improve the stability and robustness of quantum states. From Berry phases to zero-Majorana modes and non-abelian anyons, topology plays a crucial role in quantum evolution and entanglement. These insights not only shed fundamental light on the nature of quantum mechanics but also are of considerable practical importance for the development of next-generation quantum technologies. The resilience of topologically protected states to noise makes them promising candidates for the construction of scalable, fault-tolerant quantum computers, reinforcing the interplay between geometry, topology, and quantum computing.

## 2.4 Applications in Quantum Systems

Quantum systems offer a fascinating framework in which the concepts of quantum mechanics can be applied in a manifold of ways, from time optimization of dynamical evolutions to the study of non-classical correlations between particles. In this section, we explore two fundamental aspects of these systems. The first concerns the quantum brachistochrone problem [73], which involves determining the fastest evolution between two quantum states under energy constraints. This problem has applications in the field of quantum information and optimal control of quantum systems. The second concerns quantum entanglement [74] and its geometric interpretation [63,75–78]. This approach allows us to analyze this key resource of quantum mechanics from a geometric angle, in particular using the structure of Hilbert spaces and the Fubini-Study metric. It sheds new light on the profound nature of quantum correlations and their role in quantum information protocols. These themes illustrate the richness and depth of quantum phenomena, highlighting both their theoretical foundations and their practical implications.

### 2.4.1 Quantum Brachistochrone Problem

The quantum brachistochrone problem [79] is a generalization of the classical brachistochrone problem [80], which consists in determining the fastest path a particle must take under the effect of gravity to pass between two points in a vertical plane. In the quantum context, the aim is to find the minimum time required for a quantum system to evolve between two given states. The solution to this problem relies on concepts from differential geometry in the space of quantum states, notably the Fubini-Study metric, which measures the distance between two states. The optimal time is defined as the time required for the system to cover the shortest geodesic distance in this state space, maximizing the quantum evolution velocity. As we discussed earlier, the velocity of evolution of a quantum system is proportional to the energy uncertainty $\Delta E$ associated with the initial state and the Hamiltonian of the system. Greater energy uncertainty means faster evolution, reducing the time needed to reach the final state.

The geodesic distance between two states of a quantum system depends on the metric used to describe the state space. Using the Fubini-Study metric, the distance is zero when the two states are identical or when the evolution parameters lead to singular points in the state space (e.g. perfect orthogonal states). Whereas the optimal evolution time is obtained by dividing the minimum geodesic distance $\mathbf{s}_{\min}$ by the maximum velocity $\mathbf{v}_{\max}$ that the system can reach during its evolution

$$\tau = \frac{\mathbf{s}_{\min}}{\mathbf{v}_{\max}}. \tag{2.105}$$

This optimal time is independent of the intermediate parameters details and depends only on the total energy and geometric characteristics of the system. When the optimal time is reached, the system evolves along a circular or geodesic path in state space. In special cases, such as two-qubit systems, the optimal time and the ordinary time coincide, indicating that the evolution is perfectly synchronized with the dynamics of the system. In more complex systems, the optimal time is always less than the ordinary time, indicating an optimization of the evolution. In summary, the quantum brachistochrone problem provides a better understanding of optimal time evolution in quantum systems, by linking geometrical and energetic concepts to minimize the evolution time between two states.

### 2.4.2 Entanglement and Its Geometrical Interpretation

Quantum entanglement is one of the most intriguing phenomena in modern physics [81–83]. Introduced within the framework of quantum mechanics, it describes a strong correlation between the states of several particles or systems that remain connected to each other, even though they are spatially separated. This connection is such that the measurement of one of the systems instantly influences the state of the other, regardless of the distance separating them. Entanglement has played a fundamental role in understanding the fundamentals of quantum mechanics and in the development of modern quantum technologies, such as quantum cryptography and quantum computing.

What has been done so far is that quantum systems can be represented geometrically using the manifold of quantum states or the quantum phase space. And geometry provides a more intuitive way of visualizing the structure of states and their evolution. For entangled systems, these geometric representations reflect the correlations between states, as well as the underlying dynamic connections. In this context, geometric entities such as curvatures, distances and geometric phases become tools for characterizing the nature of entangled states and their evolution. The space of quantum states is often equipped with the Fubini-Study metric, a distance that measures the similarity between two quantum states. This metric gives rise to concepts such as the geodesic distance, which quantifies the minimum distance traveled between two states in this space, and is essential for evolution time optimization issues in quantum systems. Entanglement also leads to changes in the geometry of the state space, reducing the system's degrees of freedom. When a system is extremely entangled, the accessible states often form lower-dimensional subspaces in the total state space. For example, states with the same degree of entanglement may form a one- or two-dimensional curved manifold in the entire state space. This dimension reduction is crucial for quantum technologies, as it enables us to better control the evolution of quantum systems and design optimized algorithms. The influence of entanglement on geometric curvature is a key topic in quantum physics. In general, increasing quantum correlations tend to compact the structure of the state space, thus reducing the overall curvature. In specific cases, entanglement can transform a curved space into a quasi-flat or even completely flat one. Such a transformation has important implications for the way systems evolve under the effect of various interactions. Weakly entangled states are often found in regions of high curvature, while maximally entangled states are found in areas of low curvature. This indicates stabilized dynamics that are less prone to fluctuations in maximally entangled systems.

Geometric phase is another key property of entangled quantum systems. As in entangled systems, geometric phase is particularly sensitive to quantum correlations. Weakly entangled states tend to accumulate maximum geometric phase, while highly entangled states accumulate little or no phase. This behavior can be interpreted as a sign that entanglement limits the evolutionary freedom of the system, forcing it to follow geometrically optimized paths in state space. This phenomenon is particularly useful in quantum control protocols, where it is crucial to manipulate the phase of qubits with precision.

An important aspect of entanglement is its influence on the velocity of evolution of quantum systems. Weakly entangled systems evolve slowly in state space, while entanglement initially increases the speed of evolution up to a certain critical threshold. Beyond this threshold, entanglement begins to slow down the system, creating nonlinear dynamics. This observation is essential for solving problems such as the quantum brachistochrone problem. By controlling the degree of entanglement and the nature of the interactions between particles, it becomes possible to minimize the evolution time, which is crucial for quantum computing operations where the execution speed of logic gates must be optimized.

The concepts of quantum geometry linked to entanglement are now being applied in various fields of quantum technology. In quantum computing, they help to design more efficient algorithms by minimizing the distances traveled in state space or optimizing the phases acquired by qubits. In quantum cryptography, the sensitivity of entangled systems to geometric modifications enables the detection of any attempt to intercept messages. The geometry of entanglement plays a fundamental role in experiments aimed at exploring the nature of quantum reality and the limits of classical theories. By studying the geometric properties of entangled systems, physicists are better equipped to understand and manipulate quantum correlations, paving the way for new discoveries in fundamental and applied sciences.

To conclude, entanglement and its geometric interpretations offer a rich and multidimensional understanding of quantum systems. By exploring the geometry of entangled states, we gain a new perspective on the dynamics, correlations and manipulation of quantum systems. This approach is not just theoretical: it has profound practical implications for the technologies of the future, where entanglement and geometry will play a key role in the realization of advanced quantum mechanisms.

## 2.5 Summary

This chapter has unveiled the intrinsic geometric and topological foundations of quantum mechanics, anchored in the projective Hilbert space $\mathbb{C}\mathcal{P}^n$, where global phase redundancies are eliminated. Equipped with the Fubini-Study metric, this space defines distances between quantum states and governs their evolution via geodesics, while its symplectic structure mirrors classical phase space, unifying geometry and dynamics. Geometric phases such as the Berry and Aharonov-Anandan phasesemerge as curvature-driven phenomena, encoding the topology of parameter spaces and offering gauge-invariant descriptions of quantum evolution. These concepts extend to topological phases , where invariants like Chern numbers protect quantum states against perturbations, as seen in quantum Hall systems and Majorana zero modes. Applications span from the quantum brachistochrone problem , optimizing evolution times through energy uncertainty, to the geometric reinterpretation of entanglement , which constrains state-space geometry and influences quantum correlations. In the next two chapters, these geometric and topological tools will be applied to spin systems , illustrating their utility in designing robust quantum protocols, manipulating geometric phases, and harnessing topology for error-resilient quantum technologies. This geometric perspective not only deepens theoretical understanding but also paves the way for innovations in quantum control and computation.

CHAPTER 3

# Exploring Quantum Spin Systems: Insights from Dynamical Analyses

## 3.1 Overview of Spin Systems

Angular momentum is a fundamental concept in physics, relevant in both classical and quantum mechanics [84]. It describes rotational motion of a system around an axis or point. In classical mechanics, angular momentum is typically related to the spinning motion of rigid bodies, such as a top or a planet in orbit [85]. In quantum mechanics, it also encompasses intrinsic aspects of particles, such as spin [86].

In quantum mechanics, angular momentum is expressed more abstractly than in classical theory, via operators acting on wavefunctions. Two primary forms exist: orbital and spin angular momentum. The orbital component arises from a particle's motion through space, similar to its classical counterpart. It is governed by the operator $\mathbf{L}$, with components $\mathbf{L}_x, \mathbf{L}_y, \mathbf{L}_z$ that obey the commutation relations: $[\mathbf{L}_x, \mathbf{L}_y] = i\hbar\mathbf{L}_z$. These relations demonstrate that the components of orbital angular momentum cannot be measured simultaneously with perfect precision, illustrating quantum uncertainty. Nonetheless, it is possible to measure both the magnitude $\mathbf{L}^2$ and one chosen component, often $\mathbf{L}_z$, at the same time. The allowable values of orbital angular momentum are quantized in units of $\hbar$, the reduced Planck constant.

Spin is an intrinsic form of angular momentum unique to quantum particles and has no direct analogue in classical mechanics. It is an inherent property of particles, similar to mass or electric charge. For instance, an electron possesses a spin $s = \frac{1}{2}$, meaning its spin angular momentum takes discrete values proportional to $\frac{\hbar}{2}$. The spin state of an electron is typically represented by two basic states, $|+\rangle$ (up spin) and $|-\rangle$ (down spin), corresponding to $+\frac{\hbar}{2}$ and $-\frac{\hbar}{2}$ for the $S_z$ component, respectively. In quantum theory, angular momentum exhibits several core characteristics. First, it is quantizedrestricted to specific discrete values. For a spin-$\frac{1}{2}$ particle, $S_z$ only takes the values $+\frac{\hbar}{2}$ or $-\frac{\hbar}{2}$. Second, commutator relationships among the components imply that all components cannot be precisely measured simultaneously, consistent with Heisenberg's uncertainty principle. Lastly, in multi-particle systems, the total angular momentum is the sum of the individual contributions, both orbital and spin.

A particle of arbitrary spin $s$ has spin angular momentum described by the operator $\mathbf{S}$, whose components $\mathbf{S}_x, \mathbf{S}_y, \mathbf{S}_z$ satisfy the commutation rules analogous to those of orbital angular momentum:

$$[\mathbf{S}_x, \mathbf{S}_y] = i\hbar\mathbf{S}_z, \quad [\mathbf{S}_y, \mathbf{S}_z] = i\hbar\mathbf{S}_x, \quad [\mathbf{S}_z, \mathbf{S}_x] = i\hbar\mathbf{S}_y. \tag{3.1}$$

These relations also express the inability to simultaneously determine multiple spin components with arbitrary precision. Yet, as in the orbital case, the norm $\mathbf{S}^2$ and one component (typically $\mathbf{S}_z$) can be measured together. Most spin states are expanded in the eigenbasis of $\mathbf{S}_z$ and $\mathbf{S}^2$ as:

$$|\psi\rangle = \sum_{m_s=-s}^{s} c_{m_s}|s, m_s\rangle \tag{3.2}$$

where $c_{m_s}$ are complex coefficients. The operators satisfy the eigenvalue relations:

$$\begin{aligned} \mathbf{S}^2|s, m_s\rangle &= s(s+1)\hbar^2|s, m_s\rangle \\ \mathbf{S}_z|s, m_s\rangle &= m_s\hbar|s, m_s\rangle \end{aligned} \tag{3.3}$$

Here, $s$ is the spin quantum number, which can take integer or half-integer values. The magnetic spin quantum number $m_s$ ranges from $-s$ to $+s$ in integer steps, yielding $2s+1$ possible eigenvalues for $\mathbf{S}_z$.

For composite systems, the total angular momentum results from summing the individual contributions. For example, in a system of two particles with spins $s_1$ and $s_2$, the total angular momentum is defined as $\mathbf{J} = \mathbf{S}_1 + \mathbf{S}_2$. The squared norm of $\mathbf{J}$ satisfies the eigenvalue equation:

$$\mathbf{J}^2|j, m_j\rangle = j(j+1)\hbar^2|j, m_j\rangle \tag{3.4}$$

where $j$ can take values in the range:

$$j = |s_1 - s_2|, |s_1 - s_2| + 1, \ldots, s_1 + s_2. \tag{3.5}$$

This rule is central in atomic and nuclear physics, where combining angular momenta allows one to describe collective quantum states in many-particle systems.

In the quantum framework, there are two main ways to express the state of a composite system: the decoupled basis and the coupled basis. The decoupled basis consists of the tensor product of the individual angular momentum eigenstates, denoted $|s_1, m_1\rangle \otimes |s_2, m_2\rangle$, with $m_1$ and $m_2$ representing the individual spin projections along a chosen axis (commonly $z$). Conversely, the coupled basis contains states of the form $|j, m_j\rangle$, representing the eigenstates of the total angular momentum obtained by combining the subsystems according to addition rules.

In our works [75–77], we employ the decoupled basis $|s_1, m_1\rangle \otimes |s_2, m_2\rangle$ as it naturally reflects how particles interact individually before being considered collectively. In practice, many physical systems involve interactions that act on each particle's spin independently before contributing to the overall angular momentum. The transformation between decoupled and coupled bases is performed using Clebsch-Gordan coefficients, which express $|j, m_j\rangle$ states as linear combinations of $|s_1, m_1\rangle \otimes |s_2, m_2\rangle$, and vice versa.

Working in the decoupled basis has distinct benefits, particularly in calculationsboth analytical and numericalwhere interaction terms are more conveniently expressed. Moreover, in systems such as those involving spin-orbit coupling or hyperfine interactions, the temporal evolution is often clearer when states are first analyzed in the decoupled representation before transitioning, if needed, to the coupled one.

## 3.2 Dynamical Description of Two Interacting Spins Under the Anisotropic Heisenberg Model

In quantum mechanics, the study of the Heisenberg anisotropic model [87] holds particular significance, especially for analyzing the dynamics of multi-spin systems. In this analysis, we examine a two-spin system interacting through anisotropic coupling and subjected to an external magnetic field. This model serves as a framework to investigate several physical aspects, such as the system's unitary evolution and the rate at which it evolves over time. A central part of this study is to quantify how fast the system evolves using the Fubini-Study metric, which relates energy fluctuations to dynamical behavior. By evaluating the system's evolved state and the separation between states in the quantum state space, we aim to gain insight into time-related constraints and the optimal paths for state transitions. The primary goal of this work, as outlined in our first contribution [75], is to characterize the time evolution of the two-spin system, particularly the influence of the anisotropic interaction and the external magnetic field.

### 3.2.1 Unitary Evolution of the Two-Spin System

The evolution of a quantum system over time follows the Schrödinger equation. For the case of two interacting spins described by a Heisenberg anisotropic Hamiltonian with a magnetic field applied along the $z$-axis, the time evolution operator is expressed as

$$\mathbf{U}(t) = e^{-i\mathbf{H}t} \tag{3.6}$$

where $\mathbf{H}$ is the total Hamiltonian of the system:

$$\mathbf{H} = \mathbf{H}_1 + \mathbf{H}_2 + \mathbf{H}_3 \tag{3.7}$$

with

$$\mathbf{H}_1 = J(\sigma_x \otimes \sigma_x + \sigma_y \otimes \sigma_y), \qquad \mathbf{H}_2 = \nu J \sigma_z \otimes \sigma_z, \qquad \mathbf{H}_3 = \mu B_z(\sigma_z \otimes \mathbf{I} + \mathbf{I} \otimes \sigma_z). \tag{3.8}$$

with the components defined as follows:

- $\sigma_x, \sigma_y, \sigma_z$: Pauli matrices, representing spin-$\frac{1}{2}$ operators,
- $\mathbf{I}$: $2 \times 2$ identity matrix,
- $\otimes$: Tensor product acting on the two-spin (or two-qubit) Hilbert space,
- $J$: Coupling strength in the $xy$ plane (exchange interaction),
- $\nu$: Anisotropy parameter for the $z$-axis interaction,
- $B_z$: External magnetic field along the $z$-direction (in tesla),
- $\mu$: Magnetic moment associated with the spin system (e.g., $\mu = \frac{g\mu_B}{2}$).

Since the three components of the Hamiltonian commute, the time evolution operator can be separated as:

$$\mathbf{U}(t) = e^{-i\mathbf{H}_1 t} e^{-i\mathbf{H}_2 t} e^{-i\mathbf{H}_3 t}. \tag{3.9}$$

This product form offers a clear and interpretable description of the system's time evolution. The Heisenberg Hamiltonian yields the energy spectrum and corresponding eigenstates. In the standard basis $\mathcal{B} = \{|11\rangle, |10\rangle, |01\rangle, |00\rangle\}$, the eigenstates are

$$\begin{aligned}
|v_1\rangle &= |11\rangle, \\
|v_2\rangle &= \frac{1}{\sqrt{2}}(|10\rangle + |01\rangle), \\
|v_3\rangle &= \frac{1}{\sqrt{2}}(|10\rangle - |01\rangle), \\
|v_4\rangle &= |00\rangle.
\end{aligned} \tag{3.10}$$

The corresponding energy eigenvalues are:

$$E_1 = 2J + 2\mu B_z, \quad E_2 = 2J - \nu J, \quad E_3 = -2J - \nu J, \quad E_4 = 2J - 2\mu B_z. \tag{3.11}$$

The evolution characteristics depend on the coupling constant $J$, the anisotropy $\nu$, and the field intensity $\mu B_z$. The periodicity of the time evolution is governed by these quantities. If the system begins in a general state:

$$|\mathbf{\Upsilon}_i^{2,1/2}\rangle = c_i^{(11)}|11\rangle + c_i^{(10)}|10\rangle + c_i^{(01)}|01\rangle + c_i^{(00)}|00\rangle \tag{3.12}$$

where $c_i^{(kj)} \in \mathbb{C}, (j,k=0,1)$ and $|c_i^{(11)}|^2 + |c_i^{(10)}|^2 + |c_i^{(01)}|^2 + |c_i^{(00)}|^2 = 1$, then the time-evolved state becomes:

$$|\mathbf{\Upsilon}^{2,1/2}(\theta,\phi)\rangle = e^{i\frac{\nu\theta}{2}} \begin{pmatrix} c_i^{(11)} e^{-i(\phi+\nu\theta)} \\ \\ c_i^{(10)}\cos\theta - i c_i^{(01)}\sin\theta \\ \\ -i c_i^{(10)}\sin\theta + c_i^{(01)}\cos\theta \\ \\ c_i^{(00)} e^{i(\phi-\nu\theta)} \end{pmatrix}, \tag{3.13}$$

with $\eta = 2Jt, \quad \kappa = 2\mu B_z t$.

The state of the system is parametrized by two real variables $\eta$ and $\kappa$, encoding the time dependence and the magnetic field strength. Studying the periodicity of this state with respect to these parameters and the anisotropy reveals distinct dynamical scenarios with various recurrence patterns.

#### General Periodicity Criteria

If the initial state is such that the coefficients $c_i^{(10)}$ and $c_i^{(01)}$ are distinct ($c_i^{(01)} \neq \pm c_i^{(10)}$), the periodicity of the system is governed by the value of the anisotropy parameter $\nu$.

***Case (1)*: Rational $\nu = p/q$ with $p$ and $q$ coprime**

- If both $p$ and $q$ are odd, the system obeys:

$$|\mathbf{\Upsilon}^{2,1/2}(\eta,\kappa+2\pi)\rangle = |\mathbf{\Upsilon}^{2,1/2}(\eta,\kappa)\rangle, \tag{3.14}$$

$$|\mathbf{\Upsilon}^{2,1/2}(\eta+q\pi,\kappa)\rangle = e^{-i\frac{p\pi}{2}}|\mathbf{\Upsilon}^{2,1/2}(\eta,\kappa)\rangle. \tag{3.15}$$

- If one of $p$ or $q$ is even, then:

$$|\mathbf{\Upsilon}^{2,1/2}(\eta,\kappa+2\pi)\rangle = |\mathbf{\Upsilon}^{2,1/2}(\eta,\kappa)\rangle, \tag{3.16}$$

$$|\mathbf{\Upsilon}^{2,1/2}(\eta+q\pi,\kappa+\pi)\rangle = e^{-i(\frac{p}{2}+1)\pi}|\mathbf{\Upsilon}^{2,1/2}(\eta,\kappa)\rangle. \tag{3.17}$$

***Case (2)*: Irrational $\nu$**

$$|\mathbf{\Upsilon}^{2,1/2}(\eta,\kappa+2\pi)\rangle = |\mathbf{\Upsilon}^{2,1/2}(\eta,\kappa)\rangle. \tag{3.18}$$

#### Specific Cases of Evolving States

***Case (3)*: Evolution with $\eta$ only** If $c_i^{(11)} = c_i^{(00)} = 0$ and $c_i^{(10)} \neq \pm c_i^{(01)}$, then:

$$|\boldsymbol{\Upsilon}^{2,1/2}(\eta)\rangle = e^{i\frac{\nu\eta}{2}} \begin{pmatrix} 0 \\ c_i^{(10)} \cos\eta - i c_i^{(01)} \sin\eta \\ -i c_i^{(10)} \sin\eta + c_i^{(01)} \cos\eta \\ 0 \end{pmatrix}. \tag{3.19}$$

$$|\boldsymbol{\Upsilon}^{2,1/2}(\eta + \pi)\rangle = -e^{i\frac{\nu\pi}{2}} |\boldsymbol{\Upsilon}^{2,1/2}(\eta)\rangle. \tag{3.20}$$

***Case (4)*: Evolution with $\kappa$ only** If $c_i^{(11)} \neq 0$, $c_i^{(00)} \neq 0$, and $c_i^{(10)} = c_i^{(01)} = 0$, then:

$$|\boldsymbol{\Upsilon}^{2,1/2}(\kappa)\rangle = e^{-i\frac{\nu\eta}{2}} \begin{pmatrix} c_i^{(11)} e^{-i\kappa} \\ 0 \\ 0 \\ c_i^{(00)} e^{i\kappa} \end{pmatrix}. \tag{3.21}$$

$$|\boldsymbol{\Upsilon}^{2,1/2}(\kappa + \pi)\rangle = -|\boldsymbol{\Upsilon}^{2,1/2}(\kappa)\rangle. \tag{3.22}$$

***Case (5)*: Coupled evolution with $\nu = \pm 1$** If $\nu = 1$ with $c_i^{(01)} = c_i^{(10)}$, or $\nu = -1$ with $c_i^{(01)} = -c_i^{(10)}$, and at least one of $c_i^{(11)}$, $c_i^{(00)}$ is nonzero, then:

$$|\boldsymbol{\Upsilon}^{2,1/2}(\eta, \kappa)\rangle = e^{\mp\frac{i\eta}{2}} \begin{pmatrix} c_i^{(11)} e^{-i\kappa} \\ c_i^{(10)} \\ c_i^{(10)} \\ \pm c_i^{(00)} e^{i\kappa} \end{pmatrix}. \tag{3.23}$$

$$|\boldsymbol{\Upsilon}^{2,1/2}(\kappa + 2\pi)\rangle = |\boldsymbol{\Upsilon}^{2,1/2}(\kappa)\rangle. \tag{3.24}$$

***Case (6)*: Coupled evolution for general $\nu$** If $\nu \neq 1$, $c_i^{(10)} = c_i^{(01)}$ or $\nu \neq -1$, $c_i^{(10)} = -c_i^{(01)}$, and at least one of $c_i^{(11)}$, $c_i^{(00)}$ is nonzero, then:

$$|\boldsymbol{\Upsilon}^{2,1/2}(\eta, \kappa)\rangle = e^{i\frac{\nu\eta}{2}} \begin{pmatrix} c_i^{(11)} e^{-i(\kappa + \nu\eta)} \\ c_i^{(10)} e^{\mp i\eta} \\ \pm c_i^{(10)} e^{\mp i\eta} \\ c_i^{(00)} e^{i(\kappa - \nu\eta)} \end{pmatrix}. \tag{3.25}$$

$$|\boldsymbol{\Upsilon}^{2,1/2}(\eta, \kappa + 2\pi)\rangle = |\boldsymbol{\Upsilon}^{2,1/2}(\eta, \kappa)\rangle, \tag{3.26}$$

$$\left|\boldsymbol{\Upsilon}^{2,1/2}\left(\eta + \frac{\pi}{\nu \mp 1}, \kappa + \pi\right)\right\rangle = e^{-i\frac{\pi}{2(\nu \mp 1)}} |\boldsymbol{\Upsilon}^{2,1/2}(\eta, \kappa)\rangle. \tag{3.27}$$

These recurrence properties show how both the anisotropy parameter $\nu$ and the initial conditions influence the structure of the evolving state over time.

To determine the evolution speed, we use the Fubini-Study metric, which provides a measure of quantum state separation and will be explored more thoroughly in the next chapter within the context of projective Hilbert space. The metric is given by:

$$d\mathbf{s}^2 = \left[(\nu^2 - 1)A + 1 - (\nu A + F)^2\right] d\eta^2 + 2D\left[\nu - (\nu A - F)\right] d\kappa d\eta + [A - D^2]d\kappa^2, \tag{3.28}$$

where

$$A = |c_i^{(11)}|^2 + |c_i^{(00)}|^2, \quad D = |c_i^{(11)}|^2 - |c_i^{(00)}|^2, \quad F = 2\mathrm{Re}(c_i^{(10)} c_i^*).$$

The anisotropy $\nu$ and the coefficients of the initial state determine the trajectory and speed in quantum state space. The rate of evolution, obtained through the Fubini-Study distance, is essential for understanding transition times, coherence properties, and applications in quantum control and spin-based technologies.

### 3.2.2 Quantum Evolution Speed and Brachistochrone Problem

The quantum evolution of a physical system is strongly influenced by the energetic uncertainties that condition its dynamics. For the system of two coupled spins subject to the anisotropic Heisenberg model and placed in an external magnetic field, the evolution speed is given by the relation

$$\mathbf{v} = \frac{d\mathbf{s}}{dt} = 2\Delta E \tag{3.29}$$

where $\Delta E$ corresponds to the energy fluctuation associated with the system's Hamiltonian. The time-dependent state of the quantum system evolves as

$$|\boldsymbol{\Upsilon}^{2,1/2}(\eta)\rangle = e^{i\frac{\nu\eta}{2}} \begin{pmatrix} c_i^{(11)} e^{-i(k+\nu)\eta} \\ c_i^{(10)} \cos\eta - i c_i^{(01)} \sin\eta \\ -i c_i^{(10)} \sin\eta + c_i^{(01)} \cos\eta \\ c_i^{(00)} e^{i(k-\nu)\eta} \end{pmatrix} \tag{3.30}$$

with $k = \mu B_z/J$. This formulation reveals that the system's evolution traces a circular path whose radius is determined by the anisotropy parameter and the initial energy distribution. The influence of the magnetic field and the coupling strength $J$ on the velocity of evolution appears in the refined expression of $v$:

$$\mathbf{v} = 2\left[\mu B_z^2(A - D^2) - J^2((\nu A + F)(1 + \nu A + F) + (\nu^2 - 1)A + 1) + 2J\mu B_z k\nu\right]^{1/2}. \tag{3.31}$$

The magnetic field $B_z$ affects the dynamics via the parameter $k = \mu B_z/J$, while the interaction strength $J$ contributes through its quadratic terms. Their combined influence is captured by the mixed term $2J\mu B_z k\nu$, indicating interplay between external and internal features of the system.

As previously mentioned in Section (2.4.1), the quantum brachistochrone problem concerns finding the optimal evolution path of a quantum system to minimize the time needed to transition from an initial state to a target final state. This question is of major relevance in quantum control and in designing fast quantum gates.

In the considered system, optimal evolution is realized by reducing the geodesic distance traveled while increasing the evolution speed. The length of the path followed during evolution is then

$$\mathbf{s} = \mathbf{v}t \tag{3.32}$$

where $\mathbf{s}$ denotes the geodesic separation between the initial and final quantum states, and $\mathbf{v}$ is the system's instantaneous speed.

A targeted approach consists of adjusting the anisotropy parameters of the system. Suppose that $\nu$ follows a sinusoidal time profile of the form $\nu = (1/4)\sin 2\eta$, the speed then becomes

$$\mathbf{v} = 2J\left[1 - (A\sin 2\eta + F)(1 + A\sin 2\eta + F) + (\sin^2 2\eta - 1)A + 2Ak\sin 2\eta + J^2(A - D^2)\right]^{1/2}. \tag{3.33}$$

This expression illustrates how the velocity depends on the initial parameters, particularly on the anisotropy in the Heisenberg interaction. For a special choice of parameters where $A = 1$, $D = -1/2$ and $k = 1$, the velocity simplifies to

$$\mathbf{v} = J\sqrt{3 - 2\sin 2\eta} \tag{3.34}$$

and the total geodesic path is given by

$$\mathbf{s} = \int \mathbf{v}dt = \frac{\eta}{2}\sqrt{3 - 2\sin 2\eta}. \tag{3.35}$$

By identifying the value(s) of $\eta$ that maximize the velocity, we determine the minimal time needed to complete the evolution path:

$$\mathcal{T} = \frac{3\pi}{8J} \tag{3.36}$$

showing that this optimal time scales inversely with the interaction strength $J$. This result highlights how the dynamics are primarily governed by the strength of coupling between the spins, and indicates that adjusting system parameters can reduce the duration required for specific quantum operations. The solution of the quantum brachistochrone problem has concrete implications in the realization of efficient quantum circuits [88], where minimizing the execution time of unitary transformations is key to maintaining coherence and performance.

### 3.2.3 Entanglement Dynamics in the Two-Spin System

Quantum entanglement is a key feature of multi-particle systems, particularly in the context of the Heisenberg anisotropic model. Entanglement between the two spins is usually measured using the Wootters concurrence [50,74], which provides a precise quantification of the degree of quantum entanglement in the system. This measure is defined by the following expression:

$$\mathcal{C} = |2c_i^{(11)}c_i^{(00)}e^{-i2\eta(\nu+1)} + ie^{-i2\eta}(c_i^{(10)} + c_i)^2 \sin 2\eta - 2c_i^{(10)}c_i|. \tag{3.37}$$

The evolution of entanglement in a spin system is predominantly governed by the Heisenberg exchange interaction, whereas it remains unaffected by the presence of an external magnetic field. This invariance arises from the fact that the unitary evolution induced by the magnetic field acts independently on each spin, thereby preserving the initial degree of entanglement throughout the dynamics. In contrast, the exchange interaction, which couples all spins globally, drives the evolution of entanglement by altering the quantum correlations within the system. Furthermore, the interplay between these interactions and the system initial conditions reveals the sensitivity of the geometric structure of the state space to quantum correlations. In particular, the anisotropic properties of the system, controlled by the parameter $\nu$, introduce additional variations in entanglement dynamics. This analysis underscores the deep connection between spin system evolution, quantum correlations, and the underlying geometry of the state space.

This expression reveals that the concurrence depends solely on the exchange interaction between the spins and does not depend directly on the effect of the magnetic field. Indeed, the evolution operator generated by the magnetic field acts independently on each spin, which does not alter the entanglement. On the other hand, the Heisenberg interaction acts collectively on both spins, altering their degree of entanglement over time.

The analysis of entanglement in a two-spin system requires a precise characterization of the state space and its evolution under interactions. We begin by considering an initially non-entangled state defined as

$$|\mathbf{\Upsilon}_0^{2,1/2}\rangle = |+-\rangle, \tag{3.38}$$

where the basis states are given by

$$|+\rangle = \cos\frac{\chi}{2}|1\rangle + \sin\frac{\chi}{2}e^{i\gamma}|0\rangle, \quad |-\rangle = -\sin\frac{\chi}{2}|1\rangle + \cos\frac{\chi}{2}e^{i\gamma}|0\rangle. \tag{3.39}$$

These states correspond to the eigenvectors of the spin projection operator $\mathbf{S}\cdot\mathbf{n}$, where the unit vector

$$\mathbf{n} = (\sin\chi\cos\gamma, \sin\chi\sin\gamma, \cos\chi)^T \tag{3.40}$$

defines the quantization axis. The parameters $\chi \in [0,\pi]$ and $\gamma \in [0,2\pi]$ correspond to the polar and azimuthal angles, respectively, specifying the spin orientation on the Bloch sphere $S^2$.

When a strong external magnetic field $\mathbf{B}$ is applied along $\mathbf{n}$, the system enters an eigenstate of the total Hamiltonian for $\mathbf{k} \gg 1$, facilitating experimental preparation. Under time evolution, the system's state evolves into

$$|\mathbf{\Upsilon}^{2,1/2}(\eta,\kappa)\rangle = e^{i\frac{\nu\eta}{2}} \begin{pmatrix} -\frac{1}{2}\sin\chi\, e^{-i(\kappa+\nu\eta)} \\ \\ e^{i\gamma}\left(\cos^2\frac{\chi}{2}\cos\eta + i\sin^2\frac{\chi}{2}\sin\eta\right) \\ \\ e^{i\gamma}\left(-\sin^2\frac{\chi}{2}\cos\eta - i\cos^2\frac{\chi}{2}\sin\eta\right) \\ \\ \frac{1}{2}\sin\chi\, e^{i(\kappa-\nu\eta+2\gamma)} \end{pmatrix}. \tag{3.41}$$

Entanglement in the system is quantified through the concurrence, defined as

$$\mathcal{C} = \frac{1}{2}\left[\mathcal{C}_+^2\sin^4\chi + (2\sin 2\eta + \mathcal{C}_+\sin^2\chi)^2\right]^{1/2}, \tag{3.42}$$

where $\mathcal{C}_+ = \cos 2\nu\eta - \cos 2\eta$.The temporal evolution of the entanglement measure $\mathcal{C}$ exhibits distinct regimes depending on the initial parameters. When the phase parameter takes the values $\chi = 0$ or $\chi = \pi$, the entanglement undergoes purely oscillatory dynamics, following the relation

$$\mathcal{C} = |\sin 2\eta|. \tag{3.43}$$

This implies that anisotropy has no influence on the entanglement behavior. In this regime, maximum entanglement is achieved for $\eta = \pi/4$ and $\eta = 3\pi/4$, while separable states appear at $\eta = 0$ and $\eta = \pi/2$. However, when the phase is set to $\chi = \pi/2$, the entanglement measure explicitly depends on the anisotropy parameter $\nu$, following the expression

$$\mathcal{C} = |\sin(\nu+1)\eta|. \tag{3.44}$$

In this case, the locations of maximally entangled states shift to $\eta = \pi/2(\nu+1)$, whereas separable states occur at $\eta = 0$ and $\eta = \pi/(\nu+1)$. This result highlights the impact of anisotropy, which not only alters the amplitude of entanglement oscillations but also modifies the positions of entangled states in the state space. Furthermore, the time required to reach maximum entanglement is given by

$$\mathcal{T} = \frac{\pi}{4\mathrm{J}(\nu+1)}, \tag{3.45}$$

indicating that an increase in either the anisotropy parameter $\nu$ or the interaction strength J leads to a faster entanglement evolution.

The evolution of entanglement between the two spins can be analyzed over a sufficiently short time interval. To gain insights into this behavior, it is instructive to examine the concurrence $\mathcal{C}$ Eq. (3.42) by performing a second-order expansion in $\eta$, yielding

$$\mathcal{C} \simeq \eta\left[2 + (\nu-1)\sin^2\chi\right]. \tag{3.46}$$

This expression highlights two specific cases in which entanglement vanishes. First, for $\eta = 0$, the evolved state Eq. (3.41) coincides with the initial state Eq. (3.38), implying that no evolution occurs and entanglement remains unchanged. Second, when $\nu = -1$ and $\chi = \pi/2$, the system never becomes entangled. In this scenario, the initial state takes the form

$$|\boldsymbol{\Upsilon}_i^{2,1/2}\rangle = \frac{1}{2}e^{-i\eta/2}\left[-e^{-i(\varphi-\eta)}|11\rangle + e^{i(\eta+\gamma)}(|10\rangle - |01\rangle) + e^{i(\varphi+\eta+2\gamma)}|00\rangle\right], \tag{3.47}$$

which corresponds to an eigenstate of the system Hamiltonian Eq. (3.7), thereby preventing the emergence of any quantum correlations between the two spins.

Apart from these special cases, entanglement evolves linearly with time. The rate of this evolution depends not only on the initial parameters of the system but also on the degree of anisotropy characterizing the spin-spin interaction.

The evolution of entanglement in a two-spin system is investigated while maintaining all other parameters fixed. The objective is to analyze the system's dynamics by examining the evolution of its quantum correlations. Based on the approximation presented in Equation (3.46), the Fubini-Study metric, can be formulated as

$$d\mathbf{s}^2 = \frac{1}{4}\left[\frac{2(\nu^2 + \mathbf{k}^2 - 1)\sin^2\chi - (\nu-1)^2\sin^4\chi + 4}{\left(2 + (\nu-1)\sin^2\chi\right)^2}\right]d\mathcal{C}^2. \tag{3.48}$$

This result indicates that the dynamics of system unfolds on a flat, one-dimensional manifold parametrized by the competition parameter $\mathcal{C}$. Consequently, the infinitesimal distance between two neighboring entangled states can be quantified by the differential variation of entanglement between them. More generally, the geodesic distance between two entangled states $|\boldsymbol{\Upsilon}^{2,1/2}(\mathcal{C}_i)\rangle$ and $|\boldsymbol{\Upsilon}^{2,1/2}(\mathcal{C}_f)\rangle$, where $\mathcal{C}_f > \mathcal{C}_i$, is given by

$$\mathbf{s} = g_{\mathcal{C}\mathcal{C}}\left(\mathcal{C}_f - \mathcal{C}_i\right), \tag{3.49}$$

demonstrating that the evolution of the system is fundamentally governed by the difference in entanglement between these two states. Consequently, the dynamics of the two-spin system can be directly analyzed in terms of its entanglement properties. Specifically, by utilizing equation (3.48), we derive the system evolution speed as an explicit function of its entanglement, highlighting its crucial role in the system's temporal behavior. One finds

$$\mathbf{v}_{\mathcal{C}} = \frac{1}{2}\sqrt{\frac{2(\nu^2 + \mathbf{k}^2 - 1)\sin^2\chi - (\nu-1)^2\sin^4\chi + 4}{\left(2 + (\nu-1)\sin^2\chi\right)^2}}. \tag{3.50}$$

This equation characterizes speed of the system evolution in terms of dynamics of its quantum correlations, thereby emphasizing the fundamental role of entanglement in governing the system's dynamical properties. This analysis is particularly pertinent to addressing the associated quantum brachistochrone problem. To extend this investigation further, we now consider an infinitesimal evolution of the anisotropy parameter, given by $\nu = \frac{1}{4}\sin 2\eta \approx \frac{et}{2}$.. Substituting this relation into Eq. (3.46) and imposing the simplifying conditions $\chi = \pi/2$ and $k = 1$, the expression for Fubini-Study metric simplifies to

$$d\mathbf{s}_{\mathcal{C}}^2 = \frac{1}{4}\left(1 + \frac{8}{\left(1+\sqrt{1+2\mathcal{C}}\right)^2}\right)d\mathcal{C}^2. \tag{3.51}$$

In this configuration, the state space forms a curved one-dimensional manifold whose curvature varies at each point. The corresponding evolution speed is given by

$$\mathbf{v}_{\mathcal{C}} = \frac{1}{2}\sqrt{1 + \frac{8}{\left(1+\sqrt{1+2\mathcal{C}}\right)^2}}. \tag{3.52}$$

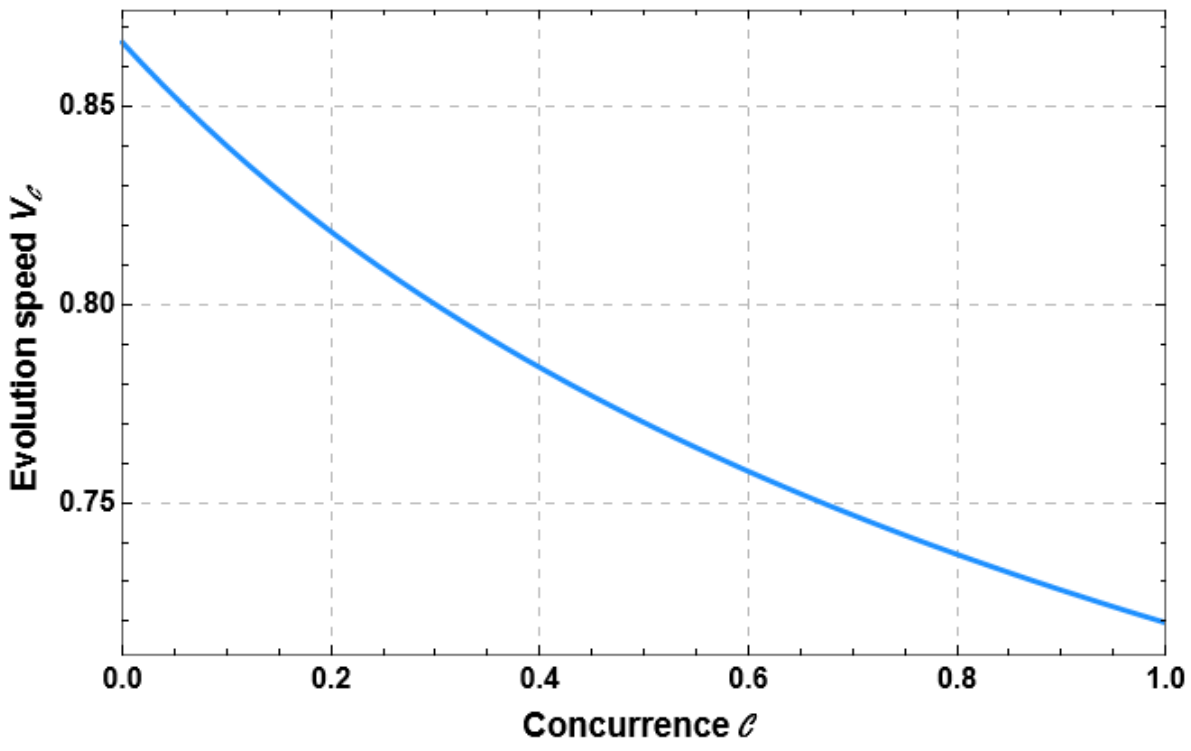

(a) Speed evolution (3.52) versus entanglement degree (3.46).

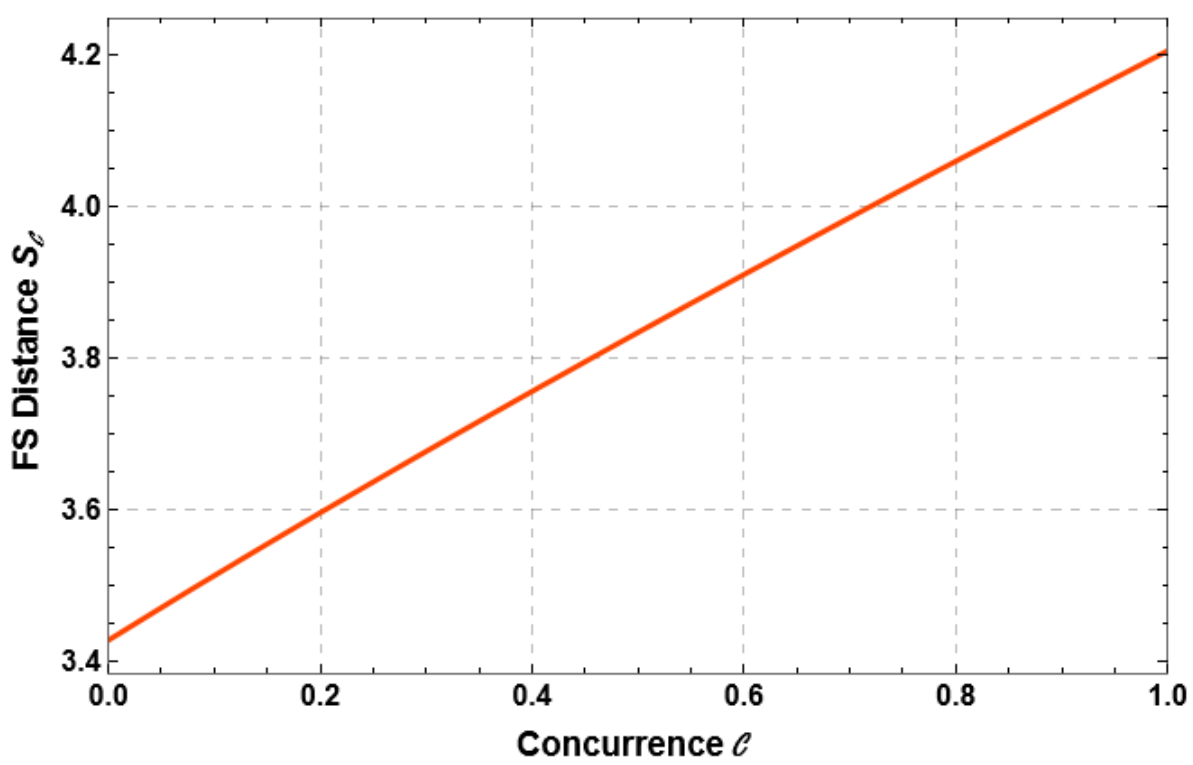

(b) Geodesic distance (3.53) versus entanglement degree (3.46).

From Figure (3.1a), we observe that this speed decreases as the degree of entanglement $\mathcal{C}$ between the two spins increases. This reduction arises from the curvature of the state space, which constrains the system's evolution. Consequently, non-entangled states evolve faster than entangled ones. The maximum speed, attained for $\mathcal{C} = 0$, is $\mathbf{v}_{\mathcal{C}} = \frac{\sqrt{3}}{2}$. By integrating equation (3.51), we obtain the geodesic distance traversed by the system as a function of entanglement:

$$\mathbf{s}_{\mathcal{C}} = \frac{(\eta_1+1)}{2\eta_2}\left[\eta_2(\eta_1-1) + 4\sqrt{2}\,\mathrm{arcsinh}\left(\frac{\eta_1+1}{2\sqrt{2}}\right) + 4\log\left(\frac{2(\eta_2+2)}{\eta_1+1}\right)\right]\mathbf{v}_{\mathcal{C}}. \tag{3.53}$$

With $\eta_1 = \sqrt{1+2\mathcal{C}}$ and $\eta_2 = \sqrt{\mathcal{C}+\eta_1+5}$, analysis of Figure (3.1b) reveals that this geodesic distance increases with entanglement (3.46). Indeed, since the state space (3.51) is entirely parameterized by the entanglement degree $\mathcal{C}$, maximally entangled states travel a greater distance than weakly entangled or non-entangled states. The minimum value of this distance is reached at $\mathcal{C} = 0$ and is $\mathbf{s}_{\mathcal{C},\min} = 3.42$. Consequently, the minimum time required to achieve time-optimal evolution is given by $\tilde{\tau} = \frac{\mathbf{s}_{\mathcal{C},\min}}{\mathbf{v}_{\mathcal{C},\max}} = 3.95$. This result shows that non-entangled states are the most suitable quantum states for this optimal evolution. Furthermore, these states can be generated via the following unitary transformation

$$|\tilde{\Upsilon}_{\rm i}\rangle \rightarrow |\tilde{\Upsilon}(\mathcal{C}=0,\tilde{\tau})\rangle = e^{-i\mathrm{H}\tilde{\tau}}|\tilde{\Upsilon}_{\rm i}\rangle, \tag{3.54}$$

where $|\tilde{\Upsilon}_{\rm i}\rangle = |\Upsilon_{\rm i}(\chi=\pi/2,\mathbf{k}=1)\rangle$. Thus, the entanglement between the two spins solved the quantum brachistochrone problem.

## 3.3 Dynamical Description of Many Interacting Qbits Under the Long-Range Ising Model

The long-range Ising model is essential to studying collective dynamics in multi-spin systems in quantum mechanics, where every spin interacts with every other spin via Ising-type couplings. These systems are especially useful for researching many-body dynamics, quantum simulation platforms, and quantum information processing phenomena. We study the unitary time evolution and the dynamical properties of a system of $N$ spin-$\frac{1}{2}$ particles that interact with each other under the all-range Ising Hamiltonian. A direct connection between energy fluctuations and the geometry of quantum state space is established by evaluating the evolution speed using the Fubini-Study metric. This method gives us a better understanding of the system's basic time limits and the best paths for changing states. This analysis is a key part of our paper [76], in which we expand the geometric description of quantum dynamics from systems with a few spins to systems with many bodies that interact over long distances.

### 3.3.1 Unitary evolution

The system under consideration is composed of a number $N$ spin-1/2 particles under the long-range Ising interaction. The Hamiltonian of the system is given by the following expression

$$\mathbf{H}^{N,1/2} = J\left(\sum_{i=1}^{N} S_i^z\right)^2 \tag{3.55}$$

where $J$ denotes the coupling strength characterizing the interaction, and $S_i^z$ is the z-axis component of the spin operator for the $i^{th}$ spin. This Hamiltonian reflects a collective interaction between all spins in the system, and its energy spectrum has a particular structure, determined by the total number of spins oriented in a given direction.

The system's quantum state can be expressed in terms of the spherical angles $\eta$ and $\phi$, which describe the direction of an orientation vector in the Bloch sphere. The system is initially in the following state

$$|+\rangle = \cos\frac{\eta}{2}|1\rangle + \sin\frac{\eta}{2}e^{i\phi}|0\rangle. \tag{3.56}$$

For a system of $N$ spins, the initial state becomes a superposition of the product states of each spin

$$|\mathbf{\Upsilon}_i^{N,1/2}\rangle = |+\rangle^{\otimes N}. \tag{3.57}$$

The time evolution of the system is governed by the unitary evolution operator, given by : $U(t) = e^{-iHt}$. When we apply this operator to the initial state $|\mathbf{\Upsilon}_i^{N,1/2}\rangle$, we obtain the evolved state of the system

$$|\mathbf{\Upsilon}_t^{N,1/2}\rangle = \sum_{p=0}^{N} C_p^N \cos^{N-p}\frac{\eta}{2}\sin^p\frac{\eta}{2} e^{-i\kappa(N-2p)^2/4}|p\rangle \tag{3.58}$$

where $C_p^N$ are the binomial coefficients and $|p\rangle$ represents a quantum state with $p$ downward-directed spins. The evolved state depends explicitly on the dynamical parameter $\kappa = Jt$, which represents the cumulative effect of coupling between the spins on the time evolution of the system. Since the initial state is a superposition of Hamiltonian eigenstates, its time evolution reveals rich, oscillatory dynamics, with periodic $2\pi$ dependence.

### 3.3.2 Quantum Evolution Speed and Optimal Path for N Qubits system

Now, we will investigate the speed of evolution and the traveled distance measured by the Fubini-Study metric - *which will be discussed in detail in the next chapter* - in order to solve the quantum brachistochrone problem, which aims to determine the minimum time needed to reach a given final state from an initial state. Under the assumption that the evolution of the system depends only on time, the metric tensor Eq. (4.27) simplifies to

$$d\mathbf{s}^2 = g_{\kappa\kappa} d\kappa^2 \tag{3.59}$$

where explicitly

$$d\mathbf{s}^2 = \frac{1}{4}N(N-1)\sin^2\eta\left(N-1-(2N-3)\sin^2\eta\right)d\kappa^2, \tag{3.60}$$

this implies that the dynamics of the system takes place on a circle of radius $\sqrt{g_{\kappa\kappa}}$. Thus, the velocity of evolution of the quantum state is expressed as :

$$\mathbf{v} = \frac{d\mathbf{s}}{dt} = 2\frac{\Delta E(t)}{\hbar}, \tag{3.61}$$

where $\Delta E$ represents the energy uncertainty of the system Hamiltonian. This relationship shows that the greater energy uncertainty, the faster system evolves. Substituting the expression of $d\mathbf{s}^2$, the evolution speed becomes

$$\mathbf{v} = J\sqrt{\frac{2}{N(N-1)}\sin^2\eta\left(N-1-(2N-3)\sin^2\eta\right)}. \tag{3.62}$$

This result shows that the evolution rate is influenced by the coupling constant $J$ and the initial parameters $(\eta, N)$. In particular, when $\eta = 0$ or $\eta = \pi$, the velocity cancels out, consistent with the fact that the wave function is not defined at these points.

After solving the quantum brachistochrone problem within the framework of the Heisenberg anisotropic model, we now analyze the optimal evolution of the system under study. To determine the minimum time required for evolution between two given states, we maximize the evolution speed $\mathbf{v}$. This amounts to solving the equation $d\mathbf{s}/d\eta = 0$, which leads to

$$\sin\eta_{\max} = \sqrt{\frac{N-1}{2N-3}}. \tag{3.63}$$

The maximum speed of the system is thus obtained as

$$\mathbf{v}_{\max} = J\frac{(N-1)}{\sqrt{N(N-1)(2N-3)}}. \tag{3.64}$$

The optimal evolution is then characterized by a Fubini-Study distance given by :

$$\mathbf{s} = \kappa\sqrt{\frac{2}{N(N-1)}\sin^2\eta\left(N-1-(2N-3)\sin^2\eta\right)} \tag{3.65}$$

Since speed $\mathbf{v}$ is independent of time, the distance $\mathbf{s}$ is a linear function of time. At singular points $\eta = 0$ and $\eta = \pi$, the distance cancels out, confirming that the system state is not defined at these points. The minimum distance the system must travel is obtained for $\eta = \pi/2$, and is equal to

$$\mathbf{s}_{min} = \kappa\sqrt{\frac{N}{2(N-1)}}, \tag{3.66}$$

therefore, the minimum time required to achieve the optimal evolution is given by :

$$\mathfrak{T} = \frac{(N-1)}{J\sqrt{2N-3}}, \quad \text{with } N > 1. \tag{3.67}$$

This result means that the optimal evolution is reached when the evolution time is as short as possible, which is an essential feature for quantum systems. The final state can be obtained by the unitary transformation

$$|\mathbf{\Upsilon}^{N,1/2}(\mathfrak{T})\rangle = e^{-i\mathbf{H}\mathfrak{T}}|\mathbf{\Upsilon}_i^{N,1/2}\rangle. \tag{3.68}$$

In this case, the Fubini-Study metric of optimal evolution takes the form

$$d\mathbf{s}^2_{\text{op}} = \frac{1}{4}N(N-1)\sin^2\eta\left(N-1-(2N-3)\sin^2\eta\right)d\kappa^2_{\min} \tag{3.69}$$

where $\kappa_{\min} = Jt_{\min}$. We note that the optimal time depends only on the number of particles $N$. For $N = 2$, $t_{\min}$ coincides with the ordinary time $t$. However, for $N \geq 3$, we observe that $t_{\min}$ is strictly less than ordinary time, meaning that the optimal time evolution is achievable. In the thermodynamic limit ($N \to \infty$), the optimal time tends towards zero ($t_{\min} \to 0$). In this case, the optimal state circle becomes a straight line, as its radius $\sqrt{g_{\kappa_{\min}\kappa_{\min}}}$ becomes infinite.

From the foregoing, we can observe that the number of particles $N$ and the energy uncertainty $\Delta E$ are crucial physical quantities for the optimization of quantum evolution in an integrable system. In addition, the link between the geometric structure of the state space and the dynamics of the system paves the way for further analysis of the role of quantum entanglement in these optimal processes.

### 3.3.3 Dynamical Effects of Quantum Correlations in Two Spin-$\frac{1}{2}$ Entanglement (N=2)

In this section, we focus on the dynamical aspect of entanglement, highlighting the link between the amount of entanglement exchanged between the two spins and certain relevant dynamical properties, such as the velocity of evolution and the geodesic distance traveled in the associated phase space. This approach allows us to tackle the quantum brachistochrone problem based on the degree of entanglement of the two-spin system. For a two-spin system, the wave function of the global quantum system (3.55) is restricted to the following form

$$|\Upsilon_t^{2,\frac{1}{2}}\rangle = e^{-i\kappa(t)}\cos^2\frac{\eta}{2}|11\rangle + \frac{1}{2}e^{i\varphi}\sin\eta(|10\rangle + |10\rangle) + e^{i(2\varphi-\kappa(t))}\sin^2\frac{\eta}{2}|00\rangle. \tag{3.70}$$

This expansion highlights the decomposition of the quantum state into a linear combination of the ground states $|11\rangle$, $|10\rangle$, $|01\rangle$ and $|00\rangle$, with complex coefficients depending on the parameters $\eta$, $\varphi$ and $\kappa(t)$. In order to identify the entanglement degree associated with the state of two spins (3.70), we use Wootters's concurrence, and after a simple calculation, we obtain

$$\mathcal{C} = \sin^2\eta|\sin\kappa|. \tag{3.71}$$

This relation shows that each pair of spins in the overall system (3.55) is entangled in the same way as any other pair. Thus, the entanglement is uniformly distributed between the spin pairs. Moreover, this entanglement evolves in a periodic manner over time, meaning that it is influenced by the system's dynamics. It also depends on the initial parameter $\eta$, indicating that the degree of entanglement can be determined by the choice of initial state.

In particular, when $\kappa = \pi/2$ and $\eta = \pi/2$, the two-spin state (3.70) reaches its maximum level of entanglement with $\mathcal{C} = 1$. On the other hand, for $\eta = 0$ or $\pi$, the spins can never be entangled ($\mathcal{C} = 0$), as the corresponding initial states $|11\rangle$ and $|11\rangle$ are eigenstates of the Hamiltonian. This absence of entanglement for these particular values of $\eta$ can also be explained from a topological and geometric point of view by the existence of a conic defect close to these singular points.

Inserting equation (3.71) into (3.62), we express the system's speed of evolution as a function of concurrence as follows:

$$\mathbf{v} = \frac{J}{2}|\sin\kappa|\left(\mathcal{C}\sqrt{2|\sin\kappa| - \mathcal{C}}\right). \tag{3.72}$$

This equation establishes an explicit link between the speed of the system and its degree of entanglement. In other words, system dynamics are directly influenced by the evolution of quantum correlations. This means that the system's dynamic characteristics can be determined by its degree of entanglement.

Two distinct regimes can be observed in the evolution of velocity (3.72) as a function of competition:

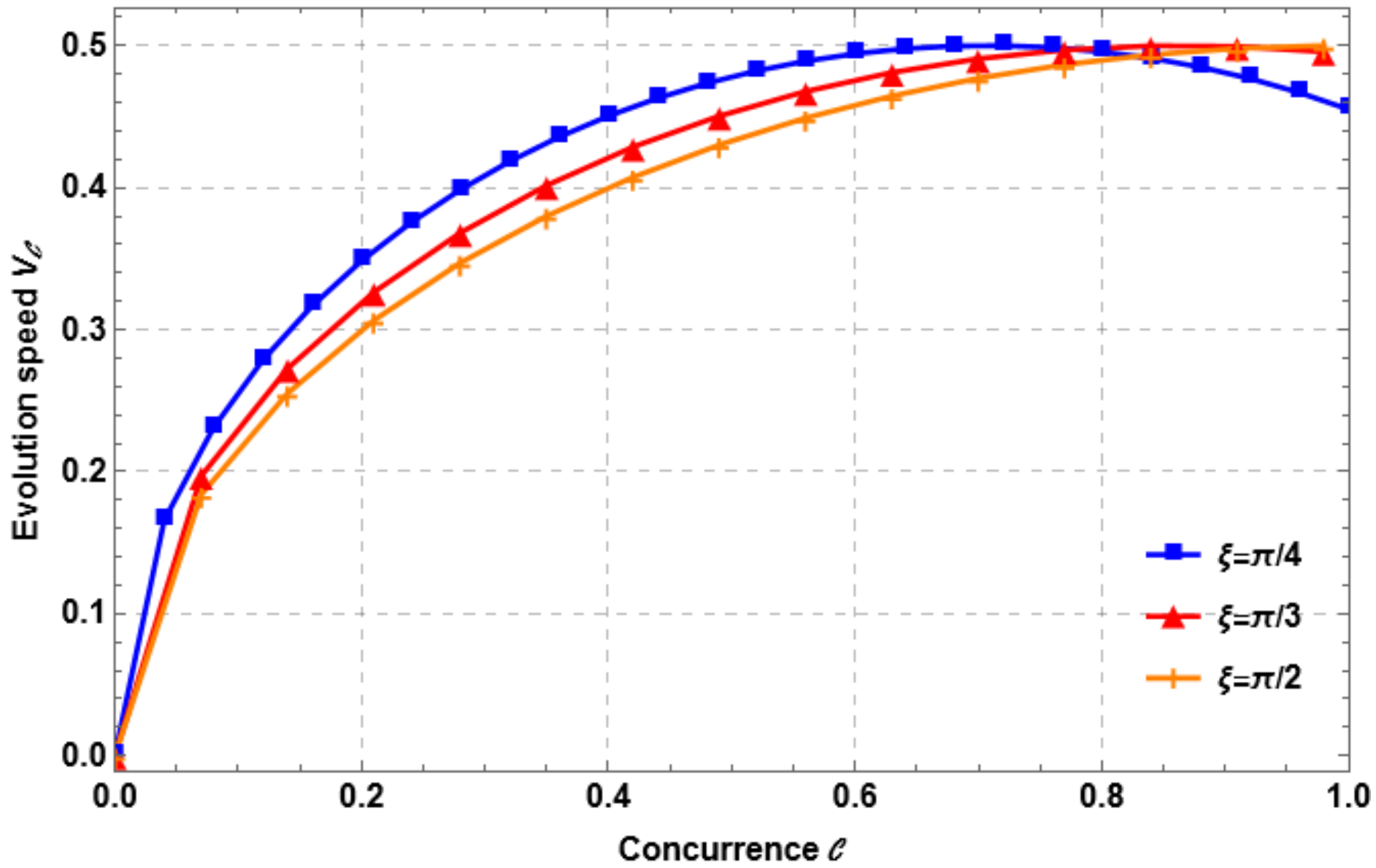

Figure 3.2: Evolution speed Eq. (3.72) versus the entanglement degree

- *First regime* $\mathcal{C} \in [0, \mathcal{C}_c]$: the evolution speed increases with the degree of entanglement until it reaches a maximum $V_{\max} = \frac{J}{2}$, corresponding to the critical entanglement level $\mathcal{C} = \mathcal{C}_c = |\sin\kappa|$. This shows that, in this stage, the presence of quantum correlations accelerates the system evolution in its phase space.

- *Second regime* $\mathcal{C} \in [\mathcal{C}_c, 1]$: beyond the critical threshold, the evolution speed changes its trend and decreases continuously until it reaches its local minimum for $\mathcal{C} = 1$. This implies that quantum correlations slow down evolution in this regime.

We therefore conclude that the dynamics of the system can be controlled by its entanglement level, a result that can be exploited in quantum information protocols. Using the Fubini-Study metric, we can express the distance traveled by the system state as a function of competition as

$$\mathbf{s} = \frac{\kappa}{2|\sin\kappa|}\sqrt{\mathcal{C}\,(2|\sin\kappa| - \mathcal{C})}. \tag{3.73}$$

This distance, which measures the separation between two quantum states in the phase space of the two-spin system, is thus related to the entanglement level and the evolution time. Analysis of Fig. (3.2) and Fig. (3.3) shows that the Fubini-Study distance Fig. (3.73) follows the same behavior as the evolution speed Fig. (3.73) as a function of the degree of entanglement. We therefore draw the same conclusions: entanglement plays a key role in system dynamics, and its properties can be investigated experimentally for applications in quantum technologies [89, 90].

Next we turn to the question of optimal evolution of the system by maximizing its speed (3.72) as a function of concurrence. The minimum time required to achieve optimal temporal evolution

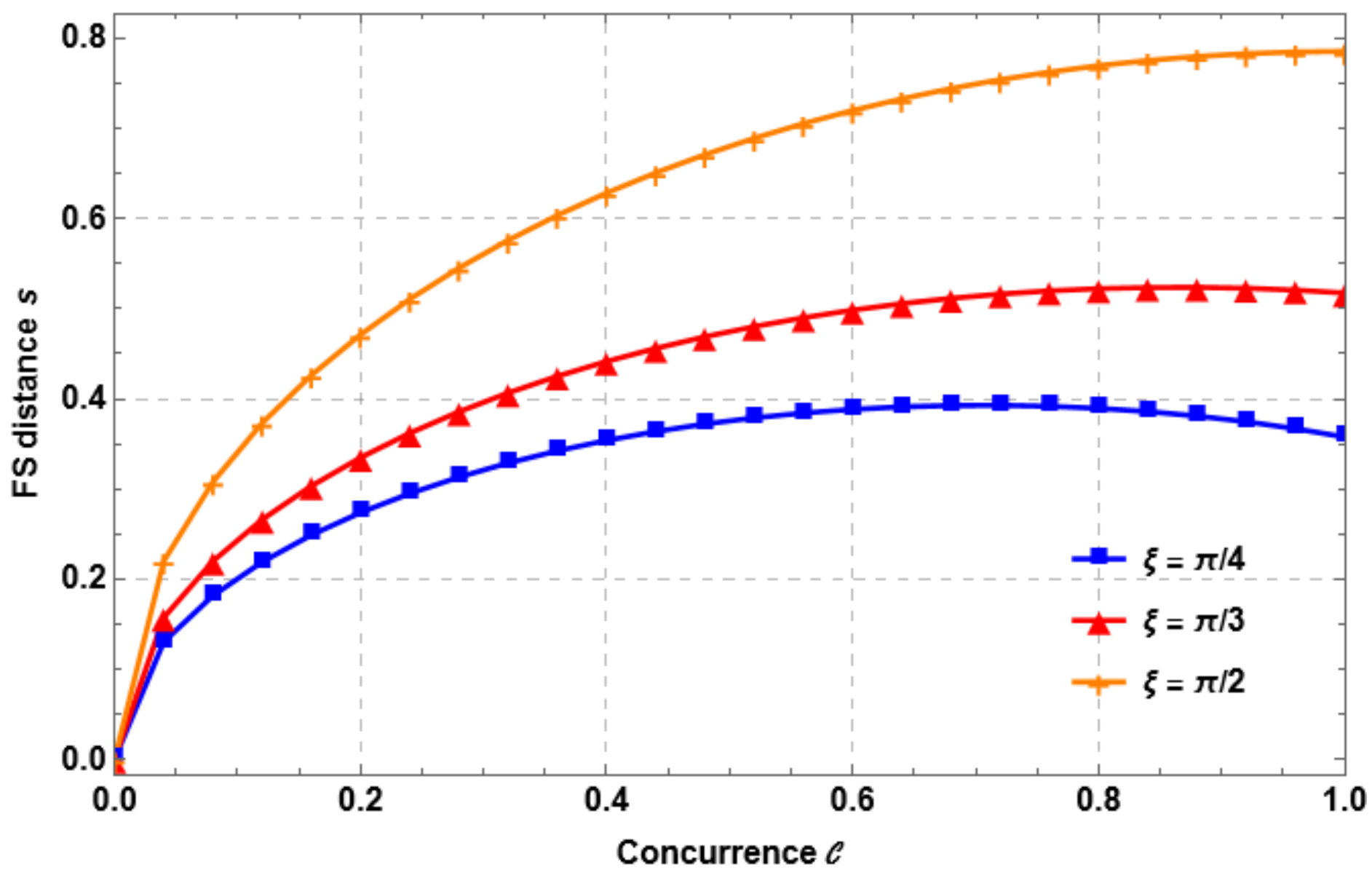

Figure 3.3: FubiniStudy Distance (3.73) vs. Concurrence (3.71) for various $\kappa$ values

in phase space is given by

$$\mathcal{T} = \frac{\mathbf{s}}{\mathbf{v}_{\max}} = \frac{\kappa}{J|\sin\kappa|}\sqrt{\mathcal{C}\left(2|\sin\kappa| - \mathcal{C}\right)}. \tag{3.74}$$

This optimal time depends on ordinary time, the coupling constant $J$, and the system's degree of entanglement. Analysis of its behavior reveals the following.

- For $\mathcal{C} = 0$, the optimal time is zero ($\mathcal{T} = 0$), because the initial state is a non-entangled state $|\mathbf{\Upsilon}^{2,\frac{1}{2}}\rangle = |++\rangle$ which undergoes no evolution.
- For $\mathcal{C} = \mathcal{C}_c$, the optimal time reaches its maximum value $\mathcal{T} = t$, which means that the optimal evolution coincides with the ordinary evolution of the system.
- For $\mathcal{C} \in ]0, \mathcal{C}_c[\cup]\mathcal{C}_c, 1]$, the optimal time is strictly less than the ordinary time ($\mathcal{T} < t$).

Thus, we conclude that ordinary time and degree of entanglement are two physical quantities that can be exploited to achieve optimal evolutions in these integrable systems. Such evolutions are crucial in quantum computing to designefficient algorithms [91–94].

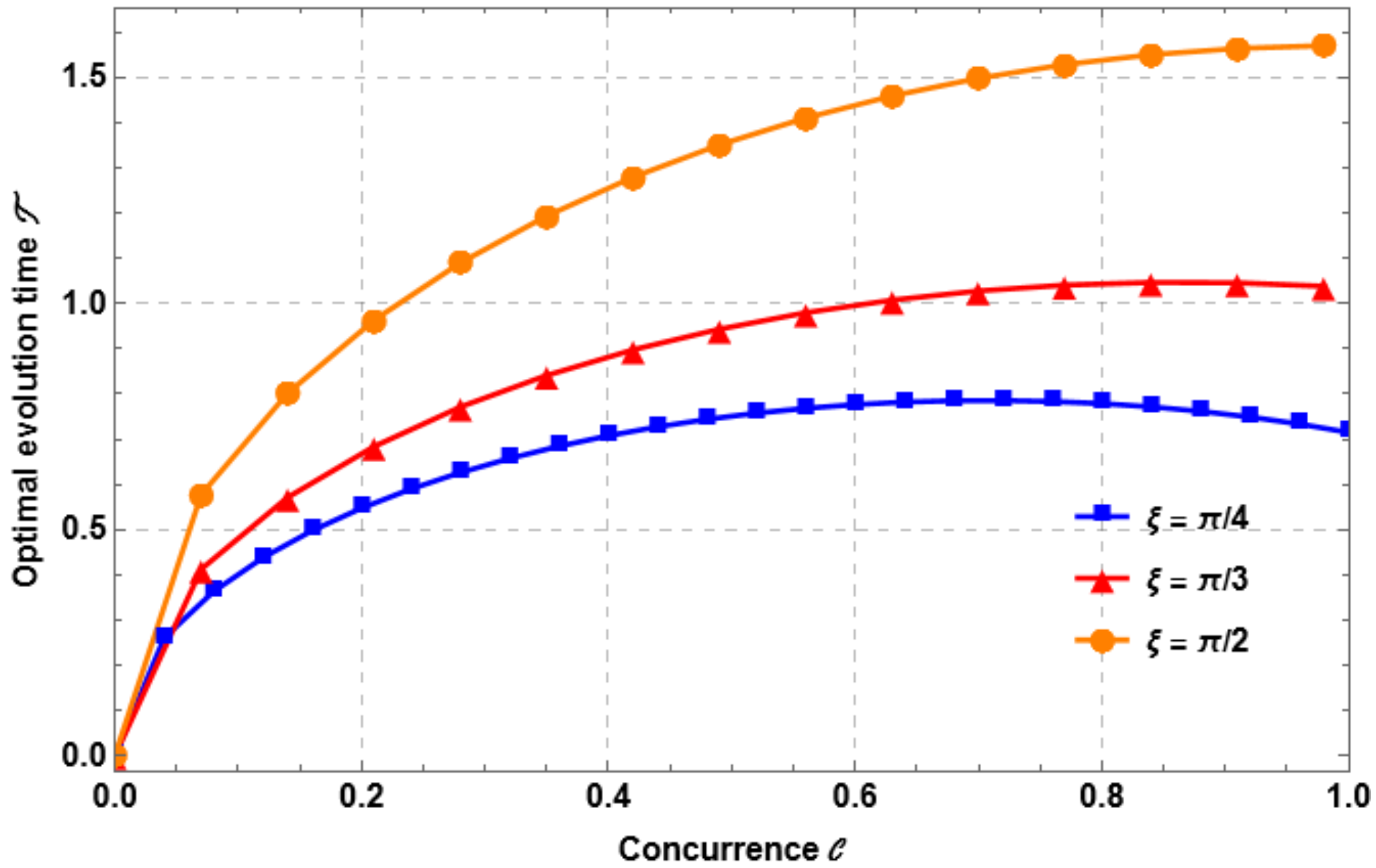

Figure 3.4: Variation of Optimal Time (3.74) with Concurrence (3.71) for various $\kappa$ values and $J = 1$

Finally, the optimal state evolution can be obtained via the following unitary operation

$$|\Upsilon\rangle \to |\Upsilon(\mathfrak{T})\rangle = e^{-iH\mathfrak{T}}|\Upsilon\rangle. \tag{3.75}$$

The set of optimal states forms a one-dimensional space in the global phase space, defined by the metric

$$dS^2_{\text{opt}} = \frac{\mathfrak{C}}{4\sin^2\kappa}\left(2|\sin\kappa| - \mathfrak{C}\right)d\kappa^2, \tag{3.76}$$

with $\kappa = J\mathfrak{T}$. Analysis of Figure 3.4 shows that as the entanglement $\mathfrak{C}$ and the ordinary time $t$ decrease ($\mathfrak{C} \to 0$ and $t \to 0$), the optimal time tends towards zero ($\mathfrak{T} \to 0$). This confirms that entanglement can be exploited to optimize the evolution of a quantum system, opening promising prospects for the development of high-performance quantum algorithms [95, 96].

## 3.4 Dynamical Description of N Interacting Spin-s particles Under Ising Model

The Ising model constitutes one of the most fundamental frameworks in quantum many-body physics, providing valuable insights into the role of spinspin interactions in collective quantum dynamics. In this study, based on our contribution [77], where we examine a system composed of $N$ qudits, each associated with a spin quantum number $s$ and local Hilbert space dimension $d = 2s+1$. The qudits interact through pairwise Ising-type couplings, enabling the exploration of dynamical phenomena in higher-spin systems beyond the usual spin-$\frac{1}{2}$ case. Our primary objective is to investigate the unitary time evolution of such systems and to characterize their evolution speed via the Fubini-Study metric, thereby connecting energy fluctuations to the geometry of the underlying quantum state space.

### 3.4.1 Unitary evolution

In this section, we consider a system made up of $N$ qudits, each represented by a spin-s ($d = 2s+1$). These qudits interact via the Ising model, governed by the following Hamiltonian

$$\mathbf{H}^{N,s} = 2J \sum_{k<l}^{N} \mathbf{S}_k^z \mathbf{S}_l^z, \tag{3.77}$$

where J represents the constant characterizing the coupling and $\mathbf{S}_k^z$ stands for the $z$ component of the spin operator $\mathbf{S}_k = (\mathbf{S}_k^x, \mathbf{S}_k^y, \mathbf{S}_k^z)^T$, associated with the $k^{th}$ qudit which satisfies the eigenvalue equation (3.3). Note that the components of the spin operators $\mathbf{S}_k^x$, $\mathbf{S}_k^y$ and $\mathbf{S}_k^z$ follow the algebraic structure of the Lie group $SU(2)$:

$$[\mathbf{S}_k^\alpha, \mathbf{S}_l^\beta] = i\delta_{kl} \sum_{\gamma=x,y,z} \epsilon_{\alpha\beta\gamma} \mathbf{S}_k^\gamma, \tag{3.78}$$

where $\delta_{kl}$ is the Kronecker symbol and $\epsilon_{alpha\beta\gamma}$ is the Levi-Civita symbol.

We assume that the system is initially prepared in a coherent state, obtained by rotating the highest-weight state $|s, s, \ldots, s\rangle$, where all spins take their maximum value, through an angle $\theta$ around the axis $\mathbf{r} = (\sin\phi, -\cos\phi, 0)$. This state can be expressed as

$$|\boldsymbol{\Upsilon}_i^{N,s}\rangle = e^{-i\theta \sum_{k=1}^{N} \mathbf{S}_k \cdot \mathbf{r}} |s, s, \ldots, s\rangle = e^{\sum_{k=1}^{N} (\zeta \mathbf{S}_k^+ - \zeta^* \mathbf{S}_k^-)} |s, s, \ldots, s\rangle, \tag{3.79}$$

where $\zeta = \frac{\kappa}{2} e^{-i\varphi}$, with $\kappa \in [0, \pi]$ and $\varphi \in [0, 2\pi]$, representing polar and azimuthal angles respectively. The initial state Eq. (3.79) can also be written as

$$|Z_k\rangle = (1 + ZZ^*)^{-Ns} \bigotimes_{k=1}^{N} |Z_k\rangle, \tag{3.80}$$

with

$$|Z_k\rangle = \sum_{m_k=-s}^{s} Z^{s+m_k} \binom{2s}{s+m_k} |m_k\rangle. \tag{3.81}$$

This state corresponds to an non-normalized $SU(2)$ coherent state for the $k^{th}$ qudit. Here, $\binom{2s}{s+m_k}$ denotes the binomial coefficient and the complex parameter $Z$ is defined by $Z = \tan\frac{\kappa}{2} e^{-i\varphi}$. This parameter represents the stereographic projection of the space of coherent states (a sphere $S^2$) onto a relevant equatorial plane via the application $CP \cong SU(2)/U(1)$ [97].

The initial state Eq. (3.79) will now evolve via the time evolution propagator

$$\mathbf{P}_t = e^{-i\mathbf{H^{N,s}}t}. \tag{3.82}$$

The evolved state of the $N-spin$ system is then written as

$$|\mathbf{\Upsilon}_t^{N,s}\rangle = (1+ZZ^*)^{-Ns} \sum_{m_1,m_2,\ldots,m_N=-s}^{s} e^{-i2\eta(t)\sum_{k<l} m_k m_l} \prod_{\nu=1}^{N} Z^{s+m_\nu} \begin{pmatrix} 2s \\ s+m_\nu \end{pmatrix} \bigotimes_{k=1}^{N} |m_k\rangle, \quad (3.83)$$

where $\eta(t) = Jt$. It is important to note that the evolution of the system is entirely governed by the parameter $\eta$, while the other parameters, namely $N$, $s$, $\kappa$ and $\varphi$, only specify the chosen initial state. Furthermore, it is noteworthy that the system state (3.83) satisfies the following periodicity conditions

$$|\mathbf{\Upsilon}^{N,s}(\eta+2\pi)\rangle = |\mathbf{\Upsilon}^{N,s}(\eta)\rangle, \quad \text{if } s \text{ is half-integer},$$

and

$$|\mathbf{\Upsilon}^{N,s}(\eta+\pi)\rangle = \pm|\mathbf{\Upsilon}^{N,s}(\eta)\rangle, \quad \text{if } s \text{ is integer}.$$

In this way, the wave function describing all $N$ particles exhibits periodicity according to the parameter $\eta$, with a period that depends on whether the particles in question are bosonic (integer spin) or fermionic (half-integer spin).

### 3.4.2 Quantum Evolution Speed and Brachistochrone Problem for N Qudits system

We aim to achieve a time-optimal evolution by maximizing the evolution speed while minimizing the corresponding geodesic distance covered by the spin-$s$ system under consideration. We assume that the evolution of the N-spin system is solely dependent on time, with all other parameters held constant. As a result, the Fubini-Study metric in Eq. (4.60) simplifies to

$$d\mathbf{s}^2 = \frac{1}{2}N(N-1)s^2\sin^2\kappa\left(1+[4s(N-1)-1]\cos^2\kappa\right)d\eta^2. \quad (3.84)$$

This indicates that the system evolves along a circular trajectory with a radius of $\sqrt{g_\eta}$ for half-integer values of $s$, and $\frac{\sqrt{g_\eta}}{2}$ for integer values of $s$. The evolution velocity of the N-spin system is given by

$$\mathbf{v} = Js\sqrt{\frac{N(N-1)\sin^2\kappa\,(1+[4s(N-1)-1]\cos^2\kappa)}{2}} \quad (3.85)$$

This velocity depends on the coupling constant $J$ between the spins, the number of particles $N$, and the spin value $s$. As these parameters increase, the evolution rate accelerates. However, at the singular points $\kappa = 0$ and $\kappa = \pi$, the velocity vanishes ($V = 0$), since the state of the N-spin system becomes undefined at these values of $\kappa$. As we have previously examined in the two systems discussed above, the evolution velocity is also tied to $\Delta E$, the energy uncertainty associated with the Ising Hamiltonian Eq. (3.77). This relationship suggests that a greater energy uncertainty leads to a shorter evolution time, and conversely, a smaller energy uncertainty results in a longer evolution time. To minimize the evolution time, it is essential to maximize the velocity $V$. This can be achieved by solving the equation

$$\frac{d\mathbf{v}}{d\kappa} = 0. \tag{3.86}$$

Upon solving, we obtain the following expression

$$\sin 2\kappa \left(2s(N-1) - [4s(N-1)-1]\sin^2\kappa\right) = 0. \tag{3.87}$$

This equation yields

$$\sin\kappa_{\max} = \sqrt{\frac{2s(N-1)}{4s(N-1)-1}}. \tag{3.88}$$

Thus, the maximum evolution speed is given by

$$\mathbf{v}_{\max} = Js^2(N-1)\sqrt{\frac{2N(N-1)}{4s(N-1)-1}}. \tag{3.89}$$

With the maximum velocity $V_{\max}$ determined, the next step is to calculate the geodesic distance corresponding to the quadratic line element Eq. (3.84) between the initial state Eq. (3.79) and the final state Eq. (3.83). Integrating the velocity equation Eq. (2.23) over the interval $[0,t]$, we obtain

$$\mathbf{s} = s\sqrt{\frac{\eta^2 N(N-1)\sin^2\kappa\,(1+[4s(N-1)-1]\cos^2\kappa)}{2}}. \tag{3.90}$$

The geodesic distance Eq. (3.90) increases linearly with time $t$ and vanishes at $\kappa = 0$ or $\kappa = \pi$ due to the singularity of the state space at these points. Conversely, at $\kappa = \pi/2$, the geodesic distance reaches its minimum value

$$\mathbf{s}_{\min} = s\sqrt{\frac{\eta^2 N(N-1)}{2}} \tag{3.91}$$

From this, we derive the optimal minimum time required for the evolution

$$\mathcal{T} = \frac{\mathbf{s}_{\min}}{\mathbf{v}_{\max}} = \frac{1}{2Js(N-1)}\sqrt{\eta^2[4s(N-1)-1]}. \tag{3.92}$$

This result highlights that the optimal evolution time is solely determined by the maximum speed and the minimum geodesic distance reached during the process, remaining independent of the parameter $\kappa$. The condition Eq. (3.92) ensures that the evolution occurs in the shortest possible time along the circular trajectory defined by the Fubini-Study metric Eq. (3.84). In this way, the optimal evolution of quantum states can be achieved via the unitary transformation as

$$|\boldsymbol{\Upsilon}_i^{N,s}\rangle \rightarrow |\Upsilon(\mathcal{T})\rangle = e^{-iH\mathcal{T}}|\Upsilon_i\rangle. \tag{3.93}$$

Expressing the minimum time $\mathcal{T}$ in terms of the ordinary time $t$, we obtain

$$\mathcal{T} = \frac{t}{2s(N-1)}\sqrt{4s(N-1)-1}. \tag{3.94}$$

It is important to note that ordinary time $t$ is associated with a circular trajectory defined by the Fubini-Study metric Eq. (3.84). Eq. (3.94) reveals that optimal and ordinary times are linked by a proportional relation. More precisely, any rise in ordinary time leads to a corresponding increase in optimal time.

In the special case of a two-qubit system ($N = 2, s = 1/2$), the optimal time and the ordinary time coincide ($\mathcal{T} = t$). On the other hand, for an $N$-qubit system ($N > 2, s > 1/2$), the optimal time is always less than the ordinary time ($\mathcal{T} < t$). In the thermodynamic limit ($N \to \infty$), the optimal time tends towards zero ($\mathcal{T} \to 0$), which means that the circle of states Eq. (3.84) becomes a straight line, as its radius becomes infinite. The same phenomenon occurs when the spin value tends towards infinity ($\to \infty$). Finally, the brachistochrone problem can also be solved by directly minimizing the evolution time with respect to the parameter $\kappa$. This leads to the trivial case where

$$\mathcal{T} = \frac{\mathbf{s}_{\min}}{\mathbf{v}_{\max}} = \frac{\mathbf{s}(\kappa = \pi/2)}{\mathbf{v}(\kappa = \pi/2)} = t. \tag{3.95}$$

This result is only valid for a two-qubit system ($N = 2, s = 1/2$) and does not apply to general values of $N$ and $s$. forthermore, it is impossible to obtain minimum-time evolution with this approach, which makes it inadequate.

### 3.4.3 Dynamical Effects of Quantum Correlations in Two spin-s Entanglement

Using I-concurrence as a reliable measure of quantum entanglement [98, 99], we will study the entanglement of a system composed of two spin-s (a two-qudits system) through a dynamical approach that explores the effect of entanglement on system dynamics. In addition, we will seek to solve the quantum brachistochrone problem in terms of the degree of quantum entanglement. The evolved state of the total quantum system can be reduced for a two-spin-$s$ system to the following form

$$|\mathbf{\Upsilon}_t^{2,s}\rangle = (1 + ZZ^*)^{-2s} \sum_{m_1,m_2=-s}^{s} e^{-i2\eta m_1 m_2} \bigotimes_{l=1}^{2} Z^{s+m_l} \sqrt{C_{2s}^{s+m_l}} |m_l\rangle. \tag{3.96}$$

The I-concurrence of a state of two qudits is defined by

$$\mathcal{C} = \sqrt{2\left(1 - \mathrm{Tr}\rho_k^2\right)}, \tag{3.97}$$

where $\rho_k$ is the partial density matrix associated with spin-$s$ $k$ ($k = 1, 2$). We first calculate the reduced density matrix $\rho_1$ corresponding to the first spin-$s$. The density matrix of the complete system is given by

$$\begin{aligned}\rho =&(1 + ZZ^*)^{-4s} \sum_{m_1,m_2=-s}^{s} \sum_{m_1',m_2'=-s}^{s} e^{i2\eta(m_1' m_2' - m_1 m_2)} \\ &\times \bigotimes_{l=1}^{2} Z^{s+m_l} \sqrt{C_{2s}^{s+m_l}} |m_l\rangle \bigotimes_{j=1}^{2} Z^{*(s+m_j)} \sqrt{C_{2s}^{s+m_j}} \langle m_j'|.\end{aligned} \tag{3.98}$$

The reduced density matrix $\rho_1$ is obtained by performing a partial trace over the second spin:

$$\rho_1 = \left(1+\tan^2\frac{\kappa}{2}\right)^{-4s} \sum_{m_1,m_1'=-s}^{s} \left[\sum_{m_2=-s}^{s} e^{-i2\eta m_1 m_2} e^{i2\eta m_1' m_2}\right] \times \left(\tan\frac{\kappa}{2}e^{-i\varphi}\right)^{2s+m_1+m_2} \left(\tan\frac{\kappa}{2}e^{i\varphi}\right)^{2s+m_1'+m_2} C_{2s}^{s+m_2}. \tag{3.99}$$

Presuming an evolution over a short period of time, we can approximate exponentials up to order two in $\eta$

$$\begin{cases} e^{-i2\eta m_1 m_2} \approx 1 - 2i\eta m_1 m_2 - 2\eta^2 (m_1 m_2)^2, \\ e^{i2\eta m_1 m_2} \approx 1 + 2i\eta m_1 m_2 - 2\eta^2 (m_1 m_2)^2 \end{cases} \tag{3.100}$$

Substituting these expressions into $\rho_1$, we obtain

$$\begin{aligned}\rho_1 =& (1+\tan^2\frac{\kappa}{2})^{-2s} \sum_{m_1,m_1'=-s}^{s} \left[1 - 2is\eta(m_1-m_1')\cos\kappa - 2\eta^2(m_1-m_1')^2(s^2 - s(2s-1)\sin^2\kappa)\right] \\ & \tan^{s+m_1}\frac{\kappa}{2} e^{-i\varphi(s+m_1)} \tan^{s+m_1'}\frac{\kappa}{2} e^{i\varphi(s+m_1')} C_{2s}^{s+m_1} C_{2s}^{s+m_1'} |m_1\rangle\langle m_1'|.\end{aligned} \tag{3.101}$$

By inserting this expression into the I-concurrence equation and using the binomial theorem and its derivatives

$$\begin{cases} \sum_{m_k=-s}^{s} m_k \tan^{2s+m_k}\frac{\kappa}{2} C_{2s}^{s+m_k} = -s(1+\tan^2\frac{\kappa}{2})^{2s}\cos\kappa, \\ \sum_{m_k=-s}^{s} m_k^2 \tan^{2s+m_k}\frac{\kappa}{2} C_{2s}^{s+m_k} = s^2 - s^2(2s-1)\sin^2\kappa(1+\tan^2\frac{\kappa}{2})^{2s}, \end{cases} \tag{3.102}$$

We obtain the final expression for the I-concurrence of the two-spin-$s$ system

$$\mathcal{C} = 2\eta s \sin^2\kappa. \tag{3.103}$$

This expression highlights the dependence of entanglement between the two spins on the dynamic parameters $(\kappa, \eta)$. Consequently, the evolution of entanglement is determined by the trajectory followed by the system in its state space. In other words, each point in the manifold of states corresponds to a specific degree of entanglement, which is uniquely defined by these parameters. By setting $\kappa = \frac{\pi}{2}$ and $\eta = \eta_{\max} = Jdt$ (with $0 < J \leq 1$ to ensure better precision), the I-concurrence reaches its maximum value

$$\mathcal{C}_{\max} = 2s\eta_{\max}. \tag{3.104}$$

However, for $\kappa = 0$ or $\kappa = \pi$, the system remains non-entangled. This follows from the fact that the initial state $|\boldsymbol{\Upsilon}_i^{2,s}\rangle = |s,s\rangle$ is an eigenstate of the two-spin-$s$ system.

We now proceed to analyze the system's dynamics in the context of entanglement between its two constituent spins. In this study, we contribute to resolving the quantum brachistochrone problem by establishing a connection between the system's evolution rate and its degree of entanglement. To this end, we substitute Eq. (3.103) into Eq. (3.85) for a system consisting of two spins ($N = 2$), enabling us to express the evolution rate in terms of the I-concurrency as follows

$$\mathbf{v} = Js\sqrt{\tilde{\eta}\,\frac{\mathcal{C}}{\mathcal{C}_{\max}}\left(4s - (4s-1)\tilde{\eta}\,\frac{\mathcal{C}}{\mathcal{C}_{\max}}\right)}. \tag{3.105}$$

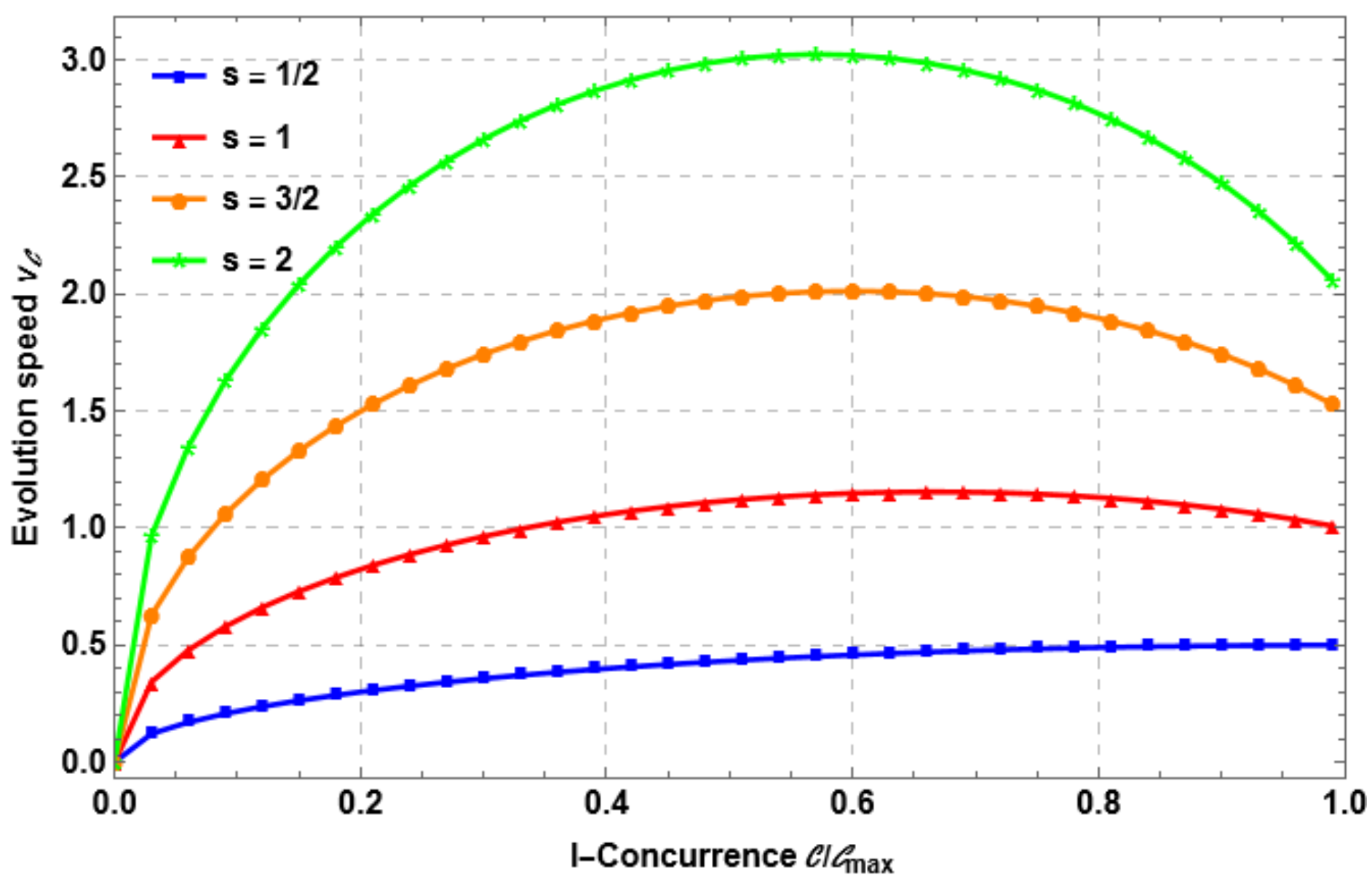

Figure 3.5: Evolution Speed vs. I-Concurrence for Various Spin Values at $\tilde{\eta} = 1$ and $J = 1$

This equation establishes a direct link between the evolution speed of a two-spin system with spin quantum number $s$ and its degree of entanglement. More specifically, Eq. (3.105) underscores the intrinsic connection between the unitary evolution of the system and dynamics of its quantum correlations. Consequently, by analyzing the entangled states of the system, one can extract valuable insights into its underlying dynamical behavior. The dependence of evolution speed on I-concurrence is further illustrated in Fig. (3.5), highlighting the role of quantum correlations in governing the system evolution. When $s > \frac{1}{2}$, the evolution speed of the system unfolds in two distinct stages. In the first phase, the evolution speed gradually increases, reaching a maximum given by $\mathbf{v}_{\max} = 2Js^2/\sqrt{4s-1}$ This peak velocity is attained when the system's entanglement reaches a critical threshold, $\mathcal{C}_* = 2s\mathcal{C}_{\max}/(4s-1)\tilde{\eta}$. During this initial phase, quantum correlations function as a dynamic accelerator, driving a rapid evolution in the space of quantum states. In the second phase, when entanglement falls within the range $\mathcal{C}_* \leq \mathcal{C} \leq \mathcal{C}_{\max}$, the speed of evolution exhibits a continuous decline, eventually reaching a local minimum at $\mathbf{v}(\mathcal{C} = \mathcal{C}_{\max})$. This deceleration reflects a phenomenon of dynamic inhibition, where quantum correlations act as a restraining force, slowing down the system's motion in phase space. A particularly notable special case arises for $s = \frac{1}{2}$. In this scenario, the second phase vanishes entirely, as entanglement instantaneously reaches its maximum value $\mathcal{C} = \mathcal{C}_{\max}$. Moreover, our findings indicate a strong dependence of evolution speed on the spin magnitude: higher spin values correspond to faster evolution. This leads to a fundamental conclusion both entanglement and spin magnitude are key physical parameters governing the speed of evolution in an interacting spin system within the Ising model.

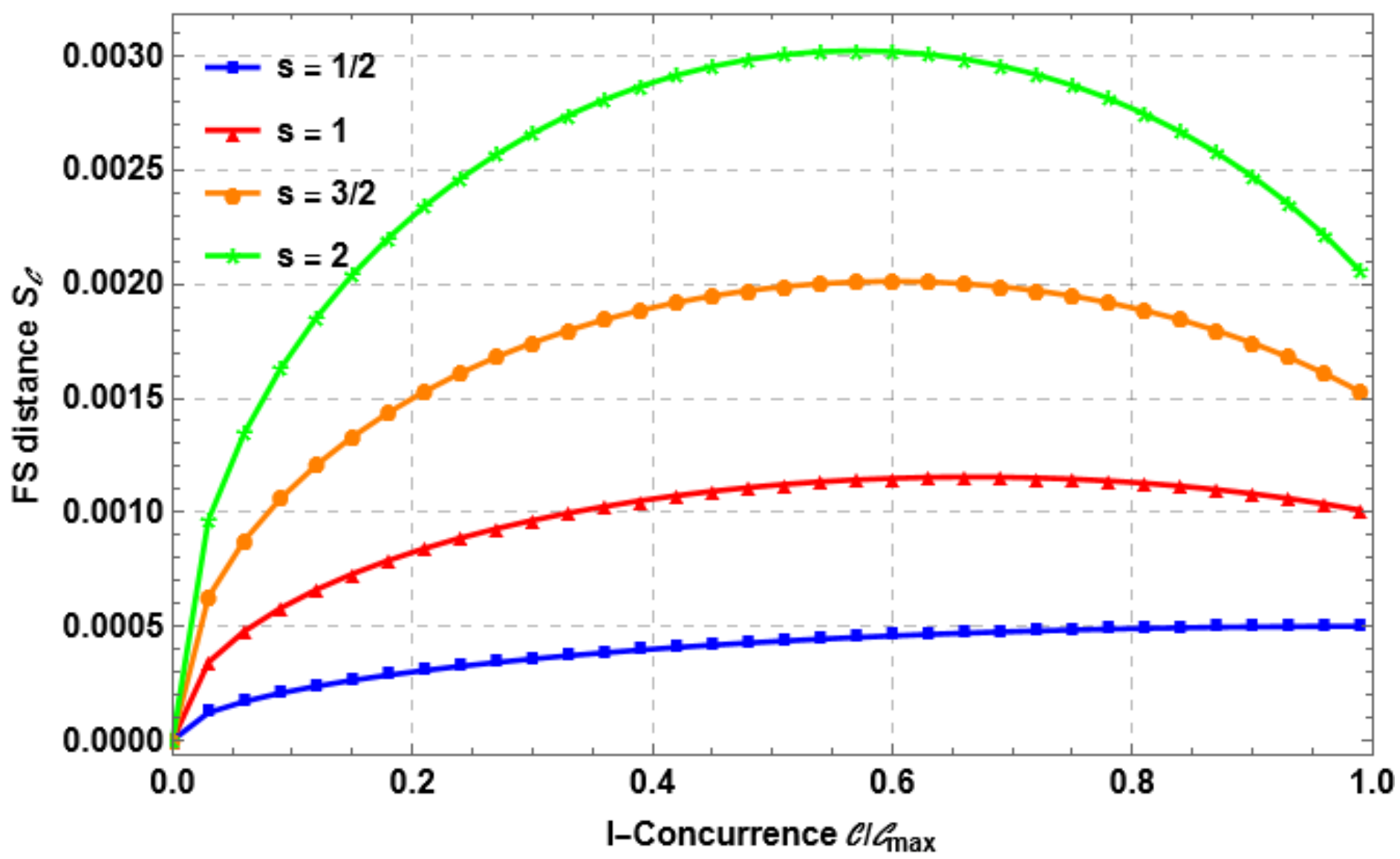

Figure 3.6: Geodesic distance (3.106) plotted against I-concurrence (3.103) for various spin values, with parameters set to $\tilde{\eta} = 1$, $J = 1$, and $\eta_{\max} = 10^{-3}$.

By leveraging Eq. (3.85), we can express the geodesic distance traversed by the two-spin system as a function of I-concurrence. This distance, measured using the Fubini-Study metric, is given by

$$\mathbf{s} = s\sqrt{\eta'_{\max}\eta\frac{\mathcal{C}}{\mathcal{C}_{\max}}\left(4s - (4s-1)\eta\frac{\mathcal{C}}{\mathcal{C}_{\max}}\right)}. \tag{3.106}$$

Eq. (3.106) thus establishes a direct connection between geodesic distance and entanglement, offering a pathway for experimentally measuring either the distance between entangled states or the system's speed of evolution within the manifold of quantum states. A comparative analysis of Fig. (3.5) and Fig. (**??**) reveals a striking similarity between the evolution of the geodesic distance Fig. (3.106) and the evolution velocity Eq. (3.105) as functions of the degree of entanglement. This correspondence suggests that quantum correlations exert a comparable influence on both system dynamics and the geometry of quantum state evolution. We now turn our attention to the quantum brachistochrone problem, which seeks to determine the minimum time required for optimal evolution of the two-spin system. By combining Eq. (3.105) and Eq. (3.106), we derive the optimal evolution time as a function of I-concurrence

$$\mathcal{T}_{\mathcal{C}} = \frac{\mathbf{s}\cdot\mathbf{v}_{\max}}{2Js}\sqrt{\eta'_{\max}\eta\frac{\mathcal{C}}{\mathcal{C}_{\max}}(4s-1)\left(4s - (4s-1)\eta\frac{\mathcal{C}}{\mathcal{C}_{\max}}\right)}. \tag{3.107}$$

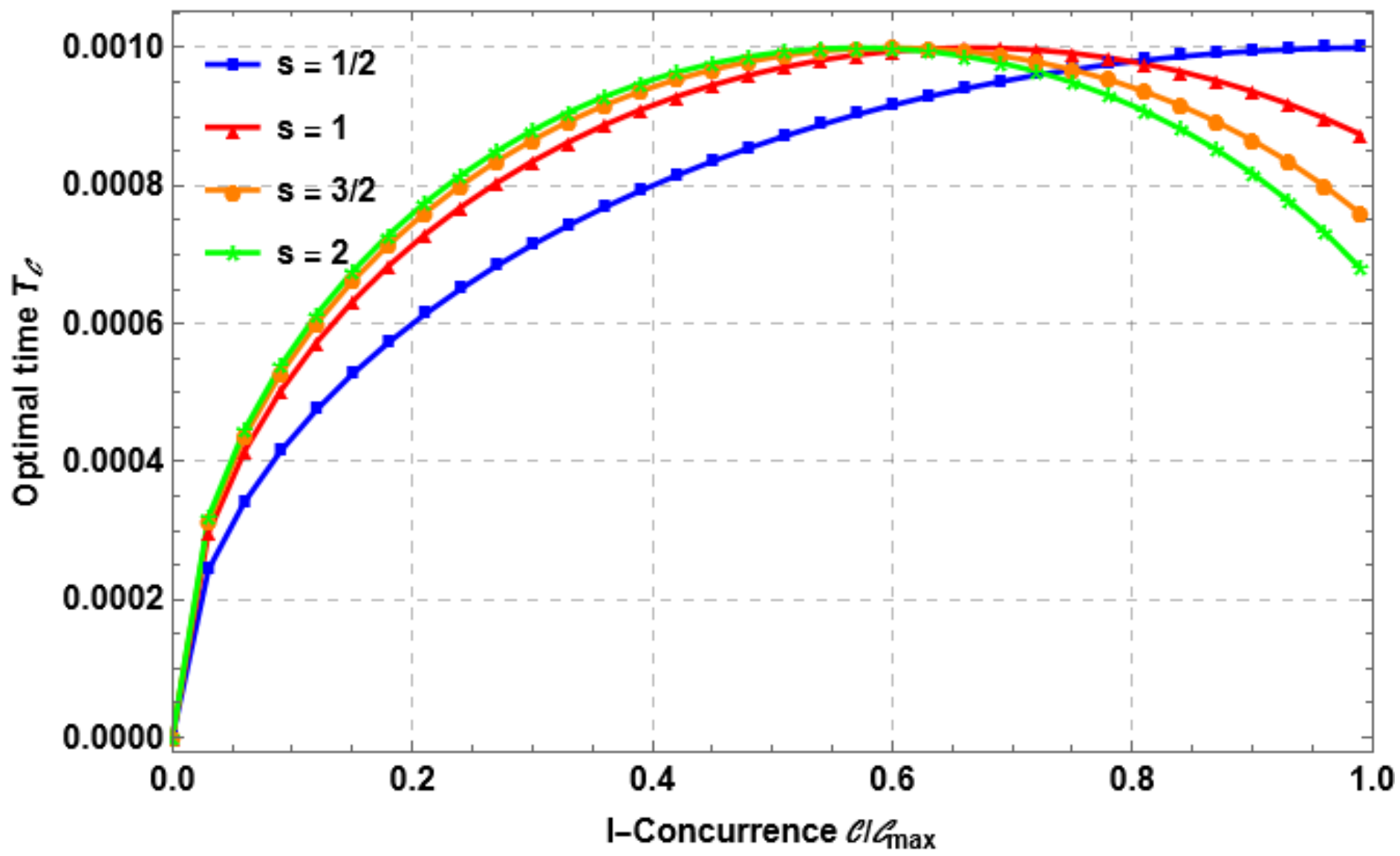

Figure 3.7: Optimal time (3.107) plotted against I-concurrence (3.103) for various spin values, with parameters set to $\tilde{\eta} = 1$, $J = 1$, and $\eta_{\max} = 10^{-3}$.

An analysis of this expression reveals several key behaviors. When $\mathcal{C} = 0$, the optimal time reduces to zero ($\mathcal{T}_c = 0$), indicating that the system does not evolve. In this scenario, the final state coincides with the initial state, which is a separable state: $|\boldsymbol{\Upsilon}_i^{2,s}\rangle = (1 + ZZ^*)^{-2s}|Z_1, Z_2\rangle$ When the entanglement reaches a critical value $\mathcal{C} = \mathcal{C}_*$, the optimal evolution time attains its maximum ($\mathcal{T} = t$), signifying that the system's optimal evolution aligns with its natural evolution. In the intervals $\mathcal{C}_{\max}[\mathcal{C} \in ]0, \mathcal{C}_*[\cup]\mathcal{C}_*, \mathcal{C}_{\max}[$, the optimal evolution time remains strictly shorter than the standard evolution time ($\mathcal{T}_{\mathcal{C}} < t$). This indicates the presence of optimized time trajectories, where entanglement plays a crucial role in accelerating evolution. The dependence of optimal evolution time on entanglement is illustrated in Fig. (3.7). Notably, as the degree of entanglement or the spin value decreases ($\mathcal{C} \to 0$), the optimal evolution time diminishes ($\mathcal{T}_{\mathcal{C}} \to 0$). The results presented in this study underscore the fundamental role of entanglement and spin in governing the optimal evolution time of a quantum system. Optimizing evolution time is a critical challenge in quantum computing, particularly in the design of quantum circuits and the implementation of quantum logic gates. By fine-tuning entanglement, it is possible to minimize computation times and enhance the execution efficiency of quantum operations. This study thus paves the way for more efficient quantum algorithms, where controlled entanglement dynamics can be strategically exploited to optimize quantum computation [88, 100, 101].

## 3.5 Summary

This chapter presents a detailed dynamical investigation of quantum spin systems, emphasizing the interplay between system interactions, quantum evolution speed, and entanglement dynamics. Beginning with an overview of spin models, it focuses on the two-spin-1/2 system governed by the anisotropic Heisenberg model. The analysis outlines the formalism for unitary time evolution, highlighting how the Hamiltonian structure dictates state trajectories. The quantum brachistochrone problem is introduced as a tool for determining minimal evolution times between states, thereby linking dynamical efficiency to system parameters. The time-dependent behavior of entanglement is examined, illustrating the sensitivity of correlation patterns to anisotropy and coupling strength.

The discussion extends to many-body systems, first addressing N interacting spin-1/2 particles within the long-range Ising framework. Analytical expressions for the unitary evolution operator and quantum speed limits are derived, revealing the scaling properties of dynamical behavior with system size. Particular attention is given to the special case N=2, where entanglement measures provide a bridge between microscopic dynamics and macroscopic observables. These results demonstrate how quantum correlations influence speed limits and evolution pathways, reinforcing the role of entanglement as both a resource and a constraint in dynamical optimization.

The final section generalizes the framework to N interacting spin-$s$ particles under the Ising model. Here, the formalism accommodates higher-spin Hilbert spaces, broadening the scope of the dynamical analysis. The impact of spin magnitude on evolution speed, minimal-time paths, and entanglement patterns is systematically explored. Across all cases, the Hamiltonian formalism and symmetry properties serve as unifying tools for characterizing dynamics. This chapter thus establishes a comprehensive picture of how system parameters, interaction topology, and quantum correlations collectively govern the temporal behavior of quantum spin systems.

CHAPTER 4

# Exploring Quantum Spin Systems: Insights from Geometrical and Topological Analyses

## 4.1 Geometrical Description of Two Interacting Spin-1/2 Under the anisotropic Heisenberg Model

### 4.1.1 Geometric Structure of the Two-Spin State Space

Exploring the state space of a system of two spins reveals a rich and intricate geometric structure. By imposing the normalization condition on the quantum state Eq. (**??**), we naturally associate it with a seven-dimensional sphere, denoted as $S^7$. This identification facilitates a geometric interpretation in terms of a principal fiber bundle, where the total space $S^7$ decomposes into a base space corresponding to the complex projective space $\mathbb{CP}^3$ and a fiber associated with the global phase symmetry $U(1)$, i.e., $S^1$. Mathematically, this relationship can be expressed locally as

$$S^7 \cong \mathbb{CP}^3 \times S^1. \tag{4.1}$$

Within this geometric framework, the quantum state Eq. (3.13) can be conveniently rewritten using local coordinates defined on the base space $\mathbb{CP}^3$. This reformulation leads to the following expression

$$|\Upsilon\rangle = \frac{1}{\sqrt{1+|z_1|^2+|z_2|^2+|z_3|^2}} \begin{pmatrix} 1 \\ z_1 \\ z_2 \\ z_3 \end{pmatrix}, \quad z_1, z_2, z_3 \in \mathbb{C}. \tag{4.2}$$

In this representation, the quantum state corresponds to a coherent configuration of the $SU(4)$ group, aligning with the findings of several studies [102–104]. The equivalence between Eq. (3.13) and Eq. (4.2) further enables us to express the local coordinates in terms of the system's physical parameters, which can be determined through the following transformations

$$\begin{aligned} z_1 &= \frac{e^{-i(\kappa+\nu\eta)}}{c_i^{(11)}}(c_i^{(10)}\cos\eta - ic_i^{(01)}\sin\eta), \\ z_2 &= \frac{e^{-i(\kappa+\nu\eta)}}{c_i^{(11)}}(-ic_i^{(10)}\sin\eta + c_i^{(01)}\cos\eta), \quad z_3 = \frac{c_i^{(00)}}{c_i^{(11)}}e^{2i\kappa}. \end{aligned} \tag{4.3}$$

The transition from the standard Hilbert space $S^7$ to the geometric basis $\mathbb{CP}^3$ reveals the intricate geometric structure underlying the manifold of states associated with this two-spin system. This structure is elegantly characterized by the Fubini-Study metric, which is defined as

$$ds^2 = \sum_{i,j=1}^{3} \frac{\partial^2 \ln K(z,z^*)}{\partial z_i \partial z_j^*} dz_i dz_j^*, \tag{4.4}$$

where the Bergmann kernel is given by

$$K(z,z^*) = 1 + z_1 z_1^* + z_2 z_2^* + z_3 z_3^*. \tag{4.5}$$

By explicitly reformulating this metric, we obtain the following refined expression

$$ds^2 = \sum_{i,j=1}^{3} \left[(1 + \delta_{ij} z z^*) - (1 + z_i^* z z_j^*)^2\right] dz_i dz_j^*. \tag{4.6}$$

Substituting the local coordinate expressions Eq. (4.3) into this formulation Eq. (4.3) yields a more explicit representation of the Fubini-Study metric

$$ds^2 = [(\nu^2-1)A + 1 - (\nu A + F)^2]d\eta^2 + 2D[\nu - (\nu A - F)]d\kappa d\eta + [A - D^2]d\kappa^2. \tag{4.7}$$

Here, the coefficients are defined as

$$A = |c_i^{(11)}|^2 + |c_i^{(00)}|^2, \quad D = |c_i^{(11)}|^2 - |c_i^{(00)}|^2, \quad F = 2\mathrm{Re}(c_i^{(10)}(c_i^{(01)})^*). \tag{4.8}$$

Thus, the Fubini-Study metric, derived from the underlying geometric structure of $\mathbb{CP}^3$, is parametrized by only two independent coordinates. Notably, the independence of the metric tensor components from the parameters $\eta$ and $\kappa$ suggests that this manifold is flat and devoid of intrinsic curvature. Moreover, the resulting geometry is intrinsically linked to the initial parameters of the system, particularly the anisotropy parameter.

In the specific case of an isotropic Heisenberg Hamiltonian ($\nu = 1$), the state space assumes the structure of a torus, aligning with previously established results in [63]. However, further analysis reveals that different initial conditions lead to distinct geometric structures, as detailed below:

- **Structure 1:** When $A = D = 0$, the Fubini-Study metric simplifies to

$$ds^2 = B(2 - B)d\eta^2, \quad \text{where } B = |c_i^{(10)} - c_i^{(01)}|^2. \tag{4.9}$$

  This corresponds to a one-dimensional flat manifold. Given the periodicity condition $\eta \in [0, \pi]$, this manifold can be interpreted as a circle with an effective radius of $\frac{1}{2}g_{\eta\eta}$. These results hold for the periodicity cases ***(3)***,***(4)*** and ***(5)***.

- **Structure 2:** The state space in this case is two-dimensional and remains flat. The periodicity conditions $\kappa \in [0, 2\pi]$ and $\eta \in [0, \frac{\pi}{\nu \pm 1}]$ indicate that the corresponding manifold forms a torus. However, this torus exhibits a twist of $\pi$ around the $\kappa$-coordinate, a behavior observed in the periodicity cases ***(4)*** and ***(5)***. Specifically, if $p$ and $q$ are both odd, the quantum manifold corresponds to a standard torus. Conversely, if one of these parameters is even while the other remains odd, the resulting structure is a torus twisted by an angle $\pi$ around $\kappa$.

- **Structure 3:** In the case ***(1)***, periodicity is present only along the parameter $\kappa \in [0, 2\pi]$, while $\eta$ is unrestricted and can take any real value ($\eta \in \mathbb{R}$). Consequently, the state manifold assumes the form of an elongated cylinder, periodic along $\kappa$ and unbounded along $\eta$.

These findings underscore the geometric richness of the two-spin quantum state space and demonstrate how its structure is influenced by system parameters. A more detailed investigation of the geometric phase accumulated by the system as it evolves within this space will be the focus of the following section.

### 4.1.2 Geometric Phases Acquired by the Two-Spin State

Having examined the geometric structure of the space of quantum states as defined by the Fubini-Study metric Eq. (4.7), we now focus on the geometric phase acquired by the evolving two-spin state Eq. (3.13). Our analysis encompasses both arbitrary and cyclic evolution scenarios, aiming to elucidate the role of geometric phases in the system's quantum dynamics.

#### Geometric Phase During Cyclic Evolution

When a quantum system undergoes cyclic evolution, it satisfies the condition $|\Upsilon(\eta_m, \kappa_m)\rangle = e^{i\kappa_{\text{tot}}}|\Upsilon(0)\rangle$, where $\eta_m$ and $\kappa_m$ represent the periods of the dynamical parameters $\eta$ and $\kappa$, respectively. In this framework, the Aharonov-Anandan (AA) connection plays a fundamental role, defined locally on the state space $\mathcal{S}$ by

$$\mathcal{A} = \frac{1}{2i}\frac{\partial \ln[K(z, z^*)]}{\partial z^j}dz^j - \frac{1}{2i}\frac{\partial \ln[K(z, z^*)]}{\partial z^{*j}}dz^{*j}, \tag{4.10}$$

where $K(z,z^*)$ is the Bergmann kernel, as defined in equation Eq. (4.5). This differential connection, explicitly written as

$$\mathcal{A} = [(1-\mathbf{A})(\nu-1)+\mathbf{B}]d\eta + (1-\mathbf{D})d\kappa, \tag{4.11}$$

reflects an additional geometric structure that governs the evolution of the quantum state. By integrating this connection over the entire cycle, we can extract the geometric phase $\Phi_g^{\mathrm{AA}}$ accumulated during the cyclic evolution

$$\Phi_g^{\mathrm{AA}} = [(1-\mathbf{A})(\nu-1)+\mathbf{B}]\eta_m + (1-\mathbf{D})\kappa_m. \tag{4.12}$$

This phase depends solely on the periods $(\eta_m, \kappa_m)$, the initial parameters, and the anisotropic properties of the system, remaining independent of the path traversed in the state space. Furthermore, expressing $\Phi_g^{\mathrm{AA}}$ as a function of the dynamic phase $\Phi_{\mathrm{dyn}}(\eta_m,\kappa_m)$, we obtain

$$\Phi_g^{\mathrm{AA}} = \Phi_{\mathrm{dyn}}(\eta_m,\kappa_m) + \frac{\nu}{2}\eta_m + (1-2\mathbf{D})\kappa_m. \tag{4.13}$$

This defines the topological phase, which is the component of $\Phi_g^{\mathrm{AA}}$ independent of any dynamic contribution. By positing $\Phi_{\mathrm{dyn}}(\eta_m,\kappa_m)=0$, we deduce

$$\Phi_{\mathrm{top}}^{\mathrm{AA}} = \frac{\nu}{2}\eta_m + (1-2\mathbf{D})\kappa_m. \tag{4.14}$$

This result coincides with the geometric phase of maximal Bell states in two-qubit systems [105]. An interesting special case arises when $\nu = 0$ and $\mathbf{D} = \frac{1}{2}$, which reduces the model to the XX-Heisenberg Hamiltonian. In this case, the topological phase $\Phi_{\mathrm{top}}^{\mathrm{AA}}$ cancels out ($\Phi_{\mathrm{top}}^{\mathrm{AA}} = 0$), leading to an equality between the AA-phase and the dynamic phase accumulated over the cycle. This property is of experimental significance, as it enables the effective measurement of the AA geometric phase by determining the time integral of the mean value of the total Hamiltonian $\mathcal{H}$ Eq. (3.7).

#### Geometric Phase During Arbitrary Evolution

We now consider the evolution of a two-spin state governed by the parametric manifold given in Eq. (4.7). The geometric phase accumulated during this arbitrary evolution is defined as [106]

$$\Phi_g(t) = \arg\langle \Upsilon_i | \Upsilon(t)\rangle - \Im \int_0^t \left\langle \Upsilon(t') \middle| \frac{\partial}{\partial t'} \middle| \Upsilon(t') \right\rangle dt'. \tag{4.15}$$

This phase can be decomposed into a global phase and a dynamic phase. The transition amplitude between the initial state Eq. (3.12) and the evolved state Eq. (3.13) is given by

$$\begin{aligned}\langle \Upsilon_i | \Upsilon(\eta,\kappa)\rangle =& |c_i^{(11)}|^2\cos(\kappa+\nu\eta) + (1-A)\cos(\eta) + |c_i^{(00)}|^2\cos(\kappa-\nu\eta) \\ &+ i\left[|c_i^{(00)}|^2\sin(\kappa-\nu\eta) - |c_i^{(11)}|^2\sin(\kappa+\nu\eta) - (1-A-B)\sin(\eta)\right].\end{aligned} \tag{4.16}$$

Substituting this expression into the first term of Eq. (4.15), we obtain the global phase acquired by the system:

$$\Phi_{\text{glob}} = \arctan\left(\frac{2|c_i^{(00)}|^2\sin(\kappa-\frac{\nu\eta}{2})-2|c_i^{(11)}|^2\sin(\kappa+\frac{\nu\eta}{2})-\mathrm{G}\sin(\nu_-\eta)+\mathrm{B}\sin(\nu_+\eta)}{2|c_i^{(11)}|^2\cos(\kappa+\frac{\nu\eta}{2})+2|c_i^{(00)}|^2\cos(\kappa-\frac{\nu\eta}{2})+\mathrm{G}\cos(\nu_-\eta)+\mathrm{B}\cos(\nu_+\eta)}\right), \quad (4.17)$$

where

$$\nu_\pm = \frac{2\pm\nu}{2}, \quad \mathrm{G} = 2(1-\mathrm{A})-\mathrm{B}. \quad (4.18)$$

The global phase depends not only on the trajectory in the state space but also on the dynamical parameters $(\eta,\kappa)$ and the initial conditions. This interdependence underscores the topological nature of the system [107, 108]. Furthermore, since the phase is defined modulo $2\pi$, it satisfies the periodicity condition:

$$\Phi_{\text{glob}}(\kappa+2\pi) = \Phi_{\text{glob}}(\kappa). \quad (4.19)$$

In contrast, the dynamic phase evolves as

$$\Phi_{\text{dyn}} = \mathrm{D}\kappa + \left[\nu\left(\mathrm{A}-\frac{1}{2}\right)+\mathrm{F}\right]\eta. \quad (4.20)$$

As opposed to the global phase, the dynamic phase varies linearly with time and does not satisfy a periodicity condition. Finally, the geometric phase is obtained by subtracting the dynamic contribution

$$\begin{aligned}\Phi_g =& \arctan\left(\frac{2|c_i^{(00)}|^2\sin(\kappa-\frac{\nu\eta}{2})-2|c_i^{(11)}|^2\sin(\kappa+\frac{\nu\eta}{2})-\mathrm{G}\sin(\nu_-\eta)+\mathrm{B}\sin(\nu_+\eta)}{2|c_i^{(11)}|^2\cos(\kappa+\frac{\nu\eta}{2})+2|c_i^{(00)}|^2\cos(\kappa-\frac{\nu\eta}{2})+\mathrm{G}\cos(\nu_-\eta)+\mathrm{B}\cos(\nu_+\eta)}\right) \\ & - \mathrm{D}\kappa - \left[\nu\left(\mathrm{A}-\frac{1}{2}\right)+\mathrm{F}\right]\eta. \end{aligned} \quad (4.21)$$

The geometric phase, inherently defined modulo $2\pi$ due to the right-hand side of Eq. (4.15), exhibits a non-linear variation with time, as expressed in Eq. (4.21). This temporal dependence underscores its intrinsic connection to the system's evolutionary trajectory and the geometric structure of the bi-parametric space Eq. (4.7), further affirming its geometric nature. Furthermore, this phase is significantly influenced by the anisotropic (and, conversely, isotropic) properties that define the physical state of spin systems. This sensitivity suggests that controlling the anisotropy parameters enables direct manipulation of the resulting geometric phase. Hence, in the specific case of the two-spin system under consideration, fine-tuning these parameters provides a practical approach for adjusting and modulating the geometric phase to meet physical and experimental requirements.

### 4.1.3 Geometric Aspects of Quantum Entanglement in a Two-Spin-1/2 System Governed by the anisotropic Heisenberg Model

Having examined the dynamical aspects of the system's entanglement in Sec. (3.2.3), we now turn our attention to its geometric aspects under the same initial conditions. In particular, we investigate the implications within the framework of the evolved state Eq. (3.41). Under these conditions, the Fubini-Study Eq. (4.7) metric takes the form

$$ds^2 = \frac{1}{4}[2(\nu^2-1)\sin^2\chi - (\nu-1)^2\sin^4\chi + 4]d^2 + \frac{1}{2}\sin^2\chi d\kappa^2. \tag{4.22}$$

The effect of anisotropy on the state-space geometry depends on the initial parameter $\chi$. In particular, as $\chi$ approaches $\pi/2$, the influence of anisotropy becomes more pronounced, significantly altering the geometric structure of the state manifold. Conversely, for $\chi = 0$ and $\chi = \pi$, the geometry remains independent of $\nu$, since in these specific cases, the initial state Eq. (3.38) simplifies to the trivial states $|11\rangle$ and $|00\rangle$, respectively. Thus, the dynamics of the two-spin system evolve on a manifold parametrized by two variables, whose structure is strongly influenced by the choice of $\chi$. This dependence highlights the crucial role of anisotropy in shaping the geometric properties of the quantum state space. Such effects have been extensively studied in various contexts, providing further insights into the interplay between anisotropy and quantum geometry. The Fubini-Study metric, formulated as a function of the degree of entanglement and the evolution time, provides a rigorous geometric framework that quantitatively connects these two fundamental physical quantities. As derived from Eq. (3.46), the Fubini-Study metric takes the form

$$ds^2 = \left[\frac{(\nu-1)(\nu+3)\sin^2\chi + 2}{2(2+(\nu-1)\sin^2\chi)^2} - 1\right] d\mathcal{C}^2 + \frac{1}{2}\sin^2\chi d\kappa^2. \tag{4.23}$$

This relation demonstrates that quantum states sharing the same degree of entanglement are confined to a subspace of reduced dimensionality. Specifically, these states trace out a circular trajectory of radius $\sqrt{g_{\kappa\kappa}}$, emphasizing the profound influence of entanglement on the geometric structure of quantum phase space. Consequently, the nature of quantum correlations plays a pivotal role in shaping the underlying geometry of the system. Entanglement also influences the geometric phase accumulated by the evolving state Eq. (3.41). Substituting Eq. (3.38), Eq. (3.41) and Eq. (3.46) into Eq. (4.15) yields

$$\begin{aligned}\Phi_g \simeq \arctan&\left[\frac{4\mathcal{C}(2+(\nu-1)\sin^2\chi)(\nu\sin^2\chi - 2\nu_+\cos^2\chi + 2\nu_-)}{\sin^2\chi[4(2+(\nu-1)\sin^2\chi)^2(\kappa^2-2)+\nu^2\mathcal{C}^2]}\right]\\ &+\frac{1}{2}\left[\frac{\nu\cos^2\chi+\sin^2\chi}{2+(\nu-1)\sin^2\chi}\right]\mathcal{C}.\end{aligned} \tag{4.24}$$

The explicit dependence of the geometric phase Eq. (4.24) on the degree of quantum entanglement, illustrated in Fig. (4.1), serves as a fundamental indicator of the sensitivity of the state-space geometry to quantum correlations. A detailed analysis of this phase as a function of the concurrence $\mathcal{C}$ and the anisotropy parameter $\nu$ reveals several distinct behavioral regimes.

In general, the geometric phase exhibits a linear evolution with respect to $\mathcal{C}$ for any given value of $\nu$, except in the particular case of $\nu = -\frac{2}{3}$, where a notable change in monotonicity occurs at $\mathcal{C} = 0.4$. This critical point marks a transition between two distinct dynamical regimes. Specifically, for $\nu = -5$ and $\nu = -3$, the geometric phase remains strictly positive over the entire interval $\mathcal{C} \in [0, 1]$, indicating that the state vector undergoes a counterclockwise rotation throughout its evolution. Conversely, for $\nu = 0.5$ and $\nu = 1.5$, the geometric phase is negative, signifying a clockwise rotation as the system transitions from the non-entangled state ($\mathcal{C} = 0$) to the maximally entangled state ($\mathcal{C} = 1$). A particularly intriguing case arises for $\nu = -\frac{2}{3}$, where the geometric phase exhibits a hybrid behavior: it increases within the range $\mathcal{C} \in [0, 0.4]$, only to decrease for $\mathcal{C} \in [0.4, 1]$. This inversion in the trend signifies a reversal in the direction of rotation of the state vector and represents a quantum phase transition. Such a phenomenon underscores the intricate interplay between quantum entanglement and anisotropy in governing the geometric properties of the system. These findings emphasize the direct control exerted by both entanglement and anisotropy on the geometric phase, thereby reaffirming the crucial role of quantum correlations in shaping the dynamical evolution of the system.

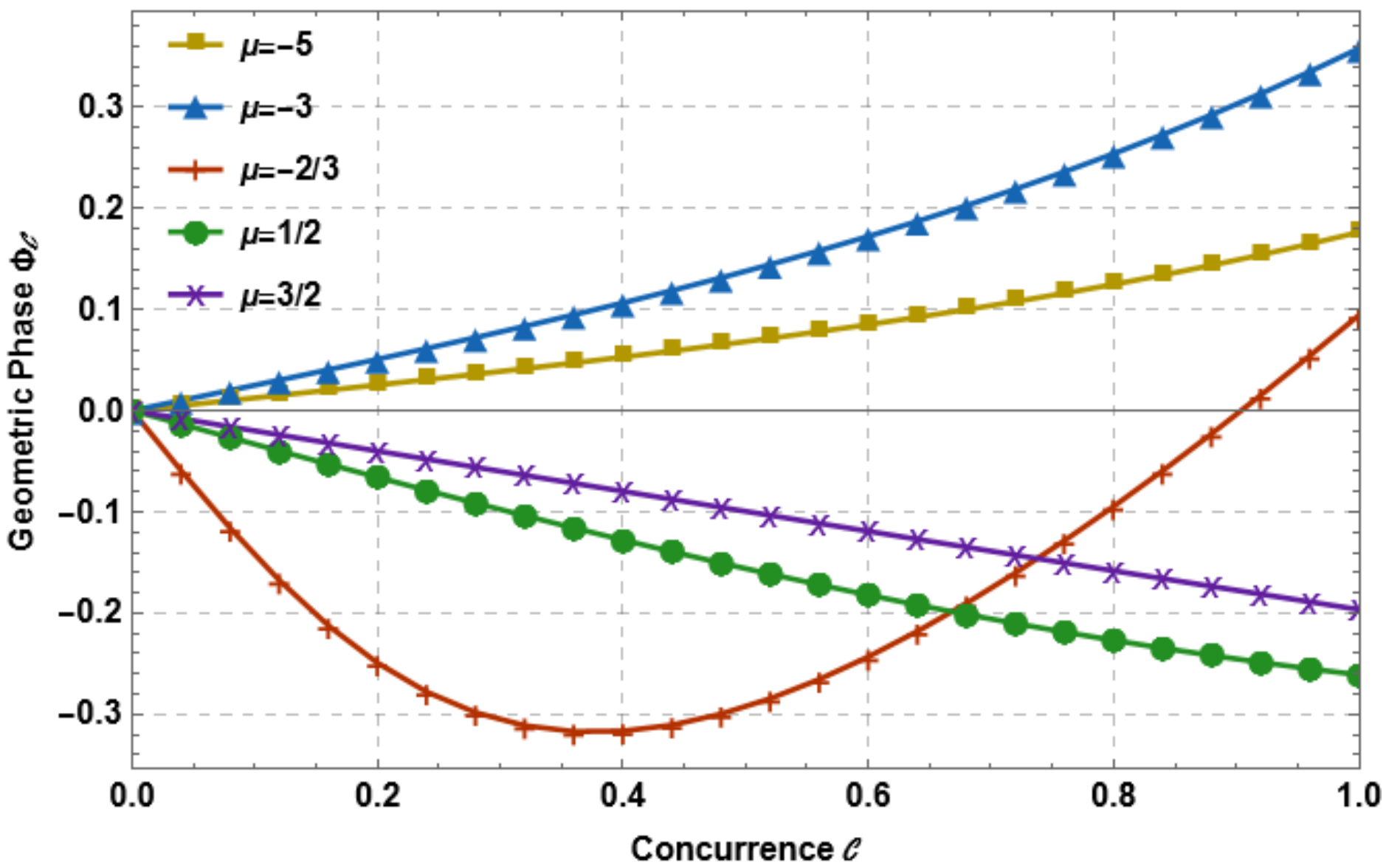

Figure 4.1: Geometrical phase (4.24) plotted against concurrence (3.46) for various values of the anisotropy parameter $\nu$, with $\chi = \pi/2$ and $B_z = 0$.

## 4.2 Geometrical Description of N Interacting Spin-1/2 Under all range Ising Model

The evolution of a spin-$\frac{1}{2}$ quantum system, governed by the unitary evolution operator, gives rise to a set of accessible states that form a quantum space. The geometric and topological structure of this space will be the subject of our analysis. To achieve this, it is essential to determine the Fubini-Study metric, which provides a measure of infinitesimal distances between neighboring quantum states. Specifically, for two quantum states $|\Upsilon(\zeta^\mu)\rangle$ and $|\Upsilon(\zeta^\mu + d\zeta^\mu)\rangle$, in this case $\zeta^{1,2} = \eta, \kappa$, the infinitesimal distance element $d\mathbf{s}$ defined in Eq. (2.18). The physical variables $\eta$, $\varphi$, and $\kappa$, which characterize the evolved state Eq. (3.58), are represented by the parameters $\zeta^{\mu}$. These parameters serve as coordinates in the state space and determine the components of the metric tensor $g_{\mu\nu}$. Based on this definition, we derive the explicit expression of the Fubini-Study metric through direct computation

$$dS^2 = dS_i^2 + \frac{1}{4}N(N-1)\sin^2\eta\left[N-1-\left(N-\frac{3}{2}\right)\sin^2\eta\right]d\kappa^2 + \frac{1}{4}N(N-1)\cos\eta\sin^2\eta d\varphi d\kappa, \quad (4.25)$$

where

$$dS_i^2 = \frac{N}{4}\left(d\eta^2 + \sin^2\eta d\varphi^2\right), \quad (4.26)$$

represents the line element that defines the sphere associated with the system's initial states. When $\kappa = 0$ (i.e., in the absence of evolution), the metric Eq. (4.25) indicates that the state space reduces to a sphere of radius $2\sqrt{N}$. However, when time evolution is incorporated, the state space takes the form of a closed three-dimensional manifold, exhibiting dependence solely on $\eta$ and $\kappa$, which implies that the geometry remains invariant with respect to the azimuthal angle $\varphi$. Consequently, the metric describing this space can be reformulated as

$$dS^2 = \frac{N}{4}d\eta^2 + \frac{1}{4}N(N-1)\sin^2\eta\left[N-1-\left(N-\frac{3}{2}\right)\sin^2\eta\right]d\kappa^2. \quad (4.27)$$

To analyze the topology of this quantum state space, it is necessary to compute the curvature $K$, which characterizes the intrinsic curvature of the manifold described by equation Eq. (4.27). This curvature is defined from the metric tensor according to [61]

$$K = \frac{1}{(g_{\eta\eta}g_{\kappa\kappa})^{1/2}}\left[\frac{\partial}{\partial\kappa}\left(\left(\frac{g_{\kappa\kappa}}{g_{\eta\eta}}\right)^{1/2}\Gamma^{\kappa}_{\eta\eta}\right) - \frac{\partial}{\partial\eta}\left(\left(\frac{g_{\kappa\kappa}}{g_{\eta\eta}}\right)^{1/2}\Gamma^{\kappa}_{\eta\kappa}\right)\right]. \quad (4.28)$$

The associated Christoffel symbols are given by

$$\Gamma^{\kappa}_{\eta\eta} = -\frac{1}{2g_{\kappa\kappa}}\left(\frac{\partial g_{\eta\eta}}{\partial\kappa}\right), \quad \text{and} \quad \Gamma^{\kappa}_{\eta\kappa} = \frac{1}{2g_{\kappa\kappa}}\left(\frac{\partial g_{\kappa\kappa}}{\partial\eta}\right). \quad (4.29)$$

Fascinatingly, the time component $g_{\kappa\kappa}$ of the metric vanishes at the critical points $\eta = 0$ and $\eta = \pi$. This vanishing leads to an indeterminacy in the curvature of $G$ precisely at these points, thereby revealing the presence of singularities in the geometric structure of the system. These singularities indicate regions where the conventional description of the curvature breaks down. However, in all other regions of the state space corresponding to a system of $N$ spin-$\frac{1}{2}$ particles, the G-curvature of remains well-defined, ensuring a consistent geometric interpretation of the quantum evolution. By utilizing the explicit expressions for the metric components $g_{\eta\eta}$ and $g_{\kappa\kappa}$ as given in Eq. (4.27), we can directly derive the G-curvature in an explicit form through the following relation:

$$K = \frac{8}{N}\left[2 - \frac{(2N-3)\cos^2\eta + N}{((2N-3)\cos^2\eta + 1)^2}\right]. \quad (4.30)$$

The analysis of the curvature of the state space reveals a fundamental dependence on the initial parameters $\eta$ and $N$, while it remains entirely independent of the parameter $\kappa$, which dictates the temporal evolution of the system. This crucial property indicates that the curvature is an intrinsic geometric feature of the state space, unaffected by the dynamical evolution of the system. Such an independence is explicitly manifested in the curvature expression $G$ given in Eq. (4.30), which satisfies the periodicity condition $K(\eta) = K(\eta + \pi)$. This periodic structure aligns with our conclusion that the quantum phase space described by Eq. (4.27) constitutes a closed two-dimensional manifold. A more detailed examination of Fig. 4.2, which depicts the evolution of the curvature $G$ as a function of the parameters $\eta$ and $N$, reveals a central symmetry about the axis $\eta = \pi/2$. This symmetry marks the location of the minimum curvature, serving as a geometric inflection point in the state space. More precisely, within the interval $\eta \in [0, \pi/2]$, the curvature $G$ exhibits a decreasing trend, imparting a concave structure to the state space. Conversely, in the complementary interval $\eta \in [\pi/2, \pi]$, the curvature increases monotonically until it attains its maximum value, thereby imparting a convex geometric character to the space. This distinct behavior strongly suggests that the quantum state space associated with a system of $N$ spin-$\frac{1}{2}$ particles acquires the global shape of a dumbbell. Furthermore, an essential aspect of the curvature behavior emerges for $N > 2$, where $G$ assumes negative values for specific values of $\eta$. This result is in direct agreement with the findings reported in Ref. [109]. The presence of negativity in the curvature signals the emergence of regions of hyperbolicity within the state space, underscoring the complex geometric structure of the manifold. Additionally, the occurrence of two singularities in the curvature expression of Eq. (4.30) suggests the existence of two conical defects in the quantum phase space described by Eq. (4.27). These singular points are localized near $\eta = 0$ and $\eta = \pi$, respectively, reinforcing the interpretation of the quantum state space as a topologically nontrivial surface.

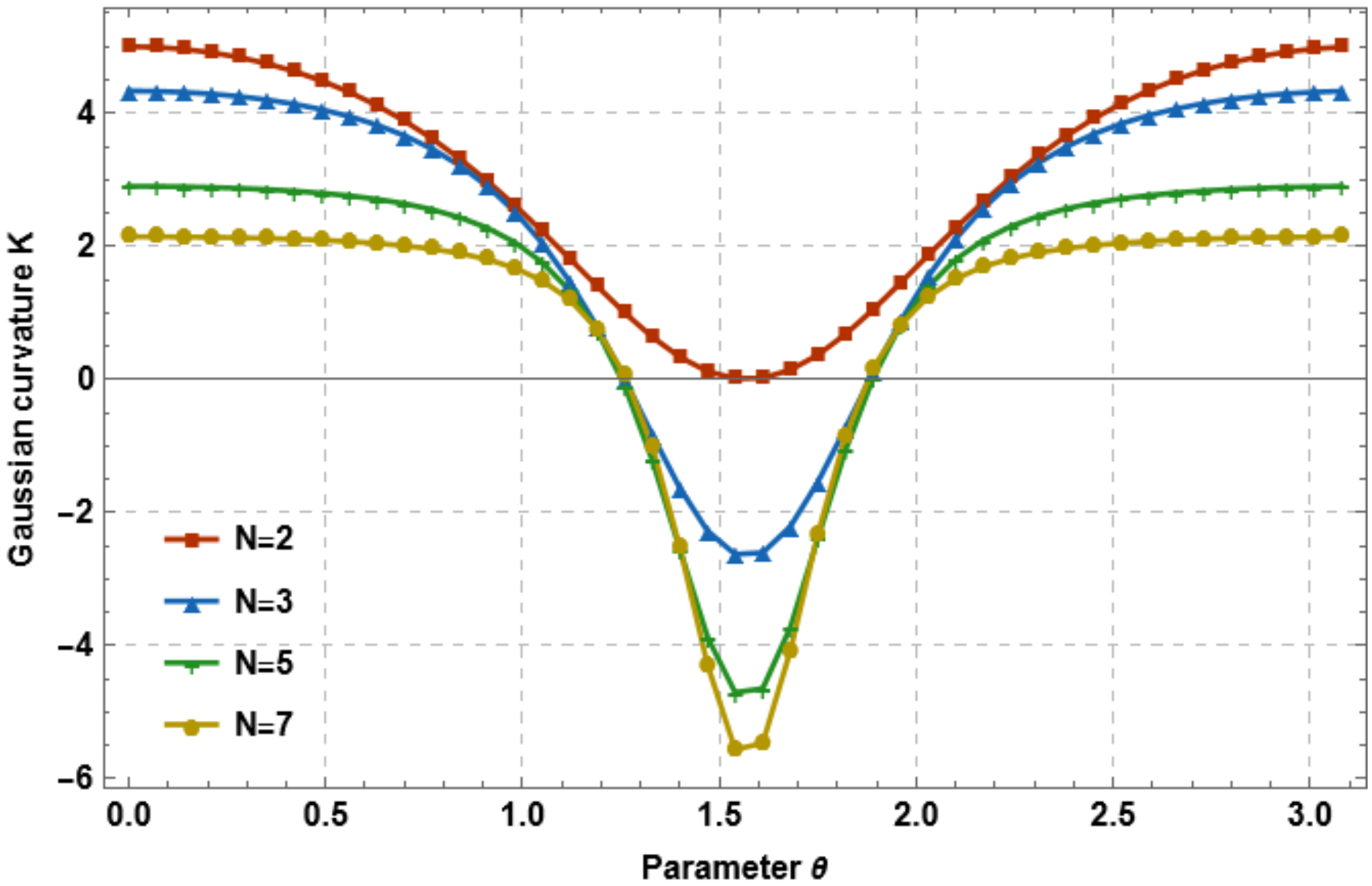

Figure 4.2: Dependence of the G-curvature (4.30) on the initial parameter $\eta$ for various spin-$\frac{1}{2}$ values.

The identification of the topology of the state space fundamentally relies on the Euler characteristic, a global invariant that encapsulates the intrinsic structure of the space. This topological quantity arises from two primary contributions: the volume contribution, determined by the integration of the Gaussian curvature over the manifold's interior, and the edge contribution, which accounts for boundary effects through the geodesic curvature. This principle is rigorously formulated by the Gauss-Bonnet theorem, which establishes a deep relationship between the intrinsic curvature of the manifold and its topological properties

$$\frac{1}{2\pi}\left[\int_M K\,dS + \oint_{\partial M} k_g\,dl\right] = \chi(M). \tag{4.31}$$

Here, $dS$ is the infinitesimal surface element over the manifold $M$, $k_g$ represents the geodesic curvature along the boundary $\partial M$, and $dl$ denotes the corresponding line element. The first integral quantifies the contribution of Gaussian curvature to the topology, encapsulating the intrinsic geometry of the bulk, while the second integral accounts for boundary effects, reflecting the curvature properties of the edge. This fundamental result expresses the Euler characteristic, a key topological invariant, entirely in terms of intrinsic geometric quantities, establishing a direct link between the global structure of the state space and its local curvature properties. By leveraging this intrinsic geometric framework, we refine the Gauss-Bonnet theorem in terms of the fundamental parameters of the state manifold

$$\int_0^{\pi}\int_0^{2\pi} K\left(g_{\eta\eta}g_{\kappa\kappa}\right)^{1/2} d\eta d\kappa + \Lambda = 2\pi\chi(M). \tag{4.32}$$

This reformulation not only deepens our understanding of the geometric properties of quantum evolution but also strengthens the connection between topological invariants and the differential structure of quantum state space. To rigorously determine the Euler characteristic of the surface, it is crucial to evaluate the Euler boundary integral, which captures the contribution of conical defects. This integral naturally emerges in the Gauss-Bonnet theorem, leading to the adapted form

$$4\pi(N-1) + \Lambda = 2\pi\chi(M). \tag{4.33}$$

Here, $\Lambda$ plays a fundamental role in characterizing the topology of the manifold $M$. Its precise computation requires analyzing the behavior of the underlying metric near singular points, specifically at $\eta = 0$ and $\eta = \pi$. Assuming that angular defects remain localized in these regions, we approximate the metric up to quadratic order in $\eta$, yielding

$$dS^2 = \frac{N}{4}d\eta^2 + \frac{1}{4}N(N-1)^2\eta^2 d\kappa^2. \tag{4.34}$$

This result highlights the intricate interplay between geometry and topology, where an explicit determination of $\Lambda$ is essential for computing the Euler characteristic $\chi(M)$ via the Gauss-Bonnet relation. Substituting

$$\Lambda = 2\left[2\pi - \frac{2\pi\sqrt{g_{\kappa\kappa}}}{\sqrt{g_{\eta\eta}}\eta}\right] = 4\pi(2-N), \tag{4.35}$$

into equation (18), we obtain

$$\chi(M) = 2. \tag{4.36}$$

This confirms that the quantum phase space associated with the spin-$\frac{1}{2}$ system exhibits a spherical topology. The prefactor 2 in $\Lambda$ directly accounts for the presence of two singular points, explicitly incorporated in the derivation.

### 4.2.1 Geometric Phases Acquired by the N Spin-1/2 State under all range Ising model

Having thoroughly examined the geometric and topological structure of the state space, as characterized by the metric tensor in Eq. (4.27), it is now essential to extend this investigation to the geometric phase that the evolving quantum state, given by Eq. (3.58), can acquire. This exploration is of fundamental significance, as it encompasses both arbitrary and cyclic evolution processes, offering deeper insights into the intrinsic geometric properties governing quantum dynamics. The study of geometric phase not only elucidates the holonomic aspects of quantum evolution but also provides a unifying perspective on the interplay between dynamical and geometric features within the system phase space.

### Geometric Phase During Cyclic Evolution

In this framework, we consider that the $N$-spin-$\frac{1}{2}$ system evolves arbitrarily along any evolution path on the two-dimensional closed manifold Eq. (4.27). The geometric phase gained through the evolved state Eq. (3.58) is given by expression Eq. (4.15), which represents the difference between the total phase and the dynamic phase [105, 106]. To calculate this geometric phase, we first need to determine the total phase acquired by the system. This total phase can be obtained by calculating the transition probability amplitude, i.e., the overlap, between the initial state Eq. (3.57) and the final state Eq. (3.58)

$$\langle \Upsilon_i | \Upsilon(t) \rangle = \sum_{p=0}^{N} C_N^P \cos^{2(N-p)}\left(\frac{\eta}{2}\right) \sin^{2p}\left(\frac{\eta}{2}\right) e^{-\frac{l\pi}{4}(N-2p)^2}. \tag{4.37}$$

We can begin by examining the expression for the total phase shift $\Phi_{\text{tot}}$ that results from inserting the overlap expression Eq. (4.37) into the first term on the right-hand side of Eq. (4.15). This total phase, which the $N$ spins-$\frac{1}{2}$ system undergoes, is then expressed as follows

$$\Phi_{\text{tot}} = -\arctan\left[\frac{\sum_{p=0}^{N} C_N^p \cos^{2(N-p)}\left(\frac{\eta}{2}\right)\sin^{2p}\left(\frac{\eta}{2}\right)\sin\left(\frac{\kappa(N-2p)^2}{4}\right)}{\sum_{p=0}^{N} C_N^p \cos^{2(N-p)}\left(\frac{\eta}{2}\right)\sin^{2p}\left(\frac{\eta}{2}\right)\cos\left(\frac{\kappa(N-2p)^2}{4}\right)}\right]. \tag{4.38}$$

It is fascinating to note that the total phase Eq. (4.38) is made up of two distinct components, one of geometric origin and the other of dynamical origin. The first, known as the geometric phase, is strongly linked to the geometric and topological characteristics of the system's quantum state space. This geometric component results from the implicit dependence of the total phase on the curvature Eq. (4.30) and the component of the metric Eq. (4.30), all of which depend on the parameters $N$ and $\eta$. On the other hand, the dynamic phase, of temporal origin, comes from the temporal evolution of the Hamiltonian eigenstates. Thus, the total phase $\Phi_{\text{tot}}$ exhibits a non-linear time dependence and satisfies the following periodic conditions

$$\Phi_{\text{tot}}(\kappa + 4\pi) = \Phi_{\text{tot}}(\kappa) \quad \text{for } N \text{ integer},$$

$$\Phi_{\text{tot}}(\kappa + 8\pi) = \Phi_{\text{tot}}(\kappa) \quad \text{pour } N \text{ semi-integer}.$$

The dynamic phase can be calculated by substituting the evolved state Eq. (3.58) into the second term on the right-hand side of Eq. (4.15). This gives us the following expression for the dynamic phase

$$\Phi_{\text{dyn}} = -\frac{\kappa N}{4}\left(N\cos^2\eta + \sin^2\eta\right), \tag{4.39}$$

which is proportional to the evolution time. This means that the dynamic phase tells us mainly about the duration of the system's evolution. On the other hand, the geometric phase accumulated by the $N$-spin-$\frac{1}{2}$ state Eq. (3.57) during any evolution in quantum phase space Eq. (4.27) can be expressed by the following relation

$$\Phi_{\rm g} = -\arctan\left[\frac{\sum_{p=0}^{N} C_N^p \cos^{2(N-p)}\left(\frac{\eta}{2}\right)\sin^{2p}\left(\frac{\eta}{2}\right)\sin\left(\frac{\kappa(N-2p)^2}{4}\right)}{\sum_{p=0}^{N} C_N^p \cos^{2(N-p)}\left(\frac{\eta}{2}\right)\sin^{2p}\left(\frac{\eta}{2}\right)\cos\left(\frac{\kappa(N-2p)^2}{4}\right)}\right] + \frac{\kappa N}{4}\left(N\cos^2\eta + \sin^2\eta\right). \tag{4.40}$$

Clearly, the resulting geometric phase Eq. (4.40) exhibits a non-linear temporal dependence, underscoring its dynamic nature. This manifests through a progressive accumulation or attenuation over time. Furthermore, its explicit dependence on the degrees of freedom $(\eta, \kappa)$, which characterize the quantum states within the phase space Eq. (4.27), signifies that the geometric phase is intimately tied to the specific trajectory traced by the system during its evolution. This trajectory-dependent behavior underscores the intrinsic geometric structure of the quantum phase, linking it directly to the topology of the state space. Additionally, the dependence of the geometric phase on the initial parameters $(N, \eta)$ highlights its sensitivity to the underlying geometry of the quantum state manifold. This suggests that the geometric phase is not merely an auxiliary quantity but rather an essential descriptor of the system's evolution, encoding fundamental aspects of its quantum dynamics. Consequently, we conclude that the geometric phase Eq. (4.40) serves as a natural parameter for classifying the set of possible evolutionary paths in the system. This property has significant implications for quantum information processing, where such phases can be exploited in the design of quantum logic gates. These gates constitute fundamental building blocks in quantum computation, facilitating the implementation of high-efficiency quantum algorithms [110,111]. Now, let us consider a specific scenario in which we examine the geometric phase acquired by a spin-$\frac{1}{2}$ state Eq. (3.58) over a very short time interval. In this regime, expanding the exponential factor in Eq. (4.37) up to second order in $\kappa$ yields

$$\langle\Upsilon_i|\Upsilon(t)\rangle \simeq 1 + \frac{\kappa^2 N(N-1)}{64}\left[4(N-1)(N+2)\cos^2\eta - (N-3)(N-2)\sin^2 2\eta + 4(3N-2)\right] - i\frac{\kappa N}{4}\left(N\cos^2\eta + \sin^2\eta\right). \tag{4.41}$$

From this, the geometric phase Eq. (4.40) can be explicitly expressed as

$$\Phi_{\rm g} \simeq -\arctan\left[\frac{\kappa N\left(N\cos^2\eta + \sin^2\eta\right)}{4 + \frac{\kappa^2 N(N-1)}{16}\left[4(N-1)(N+2)\cos^2\eta - (N-3)(N-2)\sin^2 2\eta + 4(3N-2)\right]}\right] + \frac{\kappa N}{4}\left(N\cos^2\eta + \sin^2\eta\right). \tag{4.42}$$

A particularly noteworthy observation is that for $\kappa = 0$, the geometric phase vanishes entirely, indicating that no quantum phase is acquired. This follows from the fact that the system remains in its initial state Eq. (3.57) without undergoing any evolution. Furthermore, as the number of particles increases, the dynamic phase component grows in prominence, gradually overshadowing the geometric contribution. In the thermodynamic limit ($N \to \infty$), a remarkable phenomenon emerges: the total phase completely vanishes, implying that the geometric and dynamic phases become indistinguishable at all times during the evolution. This phenomenon is of significant experimental relevance, as it provides a direct means of extracting the geometric phase through time integration of the expectation values of the energy levels of the Ising-type Hamiltonian Eq. (3.55). Such an approach could be instrumental in experimental settings, offering a viable technique for measuring geometric phases in large-scale quantum systems.

#### Geometric phase under a cyclic evolution

In this study, we focus on the geometric phase associated with the cyclic evolution of a system composed of $N$ spin-$\frac{1}{2}$ particles. The wave function, as expressed in Eq. (3.58), satisfies the cyclicity condition

$$|\Upsilon(T)\rangle = e^{i\Phi_{\text{tot}}}|\Upsilon(0)\rangle, \tag{4.43}$$

where $T$ represents the time required to complete a full evolution cycle. Within this framework, the Aharonov-Anandan (AA) geometric phase characterizes the phase acquired by the system upon completion of one cycle, corresponding to the traversal of a closed trajectory in the associated parameter space. This phase, given explicitly by Eq. (2.48), can be rewritten as

$$\Phi_{\text{g}}^{AA} = \int_0^T d\Phi_{\text{tot}} + i\int_0^T \langle\Upsilon(t)|\frac{\partial}{\partial t}|\Upsilon(t)\rangle\, dt. \tag{4.44}$$

Examining the cyclic evolution of the $N$-spin-$\frac{1}{2}$ system, a striking feature emerges: the geometric Aharonov-Anandan (AA) phase obtained is entirely independent of the dynamical aspects of the system. Instead, it depends solely on the initial parameters $\eta$ and $N$, which fundamentally govern the geometry of the state space. This observation underscores the intrinsic geometric nature of the phase, detached from the specifics of the system's temporal evolution. A direct consequence of this property is evident when we insert Eq. (4.40) and Eq. (4.39) into Eq. (4.44), yielding the expression

$$\Phi_{\text{g}}^{AA} = -\frac{\pi}{2}N(N-1)\sin^2\eta. \tag{4.45}$$

This result implies that the geometric phase cannot serve as a parameterization of the system's cyclic evolution trajectories, as these trajectories are entirely dictated by the geometric structure of the state space rather than by the nature of the evolution process itself. Such an observation highlights the predominant influence of geometric curvature in the accumulation of the cyclic phase. Consequently, a thorough analysis of the relationship between the AA geometric phase and the curvature of the state space becomes crucial. By substituting Eq. (4.30) into Eq. (4.46), we derive the refined relation

$$\Phi_{g}^{AA} = \frac{\pi N(N-1)}{2}\left[\frac{-56+3N(16-(N-1)K)}{(2N-3)(NK-16)}\right]. \tag{4.46}$$

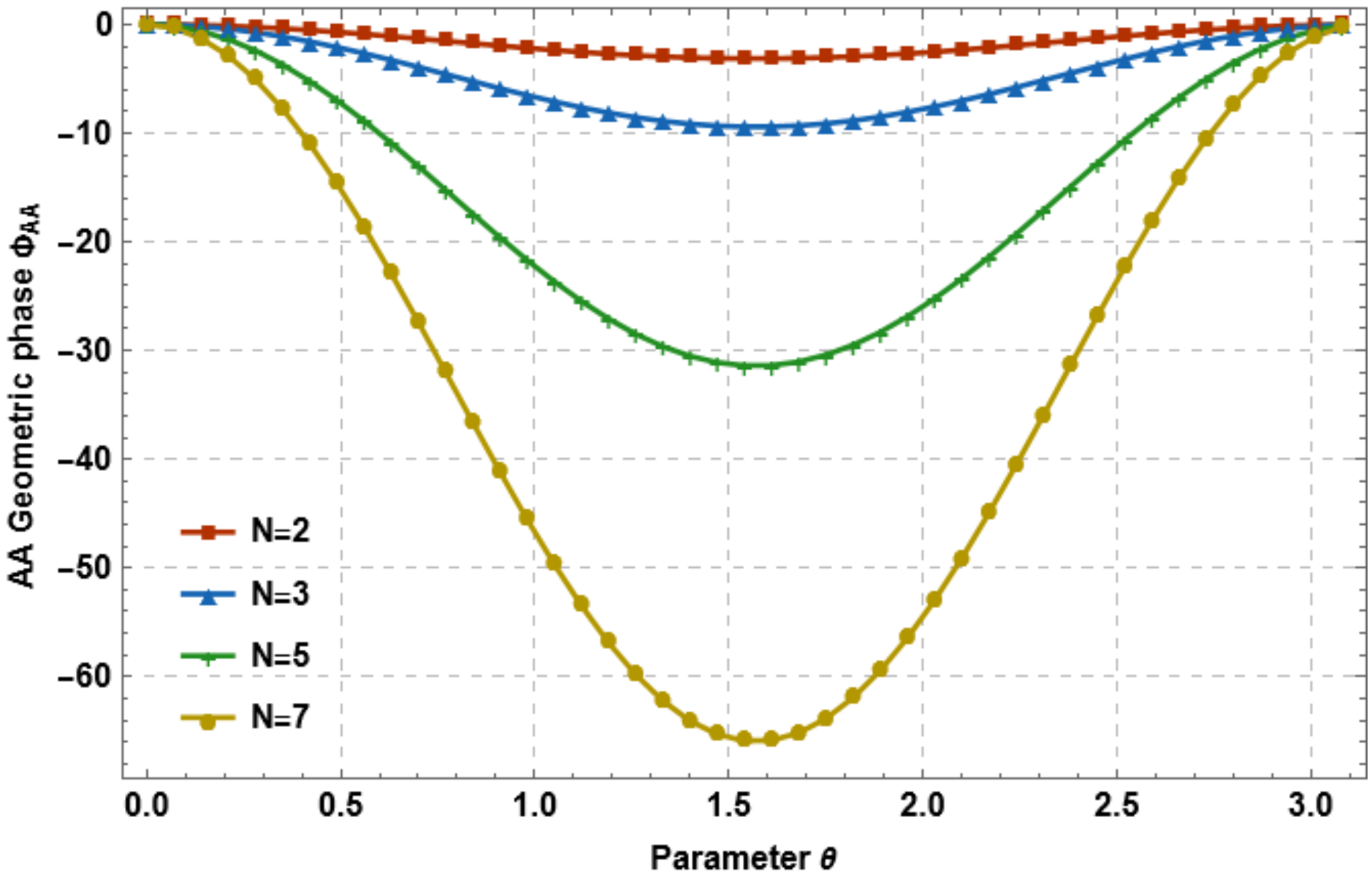

Figure 4.3: Dependence of the AA-geometric phase (4.46) on the initial parameter $\eta$ for various spin-$\frac{1}{2}$ values.

This expression explicitly establishes the dependence of the geometric phase on the geometric structure of the state space. Graphically representing $\Phi_{\mathrm{g}}^{AA}$ as a function of the parameters $(N, \eta)$ in Fig. 4.3 reveals a remarkable analogy with the behavior of the curvature $G$ in Fig. 4.2. Specifically, a decrease in $\Phi_{\mathrm{g}}^{AA}$ is observed over the interval $\eta \in [0, \pi/2]$, where the geometry of the state space exhibits a concave structure characterized by a reduction in $G$. Conversely, for $\eta \in [\pi/2, \pi]$, where the curvature becomes convex, the geometric phase $\Phi_{\mathrm{g}}^{AA}$ increases. This symmetrical behavior with respect to $\eta$ is a direct consequence of the dumbbell-like structure of the state space. Beyond the geometric phase, it is also crucial to analyze the emergence of a topological phase during the cyclic evolution of the wave function Eq. (3.58). This phase corresponds to the component of the cyclic geometric phase that remains entirely independent of any dynamical contribution. It is given by:

$$\Phi_{\mathrm{top}}^{AA} = -\frac{\pi}{2} N^2. \tag{4.47}$$

A particularly noteworthy property of this topological phase is its quadratic dependence on the number of particles $N$. Additionally, it assumes fractional values for odd $N$ and integral multiples of $\pi$ for even $N$. As a result, the topological phase Eq. (4.47) plays a fundamental role in characterizing the topology of the state space Eq. (4.27), encoding the closed evolution paths traced by the system. The implications of this property in quantum computing are especially promising, particularly in the design of highly efficient quantum circuits. This finding paves the way for an intriguing perspective in determining optimal evolution paths for the system while accounting for both the evolution speed and the Fubini-Study distance. This problem, known as the quantum brachistochrone problem, will be explored in depth in the subsequent section.

### 4.2.2 Geometric Aspects of Quantum Entanglement in a Two-Spin-1/2 System Governed by the All Range Ising Model

In order to rigorously elucidate the geometric nature of quantum correlations within the bipartite spin system under investigation, we undertake a comprehensive analysis that illustrates the fundamental interplay between quantum entanglement and the geometric structures previously derived. This approach allows us to establish a direct correspondence between the degree of entanglement and the intrinsic geometry of the system state space. The two-spin state space, which serves as the geometric arena for the system dynamics, is characterized by the following metric tensor

$$dS^2 = \frac{1}{2}d\eta^2 + \frac{1}{4}\sin^2\eta(2-\sin^2\eta)d\kappa^2. \tag{4.48}$$

This metric elegantly encodes the intrinsic geometry of the system, capturing the dependence of the infinitesimal distance element $dS^2$ on the state-space coordinates $\eta$ and $\kappa$. The first term, $\frac{1}{2}d\eta^2$, represents the contribution from variations in $\eta$, while the second term, weighted by a $\eta$-dependent coefficient, governs the role of $\kappa$ in shaping the geometry. Notably, the factor $(2-\sin^2\eta)$ introduces a nontrivial curvature, reflecting the underlying structure of the two-spin configuration space. By substituting equation Eq. (3.71) into Eq. (4.48), we obtain the Fubini-Study metric, which serves as a fundamental tool in characterizing the differential structure of the state space as a function of the competitive interplay governing the system. This metric encapsulates the intrinsic curvature properties induced by quantum correlations and is expressed explicitly as follows

$$\mathrm{d}S^2 = \frac{\mathcal{C}}{4}\left(\frac{1}{2\tan^2\kappa(|\sin\kappa|-\mathcal{C})}+\frac{2|\sin\kappa|-\mathcal{C}}{\sin^2\kappa}\right)\mathrm{d}\kappa^2-\frac{\mathrm{d}\mathcal{C}\mathrm{d}\kappa}{4\tan\kappa(|\sin\kappa|-\mathcal{C})}+\frac{\mathrm{d}\mathcal{C}^2}{8\mathcal{C}(|\sin\kappa|-\mathcal{C})}. \tag{4.49}$$

We can transform the previously obtained metric into a diagonal form, expressed as:

$$\mathrm{d}S^2 = \frac{1}{8\mathcal{C}_r(1-\mathcal{C}_r)}\mathrm{d}\mathcal{C}_r^2 + \frac{1}{4}\mathcal{C}_r(2-\mathcal{C}_r)\mathrm{d}\kappa^2, \tag{4.50}$$

where $\mathcal{C}_r = \mathcal{C}/|\sin\kappa|$ denotes the reduced competition, which varies within the interval $[0,1]$. This transformation allows us to reparameterize the relevant phase space Eq. (4.48) in terms of two experimentally accessible physical quantities: The degree of entanglement shared between the two spins as a function of the evolution time. Such a reformulation provides a robust foundation for an experimental investigation of the geometric, topological, and dynamical properties of the state space, encompassing aspects such as phase-space geometry, quantum phases, evolution velocity, and the geodesic distance traversed by the two-spin system Eq. (3.70) during its evolution. A key observation is that quantum entanglement serves as a dimensional reduction factor for the state space. Specifically, states corresponding to the same level of entanglement (i.e., $\mathcal{C}$ = constant) reside on closed one-dimensional manifolds, governed by the metric

$$\mathrm{d}S^2 = \frac{\mathcal{C}}{4}\left(\frac{1}{2\tan^2\kappa(|\sin\kappa|-\mathcal{C})}+\frac{2|\sin\kappa|-\mathcal{C}}{\sin^2\kappa}\right)\mathrm{d}\kappa^2. \tag{4.51}$$

In the complete state space Eq. (4.51), these manifolds manifest as closed curves, following the structure dictated by the metric component $g_{\kappa\kappa}$. Conversely, states characterized by the same degree of reduced entanglement (i.e., $\mathcal{C}_r$ = constant) are constrained to circular trajectories, described by

$$\mathrm{d}S^2 = \frac{1}{4}\mathcal{C}_r(2-\mathcal{C}_r)\mathrm{d}\kappa^2. \tag{4.52}$$

The radii of these circles, given by

$$\mathrm{R} = \frac{\sqrt{\mathcal{C}_r(2-\mathcal{C}_r)}}{2}, \tag{4.53}$$

depend explicitly on the level of reduced entanglement considered. This result demonstrates the crucial role of quantum correlations in reducing the dimensionality of the state space Eq. (4.51). Notably, this property extends to all integrable quantum systems, emphasizing the universality of the entanglement-induced geometric restructuring. Within the same theoretical framework, we can further investigate the influence of entanglement on the G-curvature of the phase space associated with the two-spin system Eq. (4.51). By substituting Eq. (3.70) into Eq. (4.30), we derive the explicit expression for the G-curvature as a function of entanglement

$$K = 4\left[2 + \frac{|\sin\kappa|(\mathcal{C}-3|\sin\kappa|)}{(\mathcal{C}-2|\sin\kappa|)^2}\right]. \tag{4.54}$$

This result once again underscores the explicit dependence of the state-space geometry on the degree of entanglement exchanged between the two interacting spins. From Eq. (4.54), we observe that in the case $\kappa = 0$ (i.e., in the absence of evolution), the G-curvature is independent of $\mathcal{C}$ and assumes the constant value $K = 8$, corresponding to the curvature of the initial-state sphere Eq. (3.70). However, for $\kappa > 0$ (i.e., the evolutionary regime), the G-curvature becomes explicitly entanglement-dependent, and its variation as a function of entanglement is depicted in Fig. (4.4). We observe a monotonic decrease in the G-curvature as the entanglement between the two spins increases. This phenomenon is attributed to the fact that quantum correlations act to reduce the curvature of the state space, effectively modifying its geometric structure. Furthermore, the G-curvature takes on negative values when the degree of entanglement satisfies the inequality

$$|\sin\kappa|(\mathcal{C}-3|\sin\kappa|) < -2(\mathcal{C}-2|\sin\kappa|)^2. \tag{4.55}$$

This condition reveals a fundamental aspect of quantum correlations: their capacity to induce compactification effects within the associated state space Eq. (4.51). Interestingly, separable states ($\mathcal{C} = 0$) occupy regions of maximum curvature, where $K_{\max} = 5$, whereas maximally entangled states ($\mathcal{C} = 1$) reside in regions of minimum curvature, given by

$$K_{\min} = 4\left[2 - \frac{|\sin\kappa|(3|\sin\kappa|-1)}{(2|\sin\kappa|-1)^2}\right]. \tag{4.56}$$

Thus, knowledge of the entanglement degree of the two-spin system Eq. (3.70) directly determines its position within the corresponding phase space Eq. (4.51). This conclusion highlights the deterministic nature of geometrical quantum mechanics, which relies on the geometrization of

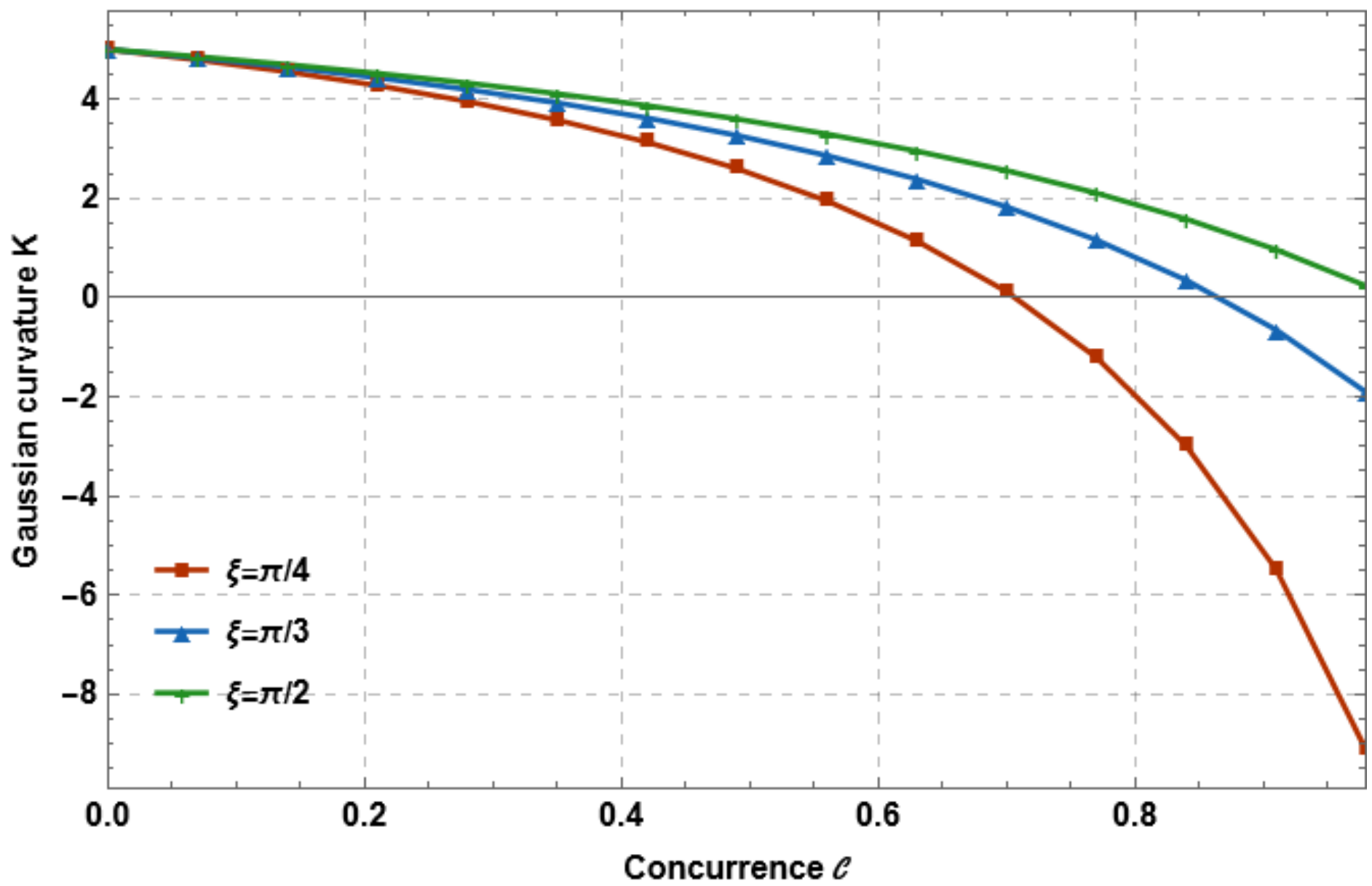

Figure 4.4: G-curvature (4.54) plotted against concurrence (3.71) for various values of $\kappa$.

Hilbert space through the introduction of a quantum phase space, analogous to its classical counterpart [102, 112].

The connection between geometric phase and quantum entanglement can also be explored in this context. By substituting Eq. (3.71) into equation Eq. (??), we derive the expression for the geometric phase $\Phi_{\rm g}$ acquired by the two-spin state Eq. (3.70) as a function of competition

$$\Phi_{\rm g} = -\arctan\left[\frac{(2|\sin\kappa| - \mathcal{C})\sin\kappa}{(2|\sin\kappa| - \mathcal{C})\cos\kappa + \mathcal{C}}\right] + \kappa\left(1 - \frac{\mathcal{C}}{2|\sin\kappa|}\right). \tag{4.57}$$

This expression reveals that the geometric phase is governed by two new physical degrees of freedom: entanglement and time. Consequently, it depends on every point (i.e., every physical state) in the underlying phase space Eq. (4.51). This implies that the geometric phase is influenced by both the path taken by the system and the geometry of the state space itself. Since the phase is expressed in terms of these two measurable quantities ($\mathcal{C}$ and $\kappa$), it is feasible to experimentally measure the geometric phase for any arbitrary evolution of the system. To better illustrate the interaction between geometric phase and entanglement, we have plotted the dependence of Eq. (4.57) on concurrence for specific values of $\kappa$ in Fig. (4.5). From this, we observe that the geometric phase Eq. (4.57) acquired by the two-spin system Eq. (3.70) during its evolution from a separable state ($\mathcal{C} = 0$) to a maximally entangled state ($\mathcal{C} = 1$) exhibits an approximately parabolic behavior. Therefore, its evolution can be divided into two distinct stages

- *First stage*: In this stage, the geometric phase decreases within the interval $\mathcal{C} \in [0, \mathcal{C}_{\rm c}]$, where $\mathcal{C}_{\rm c}$ represents the critical degree of entanglement at which this phase reaches its minimum

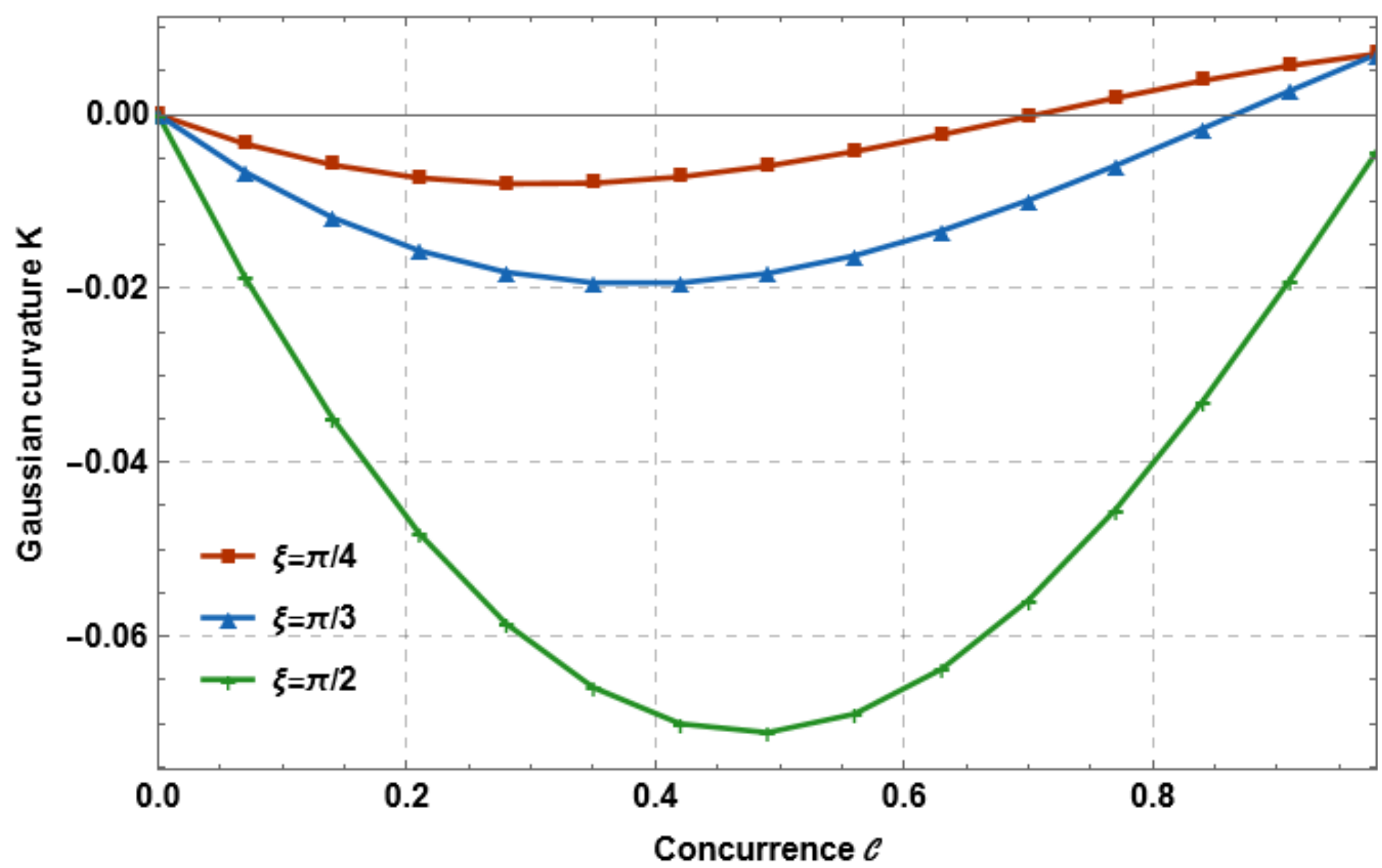

Figure 4.5: Geometric phase (4.57) versus the concurrence (3.71) for some values of $\kappa$

value (see Fig. (4.5)). This critical degree is explicitly given by

$$\mathcal{C}_{\mathrm{c}} = \sin\kappa - \cot\frac{\kappa}{2}\sqrt{\frac{\sin\kappa}{\kappa}(2-\kappa\sin\kappa-2\cos\kappa)}.$$

In this first stage, the evolving state Eq. (3.70) acquires a negative geometric phase. This can be interpreted as a loss of geometric phase by the system. Geometrically, this indicates that, during parallel transport, the state vector Eq. (3.70) undergoes a clockwise rotation (i.e., an angle of negative sign) relative to the separable state (the initial state). Thus, in the region $[0, \mathcal{C}_{\mathrm{c}}]$, quantum correlations favor the loss of geometric phase.

- *Second stage*: In this stage, the geometric phase increases in the interval $\mathcal{C} \in [\mathcal{C}_{\mathrm{c}}, 1]$ (i.e., an inverse behavior). The evolving state Eq. (3.70) accumulates a positive geometric phase, corresponding to the recovery of geometric phase by the system. Geometrically, during parallel transport, the state vector Eq. (3.70) undergoes a counter-clockwise rotation (i.e., an angle of positive sign) relative to the separable state. Thus, in the region $[\mathcal{C}_{\mathrm{c}}, 1]$, quantum correlations favor the gain of geometric phase. Consequently, the variation of the geometric phase with entanglement is approximately symmetric with respect to the critical value $\mathcal{C}_{\mathrm{c}}$, which is primarily explained by the dumbbell-shaped structure of the underlying phase space Eq. (4.51). From a practical standpoint, quantum entanglement constitutes a valuable physical resource that can be exploited experimentally to control the geometric phase induced during the evolution of the two-spin system.

In the context of cyclic evolution, the geometric phase can also be examined in relation to entanglement. By substituting Eq. (3.71) into Eq. (4.46), we derive the AA geometric phase $\Phi_{\mathrm{g}}^{AA}$

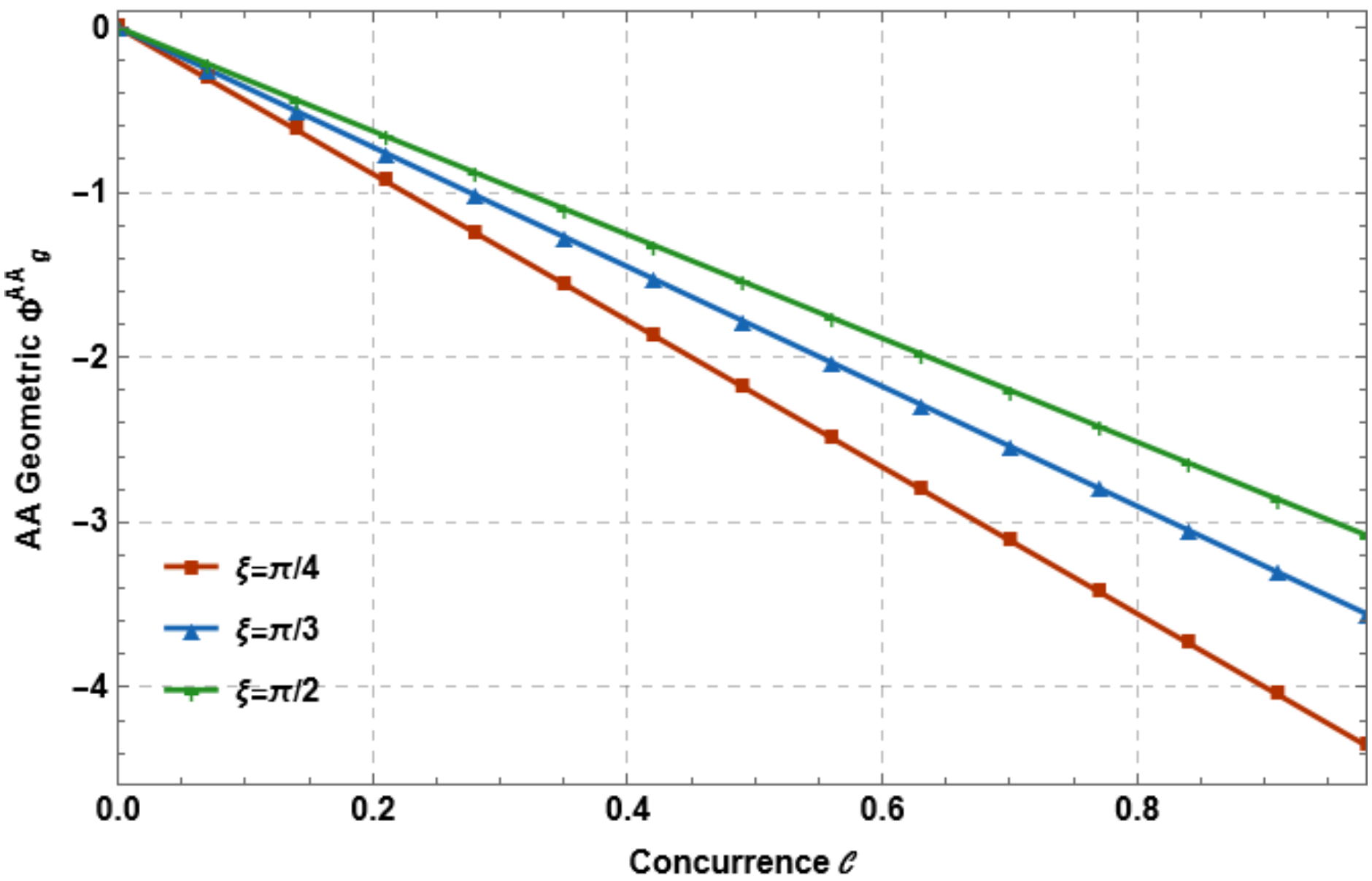

Figure 4.6: AA-geometric phase (4.58) plotted against concurrence (3.71) for various values of $\kappa$.

accumulated by the evolved state Eq. (3.70) as a function of competition, expressed as follows

$$\Phi_{\mathrm{g}}^{AA} = -\pi \frac{\mathcal{C}}{|\sin\kappa|}. \tag{4.58}$$

This expression shows that the AA geometric phase is proportional to the level of entanglement between the two spins, with a negative proportionality factor. This implies that the AA geometric phase decreases linearly as entanglement increases. To illustrate this behavior, we have plotted it in Fig. (4.6), where we observe that, as the system becomes more interleaved, it accumulates a negative-sign AA geometric phase. This behavior is similar to what was observed for the geometric phase Eq. (4.57) in its first stage (i.e., in the $[0, \mathcal{C}_{\mathrm{c}}]$ region), and thus similar interpretations can be applied to the AA geometric phase. Regarding the topological phase resulting from the cyclic evolution of the two-spin system, it is given by

$$\Phi_{\mathrm{top}}^{AA} = -2\pi. \tag{4.59}$$

This phase is unaffected by entanglement, as it represents the part of the AA-geometric phase that does not receive any contribution from the dynamical evolution.

## 4.3 Geometrical Description of N Interacting Spin-s particles Under Ising Model

### 4.3.1 Geometry and Topology of the Resulting State Space

Having established the time evolution of the $N$ particle system through the time-evolution propagator and identified the evolved state given by Eq. (3.83), we now proceed to examine the geometric and topological structure of the quantum manifold encompassing all states accessible during this evolution. To this end, we adopt the Fubini-Study metric, which provides a natural notion of distance within the projective Hilbert space of pure quantum states. This metric is defined via the infinitesimal distance $\mathrm{d}\mathbf{S}$ between two neighboring pure states $|\Upsilon(\zeta^a)\rangle$ and $|\Upsilon(\zeta^a + \mathrm{d}\zeta^a)\rangle$ given in Eq. (2.18), where the coordinates $\zeta^a$ correspond to the degrees of freedom $\kappa$, $\Phi$, and $\eta$ that parametrize the evolved state given in Eq. (3.83). We explicitly obtain the form of the metric tensor as

$$\begin{aligned} d\mathbf{S}^2 =& dS^2 + \tfrac{1}{2}\mathcal{N}(\mathcal{N}-1)\mathbf{s}^2 \sin^2\kappa \left[1 + (4\mathbf{s}(\mathcal{N}-1)-1)\cos^2\kappa\right] d\eta^2 \\ &+ \mathcal{N}(\mathcal{N}-1)\mathbf{s}^2 \cos\kappa \sin^2\kappa d\varphi d\eta, \end{aligned} \tag{4.60}$$

where

$$dS^2 = \frac{\mathcal{N}\mathrm{s}}{2}\left[d\kappa^2 + \sin^2\kappa\, d\varphi^2\right].$$

This formulation allows us to rigorously characterize the Riemannian geometry associated with the manifold of quantum states that become accessible following the evolution of the $\mathcal{N}$ spin-$\mathbf{s}$ system. It is straightforward to verify that when $\eta = 0$ (i.e., in the absence of evolution), the state manifold described by Eq. (4.60) simplifies to a sphere, as previously given in Eq. (4.3.1). This confirms that the quantum state space retains a fundamental geometric structure in the absence of evolution. A notable property of this geometry is that the metric components in Eq. (4.60) are independent of the azimuthal angle $\Phi$. This invariance implies that quantum state spaces associated with a fixed $\Phi$ exhibit an identical geometric structure. Consequently, we can infer that the state space corresponding to the $\mathcal{N}$ qudits under consideration forms a two-dimensional curved manifold, parametrized exclusively by $\kappa$ and $\eta$. The metric defining this space takes the form

$$d\mathbf{S}^2 = \tfrac{1}{2}\mathcal{N}(\mathcal{N}-1)\mathbf{s}^2 \sin^2\kappa \left(1 + [4\mathbf{s}(\mathcal{N}-1)-1]\cos^2\kappa\right) d\eta^2 + \frac{\mathcal{N}\mathbf{s}}{2} d\kappa^2. \tag{4.61}$$

We now turn our attention to the topological properties of the $\mathcal{N}$ spins-$s$ system, focusing on the geometric structure of the corresponding quantum states manifold. To this end, we analyze its curvature $K$, which is expressed in terms of the metric tensor Eq. (4.61) as follows

$$K = \frac{1}{\sqrt{g_{\kappa\kappa}g_{\eta\eta}}}\left\{\frac{\partial}{\partial\eta}\left[\left(\frac{g_{\eta\eta}}{g_{\kappa\kappa}}\right)^{1/2}\Gamma^{\eta}_{\kappa\kappa}\right] - \frac{\partial}{\partial\kappa}\left[\left(\frac{g_{\eta\eta}}{g_{\kappa\kappa}}\right)^{1/2}\Gamma^{\eta}_{\kappa\eta}\right]\right\}, \tag{4.62}$$

where the relevant Christoffel symbols are given by

$$\Gamma^{\eta}_{\kappa\kappa} = -\frac{1}{2g_{\eta\eta}}\left(\frac{\partial g_{\kappa\kappa}}{\partial \eta}\right), \quad \text{and} \quad \Gamma^{\eta}_{\kappa\eta} = \frac{1}{2g_{\eta\eta}}\left(\frac{\partial g_{\eta\eta}}{\partial \kappa}\right). \tag{4.63}$$

A crucial aspect of this analysis is the behavior of the metric component $g_{\eta\eta}$, which vanishes at the specific points $\kappa = 0$ and $\kappa = \pi$. As a direct consequence, the curvature $\mathcal{K}$ becomes singular at these points, rendering the quantum variety undefined there. This observation reveals that the state manifold possesses two topological singularities precisely at $\kappa = 0$ and $\kappa = \pi$, while at any other value of $\kappa$, the curvature remains well-defined. The presence of these singularities is indicative of an intrinsic topological structure inherent to the system. By substituting the metric components into equation Eq. (4.62), we explicitly derive the expression for the curvature

$$K = \frac{4}{\mathcal{N}\mathbf{s}}\left(2 - \frac{[4\mathbf{s}(\mathcal{N}-1)-1]\cos^2\kappa + 2\mathbf{s}(\mathcal{N}-1)+1}{([4\mathbf{s}(\mathcal{N}-1)-1]\cos^2\kappa+1)^2}\right). \tag{4.64}$$

A particularly noteworthy observation is that this geometric depends exclusively on the system's initial parameters, namely $\kappa$, $\mathcal{N}$, and $\mathbf{s}$, while exhibiting complete independence from time. This fundamental property underscores the fact that the geometric structure of the state space remains invariant under system dynamics. Moreover, further analytical investigation reveals that, for $\mathcal{N} > 2$ and $\mathbf{s} \geqslant \frac{1}{2}$, the curvature can be negative values for specific ranges of $\kappa$. This result is consistent with the findings presented in [109].

In order to analyze the topology of the quantum state space Eq. (4.61) of the $N$-spin system, we need to determine its Euler characteristic $\chi(M)$. The latter is given by the Gauss-Bonnet theorem in Eq. (2.104). The surface integral expresses the volume contribution, while the contour integral takes into account boundary contributions, notably those of conical defects. However, the curvature Eq. (4.64) of this state space presents two singularities located respectively in the vicinity of the points $\kappa = 0$ and $\kappa = \pi$. Thus, the quantum state space Eq. (4.60) contains two conical defects, which means that the associated topology will have to be determined taking these singularities into account. Applying the Fubini-Study metric Eq. (4.60), the Gauss-Bonnet equation Eq. (2.104) can be reformulated in the integral form

$$\int_0^{\pi}\int_0^{\eta^{max}} K(\mathbf{g}_{\kappa\kappa}\mathbf{g}_{\eta\eta})^{1/2} d\kappa d\eta + \Delta = 2\pi\chi(M), \tag{4.65}$$

where $\Delta$ represents the Euler integral at the boundary and takes into account contributions due to conical defects. An explicit rewriting of this relationship leads to

$$4\mathbf{s}\eta_{\max}(\mathcal{N}-1) + \Delta = 2\pi\chi(M). \tag{4.66}$$

Calculating the Euler characteristic therefore requires explicit evaluation of the Euler integral $\Delta$. To do this, we analyze the tensor metric expression near singularities. By expanding the metric expression Eq. (4.60) around the points $\kappa = 0$ and $\kappa = \pi$ to order two in $\kappa$, we obtain

$$d\mathbf{S}^2 = \frac{\mathcal{N}\mathbf{s}}{2}d\kappa^2 + 2\mathcal{N}(\mathcal{N}-1)^2\mathbf{s}^3\kappa^2 d\eta^2. \tag{4.67}$$

On the other hand, a cone of revolution with angle $2\kappa$ at its apex has a solid angle defined by $\Omega = 2\pi(1-\cos\kappa)$. This term quantifies the portion of solid angle swept by the system around the cone apex during its evolution. Exploiting the proximity of angular defects to singularities, we find the following approximation

$$2\pi\cos\kappa \approx \frac{\mathbf{S}(\eta_{\max})}{\mathbf{d}} = \frac{\sqrt{g_{ij}}\eta_{\max}}{\sqrt{g_{\alpha\alpha}}\kappa}, \tag{4.68}$$

where $\mathbf{S}(\eta_{\max})$ represents the distance covered by the system during the time period $\mathbf{t} = \eta_{\max}/\mathrm{J}$ around one of the points $\kappa = 0$ or $\pi$, and $\mathbf{d}$ corresponds to the distance between the system trajectory and the corresponding singularity. Thus, angular defects are expressed as

$$\Delta = 2\left[2\pi - \frac{\sqrt{g_{ij}}\eta_{\max}}{\sqrt{g_{\alpha\alpha}}\kappa}\right] = 2[2\pi - 2\mathbf{s}\eta_{\max}(\mathcal{N}-1)]. \tag{4.69}$$

The factor of 2 comes from the fact that we have two singular points. Inserting this expression into the Gauss-Bonnet equation Eq. (4.66), we find the Euler characteristic $\chi(M) = 2$. This means that the space of quantum states Eq. (4.60) of the spin-spin system has the topology of a sphere. In the next section, we will look at the geometric phases accumulated by the system as it evolves on this manifold of states.

### 4.3.2 Geometric phases accumulated by the N spin-s system

#### Geometric Phase in Arbitrary Evolution

In this section, we consider the $N$ spin-$s$ system evolving arbitrarily on the two-dimensional manifold Eq. (4.60). In this scenario, the geometric phase acquired by the evolving state Eq. (4.60) is given by:

$$\mathbf{\Phi}\mathrm{g}(t) = \arg\langle \Upsilon l|\Upsilon(t)\rangle - \operatorname{Im}\int_0^t \left\langle \Upsilon(t') \middle| \frac{\partial}{\partial t'} \middle| \Upsilon(t')\right\rangle dt', \tag{4.70}$$

which represents the global phase minus the dynamic phase. To evaluate the geometric phase, we first derive the global phase accumulated by the system. The transition probability amplitude (i.e., the overlap) between the initial state Eq. (3.80) and the evolving state Eq. (3.83) can be expressed as:

$$\langle \Upsilon_i|\Upsilon(t)\rangle = \left(1+\tan^2\frac{\kappa}{2}\right)^{-2\mathcal{N}\mathsf{s}} \sum_{\mathrm{m}1,\ldots,\mathrm{m}\mathcal{N}=-\mathbf{s}} e^{-i2\eta\sum_{1\leqslant i<j\leqslant\mathcal{N}}\mathsf{m}k\mathsf{m}l} \left[\prod_{\nu=1}^{\mathcal{N}}\left(\tan\frac{\kappa}{2}\right)^{2(\mathsf{s}+\mathsf{m}\nu)} C2\mathsf{s}^{\mathsf{s}+\mathsf{m}_\nu}\right]. \tag{4.71}$$

By transferring Eq. (4.71) to the first term on the right-hand side of Eq. (4.70), the total phase accumulated by the state of $N$ spin-$s$ particles is given by:

$$\Phi_{\mathrm{glob}} = -\arctan\left(\frac{\sum_{m_1,m_2,\ldots,m_N=-s}^{s}\sin\left(2\eta\sum_{1\leq k<l\leq N}m_km_l\right)\left(\prod_{v=1}^{N}\left(\tan\frac{\kappa}{2}\right)^{2(s+m_v)}C_{2s}^{s+m_v}\right)}{\sum_{m_1,m_2,\ldots,m_N=-s}^{s}\cos\left(2\eta\sum_{1\leq k<l\leq N}m_km_l\right)\left(\prod_{v=1}^{N}\left(\tan\frac{\kappa}{2}\right)^{2(s+m_v)}C_{2s}^{s+m_v}\right)}\right). \tag{4.72}$$

The global phase consists of two main components: the first is of a geometric nature, deeply linked to the geometric and topological characteristics of the system. This component arises from the explicit dependence of the global phase on the dynamical parameters $(\kappa, \eta)$ defining the evolution path on the quantum phase space Eq. (4.60). The second component is dynamic in nature, justified by the dependence of the global phase on the mean value of the Hamiltonian Eq. (3.77). Additionally, the global phase is defined modulo $2\pi$ and satisfies the periodic condition

$$\Phi_{\text{glob}}(\eta + 2\pi) = \Phi_{\text{glob}}(\eta). \tag{4.73}$$

On the other hand, the dynamic phase can be derived by substituting the evolved state Eq. (3.83) into the second term on the right-hand side of Eq. (4.70). Explicitly, we obtain

$$\Phi_{\text{dyn}} = -\eta\, \mathbf{s}^2\, \mathcal{N}(\mathcal{N}-1)\cos^2\kappa. \tag{4.74}$$

Consequently, the geometric phase accumulated by the spin-$s$ system Eq. (3.83), undergoing arbitrary evolution on the quantum state space Eq. (4.60), is given by

$$\begin{aligned}\Phi_g = -\arctan\left(\frac{\sum_{m_1,m_2,\dots,m_N=-s}^{s}\sin\left(2\eta\sum_{1\leq k<l\leq N} m_k m_l\right)\left(\prod_{v=1}^{N}\left(\tan\frac{\kappa}{2}\right)^{2(s+m_v)} C_{2s}^{s+m_v}\right)}{\sum_{m_1,m_2,\dots,m_N=-s}^{s}\cos\left(2\eta\sum_{1\leq k<l\leq N} m_k m_l\right)\left(\prod_{v=1}^{N}\left(\tan\frac{\kappa}{2}\right)^{2(s+m_v)} C_{2s}^{s+m_v}\right)}\right) \\ + \eta N(N-1)s^2\cos^2\kappa.\end{aligned} \tag{4.75}$$

Unlike the dynamic phase, the geometric phase Eq. (4.60) evolves non-linearly with time. This phase depends on the dynamic parameters $(\kappa, \eta)$, indicating its dependence on the shape of the evolution trajectory followed by the state of $N$ spin-$s$ particles Eq. (3.83) on the quantum phase space Eq. (4.60). The dependence on the initial parameters $(\mathcal{N}, \mathbf{s})$ can be interpreted as a reflection of the geometry of the state space. Let us now consider a special case in which we study the geometric phase accumulated by the state of spin-$s$ particles Eq. (4.60) over a very short period of time. Expanding the exponential factor in equation Eq. (4.71) to second order in $\eta$, we find

$$\langle\Upsilon_i|\Upsilon(t)\rangle \simeq 1 - \frac{\eta^2\,\mathbf{s}^2\,\mathcal{N}(\mathcal{N}-1)}{4}[\mathbf{s}(\mathcal{N}-1)(2\,\mathbf{s}\mathcal{N}\cos^4\kappa + \sin^2 2\kappa) + \sin^4\kappa] - i\eta\,\mathbf{s}^2\,\mathcal{N}(\mathcal{N}-1)\cos^2\kappa. \tag{4.76}$$

From this perspective, the geometric phase Eq. (4.75) reads as follows

$$\begin{aligned}\Phi_{\text{g}} \simeq -\arctan\left(\frac{4\eta\,\mathbf{s}^2\,\mathcal{N}(\mathcal{N}-1)\cos^2\kappa}{4-\eta^2\,\mathbf{s}^2\,\mathcal{N}(\mathcal{N}-1)[\mathbf{s}(\mathcal{N}-1)(2\,\mathbf{s}\mathcal{N}\cos^4\kappa + \sin^2 2\kappa) + \sin^4\kappa]}\right) \\ + \eta\,\mathbf{s}^2\,\mathcal{N}(\mathcal{N}-1)\cos^2\kappa.\end{aligned} \tag{4.77}$$

Interestingly, for $\eta = 0$, the system acquires no phase, as the evolutionary state Eq. (3.83) coincides with the initial state Eq. (3.79). Furthermore, in the thermodynamic limit $(\mathcal{N} \to \infty)$, the global phase disappears. This implies that the geometric phase and the dynamic phase are equal throughout the evolutionary process of the system, providing a pathway for experimental measurement of the geometric phase, as it can be expressed in terms of the Hamiltonian Eq. (3.77). A similar result holds for particles with large spin values $(\mathbf{s} \to \infty)$.

### Geometric Phase in Cyclic Evolution

At this stage, we investigate the geometric phase that emerges from the cyclic evolution of a spin-spin system. Within this framework, the evolved state described by equation Eq. (3.83) satisfies the cyclic condition $|\Upsilon(\mathbf{T})\rangle = e^{i\Phi_{\text{glob}}}|\Upsilon(0)\rangle$, where $\mathbf{T}$ represents the time required for a complete cyclic evolution. The geometric phase, known as the Aharonov-Anandan (AA) phase, accumulated by the system during this cyclic evolution (i.e., a closed path in the parameter space) is given in Eq. (2.48). Consequently, the AA geometric phase can be written as

$$\Phi_{\text{g}}^{\text{AA}} = \int_0^{\mathbf{T}} d\Phi_{\text{glob}} + i\int_0^{\mathbf{T}} \left\langle \Upsilon(t) \left| \frac{\partial}{\partial t} \right| \Upsilon(t) \right\rangle dt. \tag{4.78}$$

By substituting Eq. (4.72) and Eq. (4.74) into Eq. (4.78), the AA geometric phase accumulated by the spin-$s$ state Eq. (3.83), undergoing cyclic evolution on the quantum state manifold Eq. (4.60), takes the form

$$\Phi_{\text{g}}^{\text{AA}} = \eta_{\max}\mathcal{N}(\mathcal{N}-1)\,\mathbf{s}^2\cos^2\kappa. \tag{4.79}$$

It is important to note that the global phase integral cancels out ($\Phi_{\text{glob}} = 0$) due to the cyclic evolution of the $N$ spin-$s$ system. As a result, the AA geometric phase corresponds to the cyclic integral, over the interval $[0, \eta_{\max}]$, of the mean value of the Hamiltonian Eq. (3.77). In other words, during the parallel transport of the state vector Eq. (3.83) on the quantum phase space Eq. (4.60), the AA geometric phase acquired coincides with the dynamic phase. Calculating the AA geometric phase thus requires knowledge of the Ising Hamiltonian Eq. (3.77), which reflects the possibility of experimental measurement. This can be linked to the time-optimal evolution of the system by examining the evolution speed and the corresponding geodesic distance.

### 4.3.3 Geometric Representation of Entanglement

In this section, we undertake a comprehensive geometric analysis to explore the quantum correlations between the two spins. The primary objective is to establish a direct connection between the geometric properties of the system and measurable physical quantities such as entanglement levels and temporal evolution. To achieve this, we employ Eq. (3.103) and Eq. (4.60), which define the metric tensor governing the space of two-spin states as a function of the I-concurrence. We begin with the following expression for the squared distance in the state space of the two-spin system

$$d\mathbf{S}^2 = \frac{\mathbf{s}}{2\eta^2\mathcal{C}(2\mathbf{s}\eta - \mathcal{C})}\left[\frac{\eta^2}{2}d\mathcal{C}^2 - \eta\mathcal{C}d\mathcal{C}d\eta + \left\{\frac{\mathcal{C}^2}{2} + \eta'_{\max}\mathcal{C}(2\mathbf{s}\eta - \mathcal{C})\left[1 + (4\mathbf{s}-1)\left(1 - \frac{\eta'_{\max}\mathcal{C}}{2\mathbf{s}\eta^2}\right)\right]\right\}d\eta^2\right]. \tag{4.80}$$

This formulation emphasizes the interplay between the geometry of the state space, the degree of entanglement, and the system's time evolution. It facilitates a meticulous investigation of various geometric attributes, including the state-space structure, the geometric phase, the evolution rate, and the geodesic separation between entangled states. Notably, quantum entanglement induces a dimensional reduction of the state space, as exemplified by the case of two-spin states sharing an identical entanglement degree, which collectively form a one-dimensional curved manifold. The metric tensor characterizing this manifold is given by:

$$d\mathbf{S}^2 = \frac{\mathbf{s}}{2\eta^2\mathcal{C}(2\mathbf{s}\eta - \mathcal{C})}\left\{\frac{\mathcal{C}^2}{2} + \eta'_{\max}\mathcal{C}(2\mathbf{s}\eta - \mathcal{C})\left[1 + (4\mathbf{s}-1)\left(1 - \frac{\eta'_{\max}\mathcal{C}}{2\mathbf{s}\eta^2}\right)\right]\right\} d\eta^2. \tag{4.81}$$

This result underscores the pivotal role of entanglement in shaping the geometry of the state space and opens the door to an analysis of its curvilinear properties. By substituting Eq. (3.103) into equation Eq. (4.64), we express the curvature in terms of the I-concurrence:

$$K = \frac{2}{\mathbf{s}}\left[2 - \frac{(4\mathbf{s}-1)(1-\eta\frac{\mathfrak{C}}{\mathfrak{C}_{\max}}) + 2\mathbf{s} + 1}{\left[(4\mathbf{s}-1)(1-\eta\frac{\mathfrak{C}}{\mathfrak{C}_{\max}}) + 1\right]^2}\right]. \tag{4.82}$$

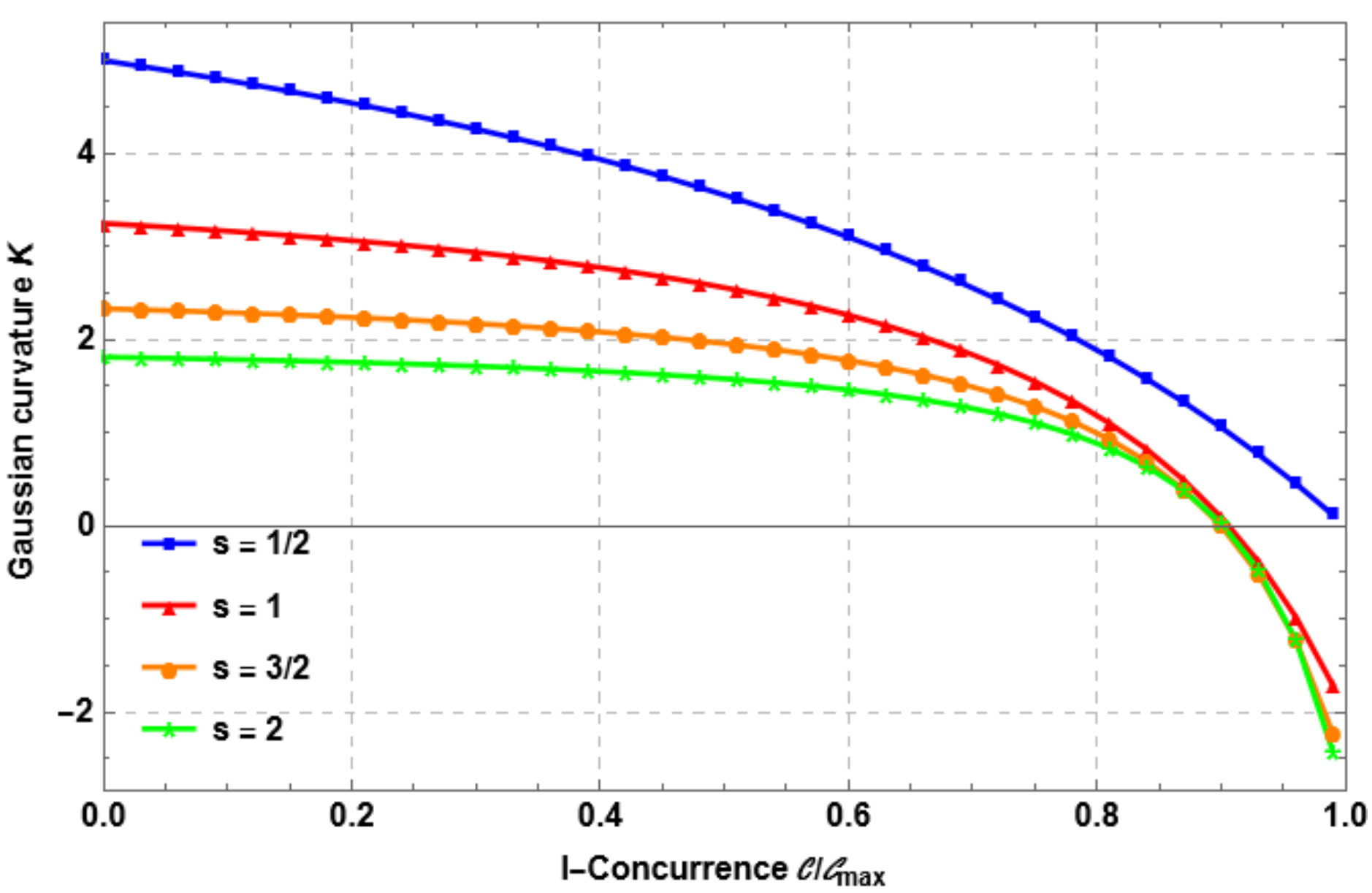

Figure 4.7: G-curvature (4.82) plotted against I-concurrence (3.103) for various spin values, with $\tilde{\eta} = 1$.

This equation confirms that the geometry of the state space is highly sensitive to quantum correlations between the two spins. To further illustrate this phenomenon, we have plotted the behavior of $K$ as a function of I-concurrence for different spin values with $\tilde{\eta} = 1$, as shown in Fig. (4.7). The results reveal that increasing the entanglement between the spins corresponds to a decrease in the curvature of the state space. Furthermore, for $\mathbf{s} > 1/2$, the curvature $K$ becomes negative, indicating that enhanced quantum correlations compress the state space. Non-entangled states ($\mathcal{C} = 0$) reside in regions of maximal curvature, whereas maximally entangled states ($\mathcal{C} = \mathcal{C}_{\max}$) appear in regions of minimal curvature. The maximum curvature values are given by

$$K_{\max} = \frac{2}{\mathbf{s}}\left[2 - \frac{3}{8\mathbf{s}}\right], \tag{4.83}$$

while the minimum curvature values are:

$$K_{\min} = \frac{2}{\mathbf{s}}\left[2 - \frac{(4\mathbf{s}-1)(1-\bar{\eta}) + 2\mathbf{s}+1}{\left[(4\mathbf{s}-1)(1-\bar{\eta})+1\right]^2}\right]. \tag{4.84}$$

These findings further solidify the central role of entanglement in characterizing the quantum phase space of the system. The established correlation between entanglement and state-space geometry enables valuable insights into the underlying geometric structure of this space. Additionally, in the asymptotic limit $\mathbf{s} \to \infty$, the curvature $K$ vanishes, signifying that the state space transitions into a flat geometry. In this context, it is also pertinent to investigate the interplay between the geometric phase and entanglement. By substituting Eq. (3.103) into equation Eq. (4.77), we derive an expression for the geometric phase acquired by the two-spin system as a function of I-concurrence

$$\Phi_g = 2\eta\mathbf{s}^2\left(1 - \bar{\eta}\frac{\mathcal{C}}{\mathcal{C}_{\max}}\right) - \arctan\left(\frac{4\eta\,\mathbf{s}^2\left(1 - \bar{\eta}\,\frac{\mathcal{C}}{\mathcal{C}_{\max}}\right)}{2 - \eta^2\mathbf{s}^2\left(2\mathbf{s}-1\right)\left[\left(2\mathbf{s}-1\right)\bar{\eta}^2\frac{\mathcal{C}^2}{\mathcal{C}_{\max}^2} - 4\mathbf{s}\bar{\eta}\frac{\mathcal{C}}{\mathcal{C}_{\max}} + 4\mathbf{s}^2\right]}\right). \tag{4.85}$$

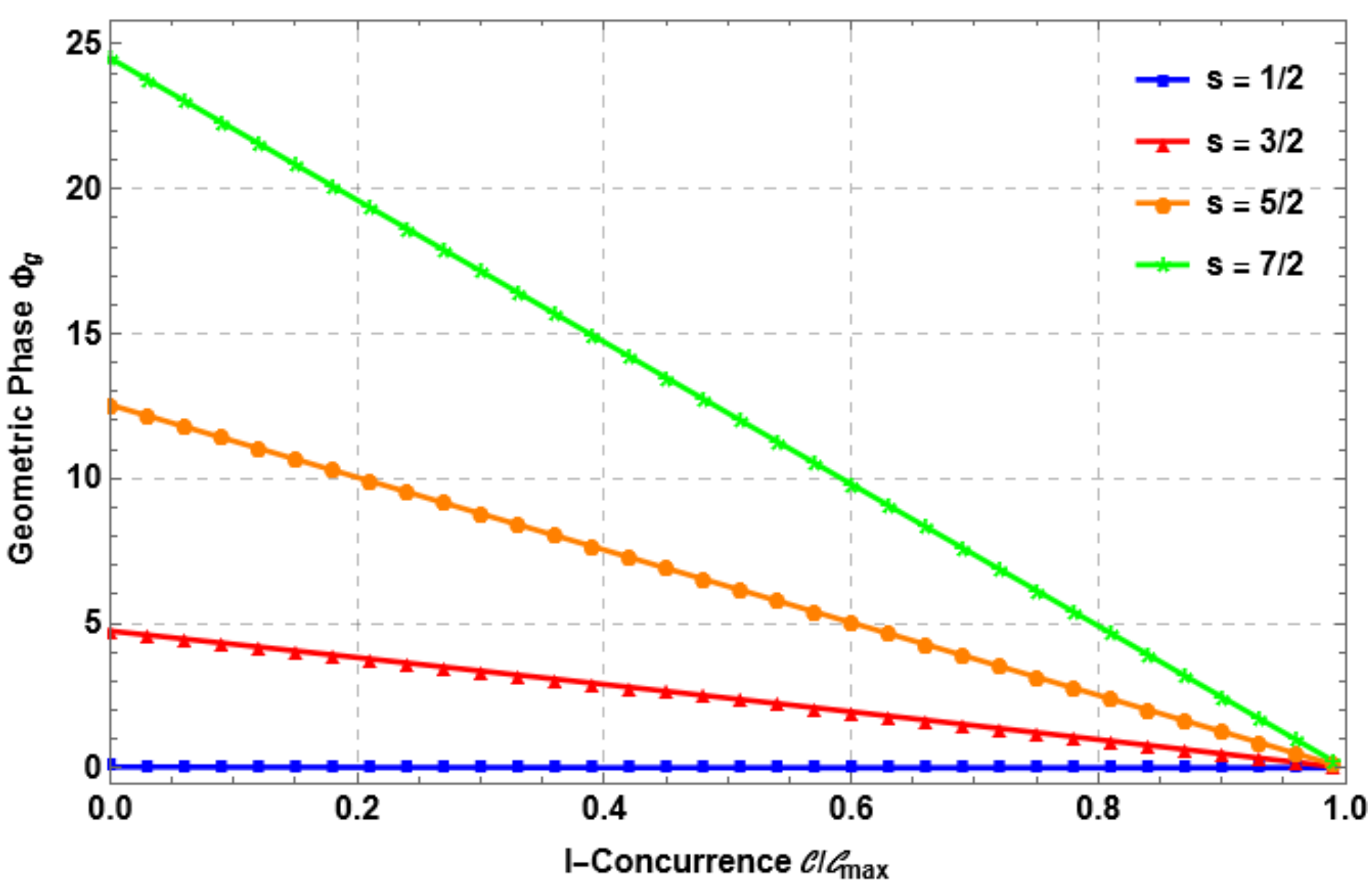

Figure 4.8: Geometrical phase (4.85) plotted against I-concurrence (3.103) for various spin values, with $\bar{\eta} = 1$.

Fig. (4.8) provides a detailed depiction of the dependence of the geometric phase on quantum entanglement for various spin values, with the parameter $\bar{\eta} = 1$. The figure clearly demonstrates

that the geometric phase exhibits an affine decrease as entanglement increases. This behavior can be understood by considering the expression of the geometric phase, where the dynamical phase contribution plays a predominant role over the global phase. Non-entangled states, which correspond to the maximum possible geometric phase, are the most favorable for accumulating this phase, whereas maximally entangled states exhibit minimal accumulation. Intermediate states occupy a transitional position between these two limiting cases. From this analysis, we conclude that quantum entanglement actively contributes to the reduction of the geometric phase throughout the evolution in state space. This finding aligns with previous results indicating that entanglement leads to a decrease in the curvature $K$ of state space [76], a fundamental quantity from which the geometric phase directly emerges. Consequently, the reduction of curvature induced by entanglement naturally translates into a diminished geometric phase. Finally, we observe that an increase in the spin value enhances the magnitude of geometric phase variation, whether as a gain or a loss. This dependence suggests a mechanism for controlling the evolution of the geometric phase by tuning both the degree of entanglement and the intrinsic spin of the system, offering potential avenues for applications in quantum information and quantum control.

## 4.4 Summary

This chapter develops a geometrical and topological perspective on quantum spin systems, building upon the dynamical foundations of the previous analysis. It begins with the two-spin-1/2 system under the anisotropic Heisenberg model, where the geometry of the joint state space is characterized via the FubiniStudy metric. The study details how evolution paths acquire geometric phases, such as Berry and AharonovAnandan phases, and how these phases depend on anisotropy and external fields. Geometric measures of entanglement are introduced, linking curvature properties of the state manifold to the degree of quantum correlations.

The treatment is then extended to N spin-1/2 particles interacting through the all-range Ising model. Here, collective geometric phases emerge from symmetric global interactions, and curvature analysis reveals how entanglement modifies the topology of accessible state spaces. The relationship between concurrence and geometric phase is examined in depth, showing that entanglement not only affects the magnitude of accumulated phases but also encodes information about the geometry of the evolution path. These results highlight the role of geometric quantities as diagnostic tools for correlation structures in many-body quantum systems.

Finally, the chapter generalizes to N interacting spin-$s$ particles under the Ising model, introducing the geometric and topological description of higher-dimensional state manifolds. The analysis covers phase accumulation under both cyclic and non-cyclic evolutions, curvatureentanglement relationships, and the impact of spin dimensionality on geometric invariants. Across all cases, the integration of Riemannian, symplectic, and topological concepts provides a unified language for characterizing quantum correlations. This framework reveals how geometry serves not merely as a descriptive tool but as a fundamental component in understanding and classifying the structure and dynamics of quantum spin systems.

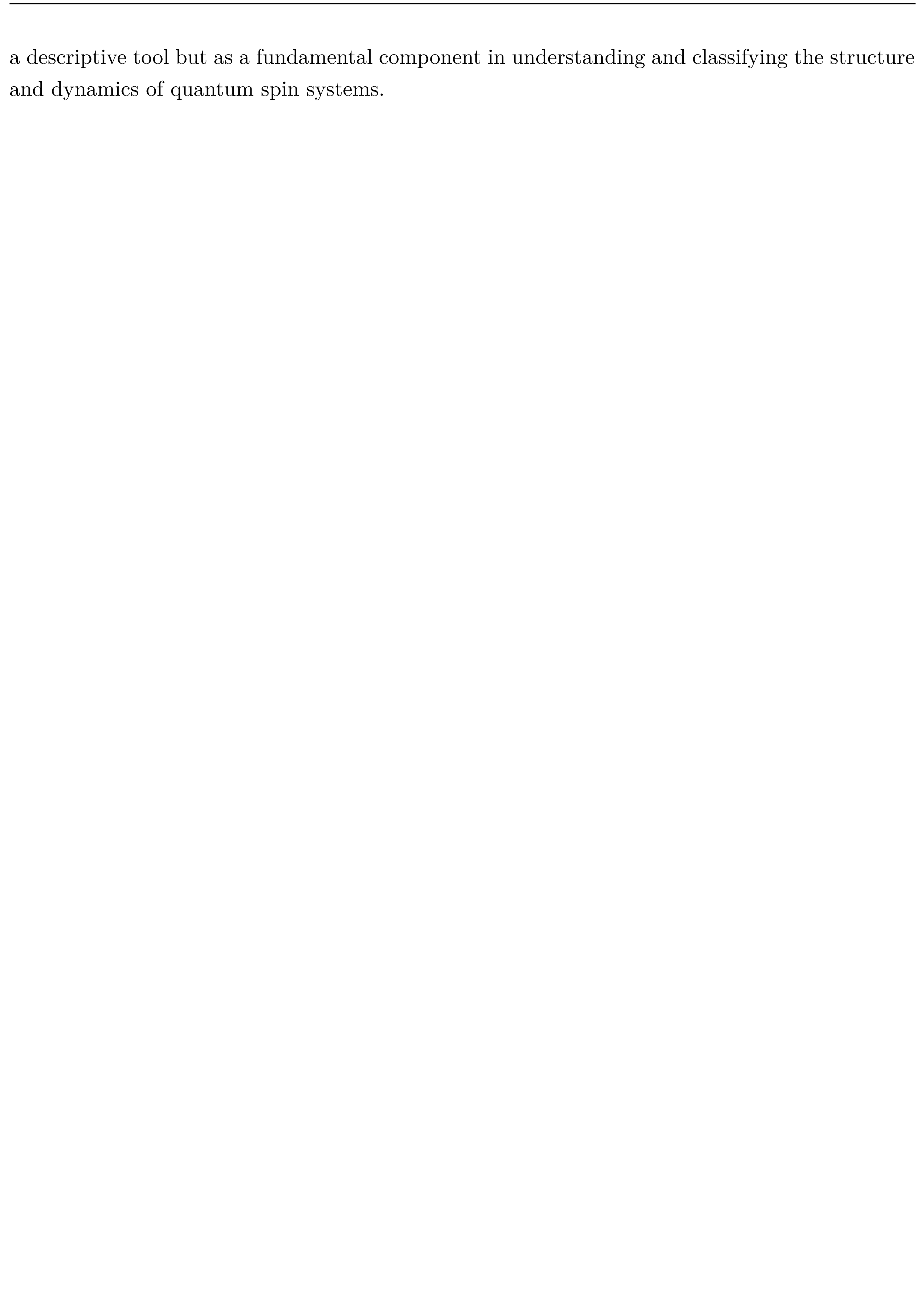

# General conclusion

In this work, we have presented a thorough study of the relationship between geometry and quantum dynamics, with particular emphasis on systems composed of interacting spin particles. Our aim was to elucidate how the mathematical structures of classical mechanics, particularly those grounded in symplectic geometry, find their natural and profound extensions in quantum theory, especially when formulated within the projective Hilbert space. By examining quantum systems from a geometric perspective, we have revealed not only the deep connections between classical and quantum formalisms but also new ways of understanding quantum evolution, entanglement, and coherence.

We began by establishing a solid foundation in both classical and quantum mechanics. The classical phase space, characterized by its symplectic structure, was introduced as a model for understanding the dynamical evolution of systems through position and momentum coordinates. Hamiltonian mechanics and the importance of symmetries provided insight into the conservation laws and trajectories of classical particles. We then transitioned to the quantum domain, where the Hilbert space replaces the classical phase space and where quantum states correspond to rays in a complex vector space. In this setting, physical quantities become Hermitian operators, and quantum state evolution is described by unitary dynamics. This transition laid the groundwork for interpreting quantum systems using tools adapted from classical geometry, reformulated within the framework of quantum theory.

We then introduced the projective Hilbert space as the proper configuration space for pure quantum states. This space is endowed with rich geometric structures, including the Fubini-Study metric, which measures the degree to which states can be distinguished, and a symplectic form that governs their unitary evolution. These geometric tools not only offer elegant formulations of quantum kinematics and dynamics but also enable the extraction of deeper physical insights, particularly in the analysis of quantum interference, coherence, and control. The appearance of geometric phasessuch as the Berry phaseillustrates how quantum systems accumulate global information about their paths in parameter space, independently of the specific dynamical details. This geometrically induced phenomenon has far-reaching implications in fields including quantum computation, condensed matter, and coherent control.

The heart of our investigation focused on spin systems, beginning with a detailed analysis of two-spin systems governed by the anisotropic Heisenberg interaction. We explored their dynamical evolution, quantum speed, and entanglement production over time. These systems served as minimal yet nontrivial models to probe fundamental quantum behaviors and test the applicability of geometric methods. We found that the interplay between interaction strength, anisotropy, and initial state configuration significantly influences entanglement evolution and optimal-time trajectories. Moreover, the brachistochrone problem revealed how geometric constraints determine the shortest duration of quantum evolution between two given states under fixed energy resources.

Building on these foundations, we extended our study to many-body spin systems. For $N$ interacting spin-$\frac{1}{2}$ particles described by the all-to-all Ising model, we analyzed the growth of entanglement and the corresponding geometric properties of the system's trajectory in state space. We then generalized our approach to systems of $N$ spin-$s$ particles, which opened the door to richer structures and a broader range of physical behaviors. These generalizations allowed us to test the scalability of geometric techniques and to identify how higher spin values influence the structure and evolution of entanglement and the emergence of quantum correlations.

The final chapter was devoted to a comprehensive geometric and topological analysis of the spin systems previously examined dynamically. We studied the geometric phase structures that arise during cyclic and non-cyclic evolutions and how these phases reflect the topological features of quantum state space. Furthermore, we investigated entanglement geometry by analyzing how different spin configurations trace non-trivial trajectories through the projective Hilbert space. This approach allowed us to highlight geometric signatures of entanglement, offering a complementary viewpoint to the more traditional algebraic measures. The topological and geometric tools used in this analysis, including holonomy, curvature, and quantum distance, demonstrated their power in describing and classifying quantum correlations, with direct implications for robust quantum control and fault-tolerant information processing.

Taken together, the results presented in this thesis offer a unified and geometrically grounded framework for understanding quantum dynamics, particularly in interacting and composite systems. By integrating tools from differential geometry, symplectic mechanics, and quantum information theory, we have constructed a versatile methodology capable of addressing both foundational questions and practical challenges in quantum science. The geometric interpretation of quantum phenomena not only enriches our grasp of the theory itself but also informs the design of new quantum protocols and technologies, including geometric quantum gates, entanglement-based sensors, and topological quantum computing architectures.

This work also opens several promising directions for future research. One possible extension is to examine open quantum systems within the same geometric framework, which would require incorporating dissipative dynamics and decoherence effects into the symplectic and metric structure of quantum state space. Another avenue lies in the investigation of quantum systems subject to time-dependent or stochastic Hamiltonians, where geometric tools could help characterize stability and controllability. Moreover, the geometric classification of entanglement in multipartite systemsbeyond spin-based modelsremains a rich and largely unexplored territory, with potential consequences for quantum communication and computational complexity.

In conclusion, this thesis underscores the central role that geometry plays in quantum theory, not just as a formal or mathematical device, but as a fundamental framework for representing quantum reality. From the formulation of physical laws to the engineering of future quantum devices, geometry provides both conceptual clarity and practical power. As quantum systems continue to grow in complexity and application, the geometric foundations developed here will remain critical for guiding innovation and understanding in the quantum domain.

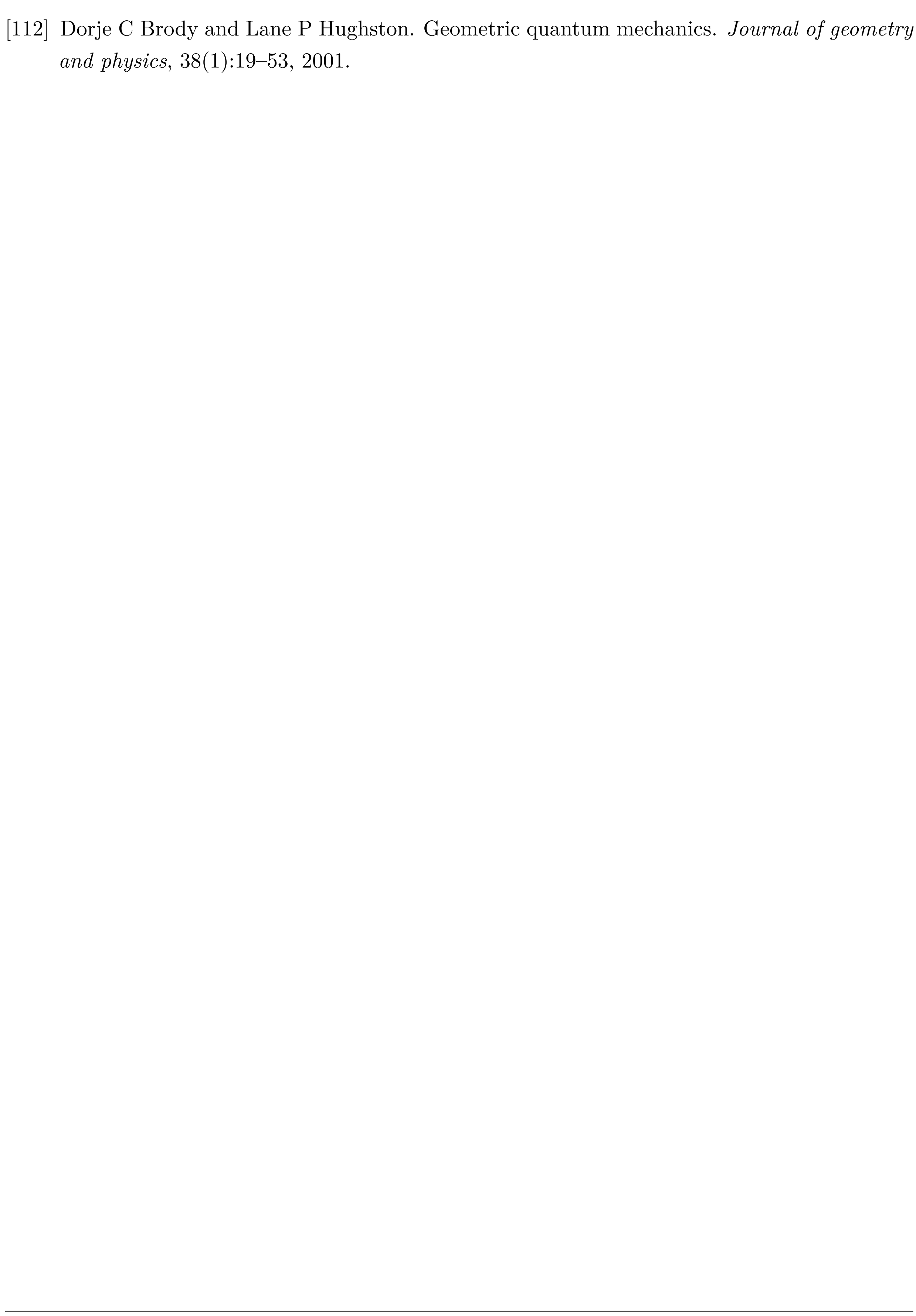

ROYAUME DU MAROC
Université Mohammed V - RABAT -
جامعة محمد الخامس – الرباط –
Faculté des sciences
كلية العلوم
CENTRE D'ÉTUDES DOCTORALES - SCIENCES ET TECHNOLOGIES

## *Résume*

*La présente thèse intitulée « Exploration des propriétés géométriques et dynamiques des systèmes de spin et de leur interaction avec l'intrication quantique », vise à explorer l'intrication et l'évolution quantique à travers une double lecture : géométrique et dynamique. Le premier volet traite de l'espace des phases classique et de son rôle central en mécanique hamiltonienne, en mettant en valeur l'importance des structures symplectiques dans la description des états mécaniques. L'étude met en évidence l'analogie formelle entre l'espace des phases classiques et l'espace de Hilbert en mécanique quantique. Le second volet est consacré à la description géométrique des états quantiques à travers la structure projective de l'espace de Hilbert. L'accent est mis sur l'interprétation géométrique de l'évolution quantique, notamment via la métrique de Fubini-Study, les structures symplectiques associées, ainsi que la phase géométrique acquise lors des évolutions unitaires. Les deux derniers volets sont consacrés à l'étude des systèmes de spins (à deux et plusieurs particules), sous différents modèles d'interaction (Heisenberg anisotropique et Ising), où l'on analyse à la fois la dynamique (vitesse d'évolution, intrication, problème du brachistochrone quantique) et les structures géométriques et topologiques associées aux états du système.*

---

## *Abstract*

*This thesis, entitled "Exploring the Geometric and Dynamical Properties of Spin Systems and Their Interplay with Quantum Entanglement", explores the quantum entanglement and evolution through both a geometric and dynamical perspective. The first part focuses on classical phase space and its central role in Hamiltonian mechanics, emphasizing the importance of symplectic structures in describing mechanical states. The study highlights the formal analogy between classical phase space and the Hilbert space used in quantum mechanics. The second part is devoted to the geometric description of quantum states through the projective structure of Hilbert space. Emphasis is placed on the geometric interpretation of quantum evolution, particularly via the Fubini-Study metric, associated symplectic structures, and the geometric phase acquired during unitary evolutions. The final two parts are dedicated to the study of spin systems (both two-body and many-body) under different interaction models (XXZ Heisenberg and all-range Ising). Both the dynamical aspects (evolution speed, entanglement, and the quantum brachistochrone problem) and the geometric and topological structures of the corresponding quantum states are analyzed.*

---

✉ Faculty of Sciences, Avenue Ibn Battouta, P.O. Box 1014 RP, Rabat – Morocco
☎ 00212 (05) 37 77 18 76 🖷 00212(05) 37 77 42 61; http://www. fsr.um5.ac.ma